# Mathematical modelling for acoustic microstreaming produced by a gas bubble undergoing asymmetric oscillations

**Claude Inserra**[1,†], **Cyril Mauger**[2], **Philippe Blanc-Benon**[2] **and Alexander A. Doinikov**[2]

[1]Univ Lyon, Université Claude Bernard Lyon 1, Centre Léon Bérard, INSERM, UMR 1032, LabTAU, F-69003 Lyon, France

[2]INSA Lyon, CNRS, École Centrale de Lyon, Université Claude Bernard Lyon 1, LMFA, UMR 5509, 69621 Villeurbanne, France

† Email address for correspondence: claude.inserra@inserm.fr

An exact solution is developed for the bubble-induced acoustic microstreaming in the case of a gas bubble undergoing asymmetric oscillations. The modeling is based on the decomposition of the solenoidal, first- and second-order, vorticity fields into poloidal and toroidal components. The result is valid for small amplitude bubble oscillations without restriction on the size of the viscous boundary layer $(2\nu/\omega)^{1/2}$ in comparison to the bubble radius. The nonspherical distortions of the bubble interface are decomposed over the set of orthonormal spherical harmonics $Y_n^m(\theta, \phi)$ of degree $n$ and order $m$. The present theory describes the steady flow produced by the nonspherical oscillations $(n, \pm m)$ that occur at a frequency different from that of the spherical oscillation as in the case of a parametrically-excited surface oscillation. The three-dimensional aspect of the streaming pattern is revealed as well as the particular flow signatures associated to different asymmetric oscillations.

# 1. Introduction

The high-frequency interfacial motion of ultrasound-driven gas bubbles can generate a steady flow in the surrounding fluid, called acoustic microstreaming. Interest in the bubble-induced streaming and the resulting stresses exerted on surrounding bodies is motivated by a variety of technological, chemical and biomedical applications. Oscillating bubbles close to or in contact with a wall can actuate the transport of particles within the viscous boundary layer at solid surfaces, with applications in the removal of contamination particles and surface cleaning (Kim *et al.* 2009; Reuter *et al.* 2017), the amplification of chemical treatments (Mason 1999), the handwashing and cleaning of sensitive surfaces (Birkin *et al.* 2016) and the micromixing of fluids (Wang *et al.* 2013). Cavitation microstreaming is also a propelling source for small water-floating objects and bubble-driven microrobots (Dijkink *et al.* 2006, Ahmed *et al.* 2015). These artificial, acoustically-driven microswimmers act as propulsion devices that can carry a payload, with a great potential for microfluidic applications and targeted drug delivery. The commonly-used geometry for these microrobots includes a gas bubble trapped into the robot body cavity, whose oscillations create an axisymmetric flow which orients and propels the microrobot (Zhou *et al.* 2022). The orientation of the propeller can be controlled by inducing one or several holes on armored bubbles (Bertin *et al.* 2015) by combining acoustic powering and magnetic steering (Aghakhani *et al.* 2020) or by designing gear-like bubble-based propellers in order to actuate rotational motion along closed trajectories (Mohanty *et al.* 2021). Biomedical applications based on bubble-induced flows include cell detachment (Ohl & Wolfrum 2003), the sorting and manipulation of biological materials (Volk *et al.* 2020), the lysis of vesicles (Marmottant *et al.* 2008) and ultrasound-mediated targeted drug delivery (Lajoinie *et al.* 2016; Pereno *et al.* 2018). The ultrasound-mediated delivery of a drug is based on the action of oscillating microbubbles nearby biological barriers that increase their permeability and allow drug and genes to penetrate into individual cells without serious consequence for the cell viability (Fan *et al.* 2014). The temporary permeabilization of biological barriers is caused by shear stresses exerted on cell tissues by the bubble-induced flows (Wu & Nyborg 2008), which are responsible for a "massage" effect on cells and the creation of transient pores on the cell membrane. Even if no consensus exists on the exact mechanism responsible for cell poration and the required bubble activity (collapsing/inertial regime or stably-oscillating regime), determining the bubble-induced flows and resulting shear stress is mandatory.

The first theoretical investigation of acoustic microstreaming was performed by Nyborg (1958). By investigating the near-boundary streaming induced by a compressible body (a gas bubble) resting on a surface, he showed how resonant bubbles produced a pronounced microstreaming in the surface vicinity. Davidson & Riley (1971) have considered the case of a spherical bubble oscillating laterally in an unbounded fluid. Their investigation covers a wide range of situations in respect to the orders of magnitude of two dimensionless parameters. The first one $\epsilon = U/R_0\omega$ is the ratio of the vibration amplitude of the velocity $U$ occuring at the angular frequency $\omega$ to the bubble radius $R_0$. The second parameter quantifies the ratio of the thickness of the Stokes layer to the bubble radius, $\gamma = (2\nu/\omega)^{1/2}/R_0 = \delta_\nu/R_0$, where $\nu$ is the kinematic viscosity and $\delta_\nu$ is the thickness of the oscillatory shear layer. A great theoretical achievement of their work is the introduction of a matching approximation between the solution within the inner boundary layer and the one in the outer boundary, performed in the case of large bubbles ($\gamma \ll 1$) driven at relatively low frequencies and with small amplitudes of lateral oscillations ($\epsilon \ll 1$). This approximation has been later used during decades by several authors, adding the contribution of small radial oscillation (Longuet-Higgins 1998), of small-amplitude axisymmetric shape oscillations in the case of the $n$th distortion mode with $n \gg 1$ (Maksimov 2007), and of any arbitrary combination of axisymmetric shape oscillations (Spelman & Lauga 2017). All the above-mentioned theoretical works are based on the matching of the inner/outer solution that assumes a small viscous penetration depth in comparison to the bubble radius. This assumption limits the findings to the case of large bubbles in low-viscosity fluids. Recently, Doinikov *et al.* (2019a) have overcome these limitations by calculating exactly the second-order mean flow induced by all possible interactions between axisymmetric shape oscillations (including the spherical and translational ones).

Yet the similarity between all these theoretical derivations is to consider initially (at rest) spherical bubbles far from any boundary. While this scenario has been recovered experimentally by using levitating, acoustically trapped bubbles (Cleve *et al.* 2019), the majority of experimental work on acoustic microstreaming is performed on substrate-attached microbubbles. The positional stability of the bubble is therefore ensured, hence facilitating the capture of the interface dynamics as well as the surrounding fluid motion. Marmottant *et al.* (2006) resolved the acoustic streaming surrounding a wall-attached bubble experiencing spherical oscillations and a translational one occurring perpendicularly to the wall. Tho *et al.* (2007) performed an extensive study of streaming patterns surrounding a substrate-attached bubble from the top view. In addition to the cases of varying translational and/or oscillating motion of the bubble interface, the authors have also extended the analysis to the case of shape oscillations, without mentioning the triggered shape instability. Marin *et al.* (2015) revealed the three-dimensional nature of the acoustic streaming flow surrounding a wall-attached cylindrical bubble by using an astigmatism particle tracking velocimetry (APTV) technique. Using the same APTV technique, Bolanos-Jimenez *et al.* (2017) captured the three-dimensional axisymmetric, "fountain-like" flow pattern surrounding a hemispherical bubble. Interestingly, the strength of the flow was used as a way of finding the bubble's lowest resonant frequency. When summarizing the literature, it becomes obvious that experimental and theoretical works on bubbles experiencing asymmetric (i.e. non-axisymmetric) oscillations are scarce, while the ease of triggering asymmetric deformations is facilitated by the contact line dynamics and the breaking of the spherical symmetry. Usually, the complexity of the asymmetric oscillations is disregarded experimentally, where undetermined shape modes are sometimes reported (Saint-Michel & Garbin 2020). The landscape of emergence of specific spherical harmonics for high-amplitude driven microbubbles was proposed by Fauconnier *et al.* (2020). The ease of triggering specific asymmetric modes near the minimum of the instability threshold, as well as the partitioning of the triggered modes within the instability region of existence for shape oscillations, was demonstrated. The only theoretical work discussing the emergence and triggering of asymmetric oscillations was performed by Maksimov (2020) in the case of a spherical bubble located in front of a wall. The splitting (partitioning) of the shape modes was theoretically recovered. Concerning the induced flows, a systematic study of the acoustic microstreaming surrounding a wall-attached bubble has been performed by Fauconnier *et al.* (2022). High-amplitude acoustic driving allowed the triggering of nonspherical shape oscillations, including asymmetric ones. The flow signatures were correlated to the bubble interface dynamics decomposed over the set of orthonormal spherical harmonics $Y_n^m(\theta, \phi)$ of degree $n$ and order $m$. The self-interaction of a given asymmetric shape oscillation results in various flower-like patterns whose number of lobes are associated to the degree $n$ and order $m$ of the triggered harmonics. Only top-view observations were performed and the three-dimensional nature of these asymmetric patterns has still to be determined.

The present paper provides the mathematical modeling of the second-order mean flow surrounding a bubble experiencing asymmetric oscillations at a frequency different from that of the spherical oscillations, so that the microstreaming comes from the interaction of the shape oscillation with itself (self-interaction). This case is relevant to parametrically-excited shape oscillations. Section 2 describes the derivation of the second-order mean flow, with a particular focus on the decomposition of the solenoidal, first- and second-order, vorticity fields into poloidal and toroidal components. The derivation is valid for small amplitude bubble oscillations with no restriction on the size of the viscous boundary layer $\delta_v = (2\nu/\omega)^{1/2}$ in comparison to the bubble radius. This means that our theory is valid for any value of the liquid viscosity $\nu$, which, in turn, means that our theory is valid for any value of the ratio $\delta_v/R_0$, where $R_0$ is the equilibrium bubble radius. We emphasize this fact because in many previous studies on acoustic streaming, the ratio $\delta_v/R_0$ is assumed small, which means that the liquid viscosity is assumed to be very low and/or $R_0$ to be big. Section 3 presents various numerical examples obtained by the proposed model, highlighting the three-dimensional nature of the streaming patterns, the reversal of the flow atop the bubble by means of the generation of an anti-fountain behavior and the possibility of creating flows with a strong azimuthal component when a travelling surface wave propagates at the interface of the bubble.

## 2. Theory

We consider a gas bubble surrounded by an infinite viscous incompressible liquid. We assume that the bubble, which is spherical at rest, undergoes asymmetric oscillations in response to an external acoustic excitation. Figure 1 shows a spherical coordinate system, originated at the equilibrium center of the bubble, which is used in our calculations. Our derivation follows the conventional procedure for calculating acoustic streaming. We assume that the amplitudes of the bubble oscillation modes are small compared to the equilibrium bubble radius. This assumption allows us to linearize the equations of liquid motion (Navier-Stokes equations) and to find their solutions, assuming the amplitudes of the bubble oscillation modes to be given quantities. These solutions give us a first-order velocity field generated by the bubble in the liquid. Then the equations of liquid motion are written with accuracy up to terms of the second order of smallness with respect to the first-order solutions and averaged over time. As a result, we obtain equations that describe the velocity field of acoustic streaming produced by the bubble.

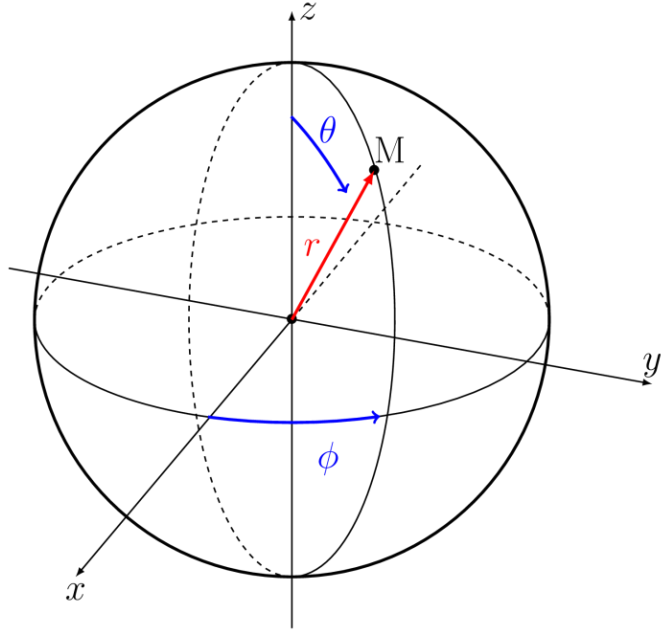

Figure 1. Coordinate system used in calculations.

### 2.1. *First-order solutions*

The surface of a bubble undergoing the oscillation modes $(n, m)$ and $(n, -m)$ can be represented by

$$r_s(\theta, \phi, t) = R_0 + e^{-i\omega t} s^{(n,m)} Y_n^m(\theta, \phi) + e^{-i\omega t} s^{(n,-m)} Y_n^{-m}(\theta, \phi), \qquad (2.1)$$

where $R_0$ is the bubble radius at rest, $\omega$ is the angular frequency of the bubble oscillations, $s^{(n,m)}$ and $s^{(n,-m)}$ are the complex amplitudes the modes $(n, m)$ and $(n, -m)$, respectively, and $Y_n^m(\theta, \phi)$ are spherical harmonics of degree $n$ and order $-n \leq m \leq n$, which are defined by (D1) in Appendix D. Here the azimuthal dependency as a complex exponential has been preferred to the cosine form (known as the real spherical harmonics) for the following reason. The cosine representation of the azimuthal part of the angular function implies an established, standing wave at the interface of the bubble. In order to allow for the most general case of two oppositely moving travelling waves at the interface of the bubble, the two components $(n, \pm m)$ are considered. When the amplitudes of the two components are equal, $s^{(n,m)} = s^{(n,-m)}$, then a stationary azimuthal wave is established. When they differ, a quasi-stationary wave exists which is composed of a partially stationary and a partially travelling component. In the general case, any three-dimensional bubble deformation can be decomposed over the set of orthonormal spherical harmonics, which means that several modes can coexist. The interaction of several modes oscillating at the same frequency (mixed-mode streaming) has been described as a dominant contribution to acoustic microstreaming in many situations, see, for example, Longuet-Higgins (1998) and Rallabandi *et al.* (2014). However, the consideration of two interacting modes $(n_1, m_1)$ and $(n_2 \neq n_1, m_2 \neq m_1)$ will lead to serious mathematical complexity in calculating the resulting flow. According to the work of Fauconnier *et al.* (2022), the possibility of triggering a single, well-identified, asymmetric mode was demonstrated. The associated flow, resulting from the interaction of the triggered asymmetric mode with itself, was captured. It is therefore worthy to consider a single asymmetric mode interacting with itself for the description of the induced acoustic microstreaming, as in (2.1).

The values of $s^{(n,\pm m)}$ and $\omega$ are considered as known quantities. They are assumed to be measured experimentally and serve as input data in our study. We also assume that $|s^{(n,\pm m)}|/R_0 \ll 1$, which allows us to linearize the equations of liquid motion. The linearized equations of a viscous incompressible liquid are given by (Landau & Lifshitz 1987)

$$\nabla \cdot \mathbf{v}_1 = 0, \tag{2.2}$$

$$\frac{\partial \mathbf{v}_1}{\partial t} = -\frac{1}{\rho}\nabla p_1 + \nu \Delta \mathbf{v}_1, \tag{2.3}$$

where $\mathbf{v}_1$ and $p_1$ are the first-order liquid velocity and pressure, respectively, $\rho$ is the constant liquid density, $\nu = \eta/\rho$ is the kinematic liquid viscosity and $\eta$ is the dynamic liquid viscosity.

The first-order velocity field generated by modes $(n, m)$ and $(n, -m)$ can be written as

$$\mathbf{v}_1 = \mathbf{v}_1^{(n,m)} + \mathbf{v}_1^{(n,-m)}, \tag{2.4}$$

where $\mathbf{v}_1^{(n,m)}$ and $\mathbf{v}_1^{(n,-m)}$ are the first-order liquid velocities generated by the modes $(n, m)$ and $(n, -m)$, respectively. Both $\mathbf{v}_1^{(n,m)}$ and $\mathbf{v}_1^{(n,-m)}$ obey (2.2) and (2.3).

The commonly-used procedure to solve (2.2) and (2.3) consists in decomposing the first-order liquid velocity $\mathbf{v}_1$ into the scalar $\varphi$ and the vector $\boldsymbol{\psi}$ velocity potentials using the Helmholtz decomposition $\mathbf{v}_1 = \nabla \varphi + \nabla \times \boldsymbol{\psi}$. In spherical coordinates associated with the basis set of vectors $(\mathbf{e}_r, \mathbf{e}_\theta, \mathbf{e}_\phi)$, the general form of these potentials is $\varphi = \varphi(r, \theta, \phi, t)$ and $\boldsymbol{\psi} = \psi_r(r, \theta, \phi, t)\mathbf{e}_r + \psi_\theta(r, \theta, \phi, t)e_\theta + \psi_\phi(r, \theta, \phi, t)e_\phi$. The Helmholtz decomposition is suitable for solving axisymmetric cases because the azimuthal invariance reduces the decomposition of the velocity field to only two unknows $\varphi(r,\theta,\phi,t)$ and $\psi_\phi(r,\theta,\phi,t)$ (Doinikov *et al.* 2019a). In the present problem, the general form of the scalar and vector potentials should be kept, leading to four unknown functions. Using the divergence-free property of the velocity field in (2.2), alternative decomposition can be proposed to reduce the number of unknown functions, such as the poloidal-toroidal decomposition sometimes called the Mie representation of a vector field. Indeed, Mie (1908) was the first to introduce this decomposition in the investigation of Maxwell's equations in a spherical geometry, after Lamb (1881) who introduced the toroidal part only in hydrodynamics. The application of the poloidal-toroidal decomposition of a solenoidal field mainly concerns terrestrial magnetism (Elsasser 1946; Bullard & Gellman 1954; Backus 1986). Chandrasekhar (1961) adapted this representation to treat a convective-flow stability problem in a spherical cavity. The validity of the poloidal-toroidal decomposition for any solenoidal fields has been demonstrated by Chadwick & Trowbridge (1967) in a bounded annular region, by Backus (1958) in a closed ball and by Padmavati & Amaranath (2002) for unbounded regions. In the context of bubble physics, Prosperetti (1977) proposed this decomposition for the vorticity field induced by bubble nonspherical oscillations in order to determine the equations of motion of the bubble interface accounting for nonspherical perturbations in viscous fluids. Therefore, we follow the approach of Prosperetti (1977) and calculate the curl of both sides of (2.3), which results in

$$\frac{\partial \boldsymbol{\omega}_1^{(n,m)}}{\partial t} = \nu \Delta \boldsymbol{\omega}_1^{(n,m)}, \quad (2.5)$$

where $\boldsymbol{\omega}_1^{(n,m)} = \nabla \times \boldsymbol{v}_1^{(n,m)}$ is called the vorticity of the velocity field $\boldsymbol{v}_1^{(n,m)}$. We then apply the poloidal-toroidal decomposition of the vorticity (Backus 1986),

$$\boldsymbol{\omega}_1^{(n,m)} = \boldsymbol{P}_1^{(n,m)} + \boldsymbol{T}_1^{(n,m)}, \quad (2.6)$$

where, in view of the equation of the bubble surface (2.1), the poloidal $\boldsymbol{P}_1^{(n,m)}$ and toroidal $\boldsymbol{T}_1^{(n,m)}$ fields are written by

$$\boldsymbol{P}_1^{(n,m)} = e^{-i\omega t}\nabla \times \nabla \times \left[\boldsymbol{e}_r P_{nm}(r)Y_n^m(\theta,\phi)\right], \quad (2.7)$$

$$\boldsymbol{T}_1^{(n,m)} = e^{-i\omega t}\nabla \times \left[\boldsymbol{e}_r T_{nm}(r)Y_n^m(\theta,\phi)\right], \quad (2.8)$$

with $\boldsymbol{e}_r = \boldsymbol{r}/r$ being the unit radial vector.

Substitution of (2.6) into (2.5) yields

$$\frac{\partial \boldsymbol{P}_1^{(n,m)}}{\partial t} + \frac{\partial \boldsymbol{T}_1^{(n,m)}}{\partial t} = \nu \Delta \boldsymbol{P}_1^{(n,m)} + \nu \Delta \boldsymbol{T}_1^{(n,m)}. \quad (2.9)$$

In view of the orthogonality of the poloidal and toroidal fields, (2.9) can be split into two equations,

$$\frac{\partial \boldsymbol{P}_1^{(n,m)}}{\partial t} = \nu \Delta \boldsymbol{P}_1^{(n,m)}, \qquad (2.10)$$

$$\frac{\partial \boldsymbol{T}_1^{(n,m)}}{\partial t} = \nu \Delta \boldsymbol{T}_1^{(n,m)}. \qquad (2.11)$$

Substituting (2.7) and (2.8) into (2.10) and (2.11) and using the identity

$$\Delta\left[\boldsymbol{e}_r F(r) Y_n^m(\theta,\phi)\right] = \boldsymbol{e}_r Y_n^m(\theta,\phi)\left[\frac{d^2 F(r)}{dr^2} - \frac{n(n+1)}{r^2} F(r)\right] + 2\nabla\left[\frac{F(r)}{r} Y_n^m(\theta,\phi)\right], \qquad (2.12)$$

where $F(r)$ is an arbitrary function, one obtains

$$-i\omega \nabla \times \nabla \times \left[\boldsymbol{e}_r P_{nm}(r) Y_n^m(\theta,\phi)\right] = \nu \nabla \times \nabla \times \left\{\boldsymbol{e}_r Y_n^m(\theta,\phi)\left[\frac{d^2 P_{nm}(r)}{dr^2} - \frac{n(n+1)}{r^2} P_{nm}(r)\right]\right\}, \qquad (2.13)$$

$$-i\omega \nabla \times \left[\boldsymbol{e}_r T_{nm}(r) Y_n^m(\theta,\phi)\right] = \nu \nabla \times \left\{\boldsymbol{e}_r Y_n^m(\theta,\phi)\left[\frac{d^2 T_{nm}(r)}{dr^2} - \frac{n(n+1)}{r^2} T_{nm}(r)\right]\right\}. \qquad (2.14)$$

It follows from (2.13) and (2.14) that $P_{nm}(r)$ and $T_{nm}(r)$ obey the following equations:

$$\left[\frac{d^2}{dr^2} - \frac{n(n+1)}{r^2} + k_v^2\right] P_{nm}(r) = 0, \qquad (2.15)$$

$$\left[\frac{d^2}{dr^2} - \frac{n(n+1)}{r^2} + k_v^2\right] T_{nm}(r) = 0, \qquad (2.16)$$

where $k_v = (1+i)/\delta_v$ is the viscous wavenumber and $\delta_v = \sqrt{2\nu/\omega}$ is the viscous penetration depth.

It is easy to check that both (2.15) and (2.16) are transformed to the Riccati-Bessel equation (Abramowitz & Stegun 1972) by multiplying by $r^2$. Solutions to the above equation are given by $k_v r z_n(k_v r)$, where $z_n$ is the spherical Bessel function of the first or second kind or the spherical Hankel function of the first or second kind. Since we are looking for solutions in the form of an outgoing wave, we get

$$P_{nm}(r) = a_{nm} k_v r h_n^{(1)}(k_v r), \qquad (2.17)$$

$$T_{nm}(r) = b_{nm} k_v r h_n^{(1)}(k_v r), \qquad (2.18)$$

where $a_{nm}$ and $b_{nm}$ are constants, called the linear scattering coefficients, that are determined by boundary conditions at the bubble surface, and $h_n^{(1)}$ is the spherical Hankel function of the first kind.

Substituting (2.7) and (2.8) into (2.6) and using (2.17) and (2.18), one obtains

$$\boldsymbol{\omega}_1^{(n,m)} = e^{-i\omega t}\left\{\nabla\times\nabla\times\left[\boldsymbol{e}_r a_{nm} k_v r h_n^{(1)}(k_v r) Y_n^m(\theta,\phi)\right] + \nabla\times\left[\boldsymbol{e}_r b_{nm} k_v r h_n^{(1)}(k_v r) Y_n^m(\theta,\phi)\right]\right\}, \quad (2.19)$$

It follows from (2.19) that $\boldsymbol{v}_1^{(n,m)}$ can be written by

$$\boldsymbol{v}_1^{(n,m)} = e^{-i\omega t}\left\{\nabla\times\left[\boldsymbol{e}_r a_{nm} k_v r h_n^{(1)}(k_v r) Y_n^m(\theta,\phi)\right] + \boldsymbol{e}_r b_{nm} k_v r h_n^{(1)}(k_v r) Y_n^m(\theta,\phi) - \nabla\varphi\right\}, \quad (2.20)$$

where a scalar function $\varphi$ is introduced in order to satisfy (2.2).

Substitution of (2.20) into (2.2) yields

$$\Delta\varphi = \nabla\cdot\left[\boldsymbol{e}_r b_{nm} k_v r h_n^{(1)}(k_v r) Y_n^m(\theta,\phi)\right]. \quad (2.21)$$

Equation (2.21) is the inhomogeneous Laplace equation, which is also called Poisson's equation. Since we are looking for a solution for $r > R_0$, the solution to (2.21) can be represented by

$$\varphi(r,\theta,\phi) = c_{nm} r^{-(n+1)} Y_n^m(\theta,\phi) + \varphi_{nm}(r) Y_n^m(\theta,\phi), \quad (2.22)$$

where the first term on the right-hand side is the solution to the Laplace equation ($\Delta\varphi = 0$), $c_{nm}$ being a constant coefficient, and the second term is a particular solution to Poisson's equation (2.21), $\varphi_{nm}(r)$ being a function to be found.

To find $\varphi_{nm}(r)$, we substitute (2.22) into (2.21) and calculate the right-hand side of (2.21). As a result, we obtain the following equation:

$$\left[\frac{d^2}{dx^2} + \frac{2}{x}\frac{d}{dx} - \frac{n(n+1)}{x^2}\right]\varphi_{nm}(x) = \frac{b_{nm}}{k_v}\left[3h_n^{(1)}(x) + x h_n^{(1)\prime}(x)\right] = \frac{b_{nm}}{k_v}\left[(n+3)h_n^{(1)}(x) - x h_{n+1}^{(1)}(x)\right], \quad (2.23)$$

where $x = k_v r$ and the prime denotes the derivative with respect to an argument in brackets.

To solve (2.23), we use the following identities (Abramowitz & Stegun 1972):

$$\frac{n}{x}h_n^{(1)}(x) - h_n^{(1)\prime}(x) = h_{n+1}^{(1)}(x), \quad (2.24)$$

$$\frac{n+1}{x}h_n^{(1)}(x) + h_n^{(1)\prime}(x) = h_{n-1}^{(1)}(x). \quad (2.25)$$

With the help of (2.24) and (2.25), we obtain

$$\left[\frac{d^2}{dx^2} + \frac{2}{x}\frac{d}{dx} - \frac{n(n+1)}{x^2}\right]\left[\alpha h_n^{(1)}(x) + \beta x h_{n+1}^{(1)}(x)\right] = (2\beta - \alpha)h_n^{(1)}(x) - \beta x h_{n+1}^{(1)}(x). \quad (2.26)$$

Comparison of (2.23) with (2.26) reveals that the solution to (2.23) is obtained by setting $\alpha = -(n+1)b_{nm}/k_v$ and $\beta = b_{nm}/k_v$,

$$\varphi_{nm}(x) = \frac{b_{nm}}{k_v}\left[xh_{n+1}^{(1)}(x) - (n+1)h_n^{(1)}(x)\right]. \tag{2.27}$$

It follows that the general solution to (2.21) is given by

$$\varphi(r,\theta,\phi) = \left\{c_{nm}r^{-(n+1)} + \frac{b_{nm}}{k_v}\left[k_v r h_{n+1}^{(1)}(k_v r) - (n+1)h_n^{(1)}(k_v r)\right]\right\}Y_n^m(\theta,\phi). \tag{2.28}$$

The coefficients $a_{nm}$, $b_{nm}$ and $c_{nm}$ are calculated in Appendix A. In the process of this calculation, the following boundary conditions at the bubble surface have been used: the condition that the normal component of the first-order liquid velocity is equal to the normal component of the bubble surface velocity and the condition that the tangential stress generated by the liquid motion vanishes at the bubble surface because the gas viscosity is much lower than the liquid viscosity. By using the results obtained in Appendix A, $\boldsymbol{v}_1^{(n,m)}$ is represented by

$$\boldsymbol{v}_1^{(n,m)} = v_{1r}^{(n,m)}\boldsymbol{e}_r + v_{1\theta}^{(n,m)}\boldsymbol{e}_\theta + v_{1\phi}^{(n,m)}\boldsymbol{e}_\phi, \tag{2.29}$$

$$v_{1r}^{(n,m)} = e^{-i\omega t}s^{(n,m)}V_n(r)Y_n^m(\theta,\phi), \tag{2.30}$$

$$v_{1\theta}^{(n,m)} = e^{-i\omega t}s^{(n,m)}W_n(r)\frac{\partial Y_n^m(\theta,\phi)}{\partial \theta}, \quad n \geq 1, \tag{2.31}$$

$$v_{1\phi}^{(n,m)} = e^{-i\omega t}s^{(n,m)}\frac{W_n(r)}{\sin\theta}\frac{\partial Y_n^m(\theta,\phi)}{\partial \phi}, \quad n,m \geq 1, \tag{2.32}$$

where the functions $V_n(r)$ and $W_n(r)$ are calculated by

$$V_n(r) = \frac{(n+1)\alpha_n}{r^{n+2}} + \frac{n(n+1)\beta_n h_n^{(1)}(k_v r)}{k_v r}, \tag{2.33}$$

$$W_n(r) = -\frac{\alpha_n}{r^{n+2}} + \frac{\beta_n}{k_v r}\left[(n+1)h_n^{(1)}(k_v r) - k_v r h_{n+1}^{(1)}(k_v r)\right], \tag{2.34}$$

and the coefficients $\alpha_n$ and $\beta_n$ are given by

$$\alpha_n = \frac{i\omega R_0^{n+2}[(2-n-n^2)h_n^{(1)}(\bar{x}) - \bar{x}^2 h_n^{(1)\prime\prime}(\bar{x})]}{(n+1)[\bar{x}^2 h_n^{(1)\prime\prime}(\bar{x}) - (n^2+3n+2)h_n^{(1)}(\bar{x})]}, \quad n \geq 0, \tag{2.35}$$

$$\beta_n = \frac{2i(n+2)\bar{x}\omega}{(n+1)[\bar{x}^2 h_n^{(1)\prime\prime}(\bar{x}) - (n^2+3n+2)h_n^{(1)}(\bar{x})]}, \quad n \geq 1, \tag{2.36}$$

where $\bar{x} = k_v R_0$. Note that (2.33) – (2.36) follow from (A8), (A9), (A22) and (A23) by setting $c_{nm} = s^{(n,m)} \alpha_n$ and $b_{nm} = s^{(n,m)} \beta_n$.

The first-order velocity generated by the mode $(n, -m)$ is obtained by replacing $m$ with $-m$ in (2.29) – (2.32),

$$\boldsymbol{v}_1^{(n,-m)} = v_{1r}^{(n,-m)} \boldsymbol{e}_r + v_{1\theta}^{(n,-m)} \boldsymbol{e}_\theta + v_{1\phi}^{(n,-m)} \boldsymbol{e}_\phi, \tag{2.37}$$

$$v_{1r}^{(n,-m)} = e^{-i\omega t} s^{(n,-m)} V_n(r) Y_n^{-m}(\theta, \phi), \tag{2.38}$$

$$v_{1\theta}^{(n,-m)} = e^{-i\omega t} s^{(n,-m)} W_n(r) \frac{\partial Y_n^{-m}(\theta, \phi)}{\partial \theta}, \quad n \geq 1, \tag{2.39}$$

$$v_{1\phi}^{(n,-m)} = e^{-i\omega t} s^{(n,-m)} \frac{W_n(r)}{\sin\theta} \frac{\partial Y_n^{-m}(\theta, \phi)}{\partial \phi}, \quad n, m \geq 1. \tag{2.40}$$

With the help of (D1) and (D3) from Appendix D, (2.38) – (2.40) are represented by

$$v_{1r}^{(n,-m)} = e^{-i\omega t} s^{(n,-m)} V_n(r) (-1)^m Y_n^{m*}(\theta, \phi), \tag{2.41}$$

$$v_{1\theta}^{(n,-m)} = e^{-i\omega t} s^{(n,-m)} W_n(r) (-1)^m \frac{\partial Y_n^{m*}(\theta, \phi)}{\partial \theta}, \quad n \geq 1, \tag{2.42}$$

$$v_{1\phi}^{(n,-m)} = e^{-i\omega t} s^{(n,-m)} \frac{W_n(r)}{\sin\theta} (-1)^m \frac{\partial Y_n^{m*}(\theta, \phi)}{\partial \phi}, \quad n, m \geq 1, \tag{2.43}$$

where the asterisk denotes the complex conjugate.

## 2.2. *Solutions of the equations of acoustic streaming*

The equations of acoustic streaming are given by (Doinikov *et al.* 2019a)

$$\nabla \cdot \boldsymbol{v}_E = 0, \tag{2.44}$$

$$\Delta \nabla \times \boldsymbol{v}_E = \frac{1}{\nu} \nabla \times \langle \boldsymbol{v}_1 \cdot \nabla \boldsymbol{v}_1 \rangle, \tag{2.45}$$

where $\boldsymbol{v}_E$ is the Eulerian streaming velocity, i.e. the time-averaged second-order liquid velocity, and $\langle \ \rangle$ means the time average.

Substitution of (2.4) into (2.45) yields

$$\Delta \nabla \times \boldsymbol{v}_E = \frac{1}{\nu} \nabla \times \langle \boldsymbol{v}_1^{(n,m)} \cdot \nabla \boldsymbol{v}_1^{(n,m)} + \boldsymbol{v}_1^{(n,-m)} \cdot \nabla \boldsymbol{v}_1^{(n,-m)} + \boldsymbol{v}_1^{(n,m)} \cdot \nabla \boldsymbol{v}_1^{(n,-m)} + \boldsymbol{v}_1^{(n,-m)} \cdot \nabla \boldsymbol{v}_1^{(n,m)} \rangle. \tag{2.46}$$

This equation can be divided into three equations,

$$\Delta \nabla \times \boldsymbol{v}_E^{(n,m)} = \frac{1}{\nu} \nabla \times \left\langle \boldsymbol{v}_1^{(n,m)} \cdot \nabla \boldsymbol{v}_1^{(n,m)} \right\rangle, \tag{2.47}$$

$$\Delta \nabla \times \boldsymbol{v}_E^{(n,-m)} = \frac{1}{\nu} \nabla \times \left\langle \boldsymbol{v}_1^{(n,-m)} \cdot \nabla \boldsymbol{v}_1^{(n,-m)} \right\rangle, \tag{2.48}$$

$$\Delta \nabla \times \boldsymbol{v}_E^{(\times)} = \frac{1}{\nu} \nabla \times \left\langle \boldsymbol{v}_1^{(n,m)} \cdot \nabla \boldsymbol{v}_1^{(n,-m)} + \boldsymbol{v}_1^{(n,-m)} \cdot \nabla \boldsymbol{v}_1^{(n,m)} \right\rangle, \tag{2.49}$$

where (2.47) describes the acoustic streaming generated by the mode $(n,m)$ alone, (2.48) describes the acoustic streaming generated by the mode $(n,-m)$ alone and (2.49) describes the acoustic streaming due to the interaction of the above modes (streaming due to cross terms).

Let us first consider (2.47). Since $\nabla \times \boldsymbol{v}_E^{(n,m)}$ is a divergence-free vector field, it can be represented by a poloidal-toroidal decomposition (Backus 1986),

$$\nabla \times \boldsymbol{v}_E^{(n,m)} = \nabla \times \nabla \times \left[ \boldsymbol{e}_r \sum_{k=1}^{\infty} \sum_{l=-k}^{k} P_{kl}^{(n,m)}(r) Y_k^l(\theta,\phi) \right] + \nabla \times \left[ \boldsymbol{e}_r \sum_{k=1}^{\infty} \sum_{l=-k}^{k} T_{kl}^{(n,m)}(r) Y_k^l(\theta,\phi) \right]$$

$$= \sum_{k=1}^{\infty} \sum_{l=-k}^{k} \left\{ \boldsymbol{e}_r \frac{k(k+1) P_{kl}^{(n,m)}(r)}{r^2} Y_k^l(\theta,\phi) + \frac{P_{kl}^{(n,m)\prime}(r)}{r} \left[ \boldsymbol{e}_\theta \frac{\partial Y_k^l(\theta,\phi)}{\partial \theta} + \frac{\boldsymbol{e}_\phi}{\sin\theta} \frac{\partial Y_k^l(\theta,\phi)}{\partial \phi} \right] \right.$$

$$\left. + \frac{T_{kl}^{(n,m)}(r)}{r} \left[ \frac{\boldsymbol{e}_\theta}{\sin\theta} \frac{\partial Y_k^l(\theta,\phi)}{\partial \phi} - \boldsymbol{e}_\phi \frac{\partial Y_k^l(\theta,\phi)}{\partial \theta} \right] \right\}. \tag{2.50}$$

Recall that the prime in the superscript denotes the derivative with respect to an argument in brackets. Substitution of (2.50) into (2.47) and the calculation of $\Delta \nabla \times \boldsymbol{v}_E^{(n,m)}$ yield

$$\Delta \nabla \times \boldsymbol{v}_E^{(n,m)} = \sum_{k=1}^{\infty} \sum_{l=-k}^{k} \left\{ \boldsymbol{e}_r \frac{k(k+1)}{r^2} \left[ P_{kl}^{(n,m)\prime\prime}(r) - \frac{k(k+1) P_{kl}^{(n,m)}(r)}{r^2} \right] Y_k^l(\theta,\phi) \right.$$

$$+ \frac{1}{r} \left[ P_{kl}^{(n,m)\prime\prime\prime}(r) - \frac{k(k+1) P_{kl}^{(n,m)\prime}(r)}{r^2} + \frac{2k(k+1) P_{kl}^{(n,m)}(r)}{r^3} \right] \left[ \boldsymbol{e}_\theta \frac{\partial Y_k^l(\theta,\phi)}{\partial \theta} + \frac{\boldsymbol{e}_\phi}{\sin\theta} \frac{\partial Y_k^l(\theta,\phi)}{\partial \phi} \right]$$

$$\left. + \frac{1}{r} \left[ T_{kl}^{(n,m)\prime\prime}(r) - \frac{k(k+1) T_{kl}^{(n,m)}(r)}{r^2} \right] \left[ \frac{\boldsymbol{e}_\theta}{\sin\theta} \frac{\partial Y_k^l(\theta,\phi)}{\partial \phi} - \boldsymbol{e}_\phi \frac{\partial Y_k^l(\theta,\phi)}{\partial \theta} \right] \right\} = \frac{1}{\nu} \nabla \times \left\langle \boldsymbol{v}_1^{(n,m)} \cdot \nabla \boldsymbol{v}_1^{(n,m)} \right\rangle.$$

$$\tag{2.51}$$

Equating the *r*-components of both sides of (2.51), one obtains

$$\sum_{k=1}^{\infty} \sum_{l=-k}^{k} \frac{k(k+1)}{r^2} \left[ P_{kl}^{(n,m)\prime\prime}(r) - \frac{k(k+1) P_{kl}^{(n,m)}(r)}{r^2} \right] Y_k^l(\theta,\phi) = \frac{1}{\nu} \boldsymbol{e}_r \cdot \left[ \nabla \times \left\langle \boldsymbol{v}_1^{(n,m)} \cdot \nabla \boldsymbol{v}_1^{(n,m)} \right\rangle \right]. \tag{2.52}$$

Calculating the curl of both sides of (2.51) and taking the *r*-component, one finds

$$\sum_{k=1}^{\infty}\sum_{l=-k}^{k}\frac{k(k+1)}{r^2}\left[T_{kl}^{(n,m)//}(r)-\frac{k(k+1)T_{kl}^{(n,m)}(r)}{r^2}\right]Y_k^l(\theta,\phi)=\frac{1}{\nu}\boldsymbol{e}_r\cdot\left[\nabla\times\nabla\times\left\langle\boldsymbol{v}_1^{(n,m)}\cdot\nabla\boldsymbol{v}_1^{(n,m)}\right\rangle\right]. \qquad (2.53)$$

Equations (2.52) and (2.53) make it possible to calculate $P_{kl}^{(n,m)}(r)$ and $T_{kl}^{(n,m)}(r)$ and hence $\boldsymbol{v}_E^{(n,m)}$. In view of the cumbersome nature of these calculations, they are performed in Appendix B. As a result, we obtain the components of the Eulerian streaming velocity in the following form:

$$v_{Er}^{(n,m)}(r,\theta)=\frac{1}{2\sqrt{\pi}}\sum_{k=1}^{\infty}\sqrt{2k+1}\left[T_{k0}^{(n,m)}(r)+\Phi_{k0}^{(n,m)/}(r)\right]P_k(\cos\theta), \qquad (2.54)$$

$$v_{E\theta}^{(n,m)}(r,\theta)=\frac{1}{2\sqrt{\pi}r}\sum_{k=1}^{\infty}\sqrt{2k+1}\Phi_{k0}^{(n,m)}(r)P_k^1(\cos\theta), \qquad (2.55)$$

$$v_{E\phi}^{(n,m)}(r,\theta)=-\frac{1}{2\sqrt{\pi}r}\sum_{k=1}^{\infty}\sqrt{2k+1}P_{k0}^{(n,m)}(r)P_k^1(\cos\theta), \qquad (2.56)$$

where $P_k$ is the Legendre polynomial of degree $k$, $P_k^1$ is the associated Legendre polynomial of the first order and degree $k$ and the functions $P_{k0}^{(n,m)}(r)$, $T_{k0}^{(n,m)}(r)$, $\Phi_{k0}^{(n,m)}(r)$ and $\Phi_{k0}^{(n,m)/}(r)$ are calculated by (B18), (B24), (B33) and (B45).

In the process of the calculation of $\boldsymbol{v}_E^{(n,m)}$ in Appendix B, we also calculate the Stokes drift velocity $\boldsymbol{v}_S^{(n,m)}$ (Longuet-Higgins 1998), whose components are given by

$$v_{Sr}^{(n,m)}(r,\theta)=\frac{|s^{(n,m)}|^2}{4\sqrt{\pi}\omega}\sum_{k=1}^{\infty}\sqrt{2k+1}A_{k0}^{nmnm}\,\text{Re}\left\{iV_n^*(r)\left[\frac{2n(n+1)-k(k+1)}{2r}W_n(r)-V_n^{'}(r)\right]\right\}P_k(\cos\theta),$$

(2.57)

$$v_{S\theta}^{(n,m)}(r,\theta)=\frac{|s^{(n,m)}|^2}{8\sqrt{\pi}\omega}\text{Re}\left\{iV_n^*(r)\left[\frac{W_n(r)}{r}-W_n^{'}(r)\right]\right\}\sum_{k=1}^{\infty}\sqrt{2k+1}A_{k0}^{nmnm}P_k^1(\cos\theta), \qquad (2.58)$$

$$v_{S\phi}^{(n,m)}(r,\theta)=\sum_{k=1}^{\infty}E_k^{(n,m)}(r)P_k^1(\cos\theta), \qquad (2.59)$$

where Re means "the real part of", the coefficients $A_{k0}^{nmnm}$ are calculated by (D20) and the function $E_k^{(n,m)}$ is defined by (B63). Note that (2.57) and (2.58) follow from (B54) and (B56) using that $f_{nm}(r)=s^{(n,m)}V_n(r)$ and $g_{nm}(r)=s^{(n,m)}W_n(r)$.

Knowing $\mathbf{v}_E^{(n,m)}$ and $\mathbf{v}_S^{(n,m)}$, we can calculate the Lagrangian streaming velocity produced by the mode $(n,m)$, which is defined by $\mathbf{v}_L^{(n,m)} = \mathbf{v}_E^{(n,m)} + \mathbf{v}_S^{(n,m)}$.

Due to symmetry, solutions to (2.48) are expressed in terms of the solutions to (2.47) as follows:

$$v_{Lr}^{(n,-m)}(r,\theta) = \frac{|S^{(n,-m)}|^2}{|S^{(n,m)}|^2} v_{Lr}^{(n,m)}(r, \pi - \theta), \tag{2.60}$$

$$v_{L\theta}^{(n,-m)}(r,\theta) = -\frac{|S^{(n,-m)}|^2}{|S^{(n,m)}|^2} v_{L\theta}^{(n,m)}(r, \pi - \theta), \tag{2.61}$$

$$v_{L\phi}^{(n,-m)}(r,\theta) = -\frac{|S^{(n,-m)}|^2}{|S^{(n,m)}|^2} v_{L\phi}^{(n,m)}(r, \pi - \theta). \tag{2.62}$$

To solve (2.49), by analogy with the solution of (2.47), $\nabla \times \mathbf{v}_E^{(\times)}$ is represented by a poloidal-toroidal decomposition,

$$\nabla \times \mathbf{v}_E^{(\times)} = \mathrm{Re}\left\{\nabla \times \nabla \times \left[\mathbf{e}_r \sum_{k=1}^{\infty}\sum_{l=-k}^{k} P_{kl}^{(\times)}(r) Y_k^l(\theta,\phi)\right] + \nabla \times \left[\mathbf{e}_r \sum_{k=1}^{\infty}\sum_{l=-k}^{k} T_{kl}^{(\times)}(r) Y_k^l(\theta,\phi)\right]\right\}$$

$$= \mathrm{Re}\sum_{k=1}^{\infty}\sum_{l=-k}^{k}\left\{\mathbf{e}_r \frac{k(k+1)P_{kl}^{(\times)}(r)}{r^2} Y_k^l(\theta,\phi) + \frac{P_{kl}^{(\times)\prime}(r)}{r}\left[\mathbf{e}_\theta \frac{\partial Y_k^l(\theta,\phi)}{\partial \theta} + \frac{\mathbf{e}_\phi}{\sin\theta}\frac{\partial Y_k^l(\theta,\phi)}{\partial \phi}\right]\right.$$

$$\left. + \frac{T_{kl}^{(\times)}(r)}{r}\left[\frac{\mathbf{e}_\theta}{\sin\theta}\frac{\partial Y_k^l(\theta,\phi)}{\partial \phi} - \mathbf{e}_\phi \frac{\partial Y_k^l(\theta,\phi)}{\partial \theta}\right]\right\}, \tag{2.63}$$

where the functions $P_{kl}^{(\times)}(r)$ and $T_{kl}^{(\times)}(r)$, by analogy with (2.52) and (2.53), obey the following equations:

$$\mathrm{Re}\sum_{k=1}^{\infty}\sum_{l=-k}^{k}\frac{k(k+1)}{r^2}\left[P_{kl}^{(\times)\prime\prime}(r) - \frac{k(k+1)P_{kl}^{(\times)}(r)}{r^2}\right]Y_k^l(\theta,\phi)$$

$$= \frac{1}{\nu}\mathbf{e}_r \cdot \left[\nabla \times \left\langle \mathbf{v}_1^{(n,m)} \cdot \nabla \mathbf{v}_1^{(n,-m)} + \mathbf{v}_1^{(n,-m)} \cdot \nabla \mathbf{v}_1^{(n,m)}\right\rangle\right], \tag{2.64}$$

$$\mathrm{Re}\sum_{k=1}^{\infty}\sum_{l=-k}^{k}\frac{k(k+1)}{r^2}\left[T_{kl}^{(\times)\prime\prime}(r) - \frac{k(k+1)T_{kl}^{(\times)}(r)}{r^2}\right]Y_k^l(\theta,\phi)$$

$$= \frac{1}{\nu}\mathbf{e}_r \cdot \left[\nabla \times \nabla \times \left\langle \mathbf{v}_1^{(n,m)} \cdot \nabla \mathbf{v}_1^{(n,-m)} + \mathbf{v}_1^{(n,-m)} \cdot \nabla \mathbf{v}_1^{(n,m)}\right\rangle\right]. \tag{2.65}$$

Equations (2.64) and (2.65) are solved in Appendix C. As a result, we obtain the components of $\boldsymbol{v}_E^{(\times)}$ in the following form:

$$v_{Er}^{(\times)}(r,\theta,\phi) = \mathrm{Re} \sum_{k=1}^{\infty} \sum_{l=-k}^{k} \left[ T_{kl}^{(\times)}(r) + \Phi_{kl}^{(\times)\prime\prime}(r) \right] Y_k^l(\theta,\phi), \qquad (2.66)$$

$$v_{E\theta}^{(\times)}(r,\theta,\phi) = \mathrm{Re} \sum_{k=1}^{\infty} \sum_{l=-k}^{k} \frac{\Phi_{kl}^{(\times)}(r)}{2r}$$

$$\times \left[ \sqrt{k(k+1)-l(l+1)} Y_k^{l+1}(\theta,\phi) e^{-i\phi} - \sqrt{k(k+1)-l(l-1)} Y_k^{l-1}(\theta,\phi) e^{i\phi} \right], \qquad (2.67)$$

$$v_{E\phi}^{(\times)}(r,\theta,\phi) = -\mathrm{Re} \sum_{k=1}^{\infty} \sum_{l=-k}^{k} \frac{il\Phi_{kl}^{(\times)}(r) e^{i\phi}}{r} \sqrt{\frac{(2k+1)(k-l)!}{(k+l)!}}$$

$$\times \sum_{s=1}^{[(k-l+2)/2]} \sqrt{\frac{(2k-4s+3)(k+l-2s)!}{(k-l-2s+2)!}} Y_{k-2s+1}^{l-1}(\theta,\phi), \qquad (2.68)$$

where the functions $T_{kl}^{(\times)}(r)$, $\Phi_{kl}^{(\times)}(r)$ and $\Phi_{kl}^{(\times)\prime}(r)$ are calculated by (C13), (C22) and (C33).

The components of the Stokes drift velocity produced by the interaction of the modes $(n,m)$ and $(n,-m)$ are given by

$$v_{Sr}^{(\times)}(r,\theta,\phi) = \mathrm{Re} \sum_{k=1}^{\infty} S_k^{(\times)}(r) \sum_{l=-k}^{k} D_{kl}^{nmnm} Y_k^l(\theta,\phi), \qquad (2.69)$$

$$v_{S\theta}^{(\times)}(r,\theta,\phi) = \frac{1}{2} \mathrm{Re} \Big\{ U_{nm}(r)$$

$$\times \sum_{k=1}^{\infty} \sum_{l=-k}^{k} D_{kl}^{nmnm} \left[ \sqrt{k(k+1)-l(l+1)} Y_k^{l+1}(\theta,\phi) e^{-i\phi} - \sqrt{k(k+1)-l(l-1)} Y_k^{l-1}(\theta,\phi) e^{i\phi} \right] \Big\}, \qquad (2.70)$$

$$v_{S\phi}^{(\times)}(r,\theta,\phi) = -\mathrm{Re} \Big\{ iU_{nm}(r) e^{i\phi} \sum_{k=1}^{\infty} \sum_{l=-k}^{k} l \sqrt{\frac{(2k+1)(k-l)!}{(k+l)!}} D_{kl}^{nmnm}$$

$$\times \sum_{s=1}^{[(k-l+2)/2]} \sqrt{\frac{(2k-4s+3)(k+l-2s)!}{(k-l-2s+2)!}} Y_{k-2s+1}^{l-1}(\theta,\phi) \Big\}, \qquad (2.71)$$

where the coefficients $D_{kl}^{nmnm}$ are calculated by (D27) and the functions $S_k^{(\times)}(r)$ and $U_{nm}(r)$ are defined by (C43) and (C44).

Knowing $\boldsymbol{v}_E^{(\times)}$ and $\boldsymbol{v}_S^{(\times)}$, we can calculate the corresponding Lagrangian streaming velocity: $\boldsymbol{v}_L^{(\times)} = \boldsymbol{v}_E^{(\times)} + \boldsymbol{v}_S^{(\times)}$.

Finally, summing up the three components of the acoustic streaming, we get the total flow:

$$\mathbf{v}_L^{(total)}(r,\theta,\phi) = \mathbf{v}_L^{(n,m)}(r,\theta) + \mathbf{v}_L^{(n,-m)}(r,\theta) + \mathbf{v}_L^{(\times)}(r,\theta,\phi). \qquad (2.72)$$

In the process of the calculations described in this section, we have applied the boundary conditions for the acoustic streaming at the bubble surface that assume that the normal velocity and the tangential stress produced by the Lagrangian streaming vanish at the equilibrium bubble surface.

It is worth noting that we do not provide expressions for the first- and second-order pressures because they are not used in the calculation of acoustic streaming. The above pressures are needed, for example, for the calculation of the acoustic radiation force on the bubble. However, this problem is beyond the scope of our paper. It requires an individual consideration somewhere else because, in particular, the calculation of the second-order pressure is not a trivial mathematical problem.

## 3. Results and discussion

### 3.1 Three classes of spherical harmonics

Let us first consider how the asymmetric shape deformations are described mathematically. In the general case, the shape of a deformed three-dimensional body can always be decomposed over the set of orthonormal spherical harmonics $Y_n^m(\theta,\phi)$ of degree $n$ and order $m$. These two indexes are related to the spatial evolution of the spherical harmonics along the spherical angular coordinates, the colatitude $\theta \in [0\ \pi]$ and the longitude $\phi \in [0\ 2\pi]$. For a given degree $n$, three classes of spherical harmonics are usually considered: the zonal harmonics when $m = 0 < n$, the tesseral harmonics when $0 < m < n$ and the sectoral harmonics when $m = n$. The three classes of spherical harmonics are represented in figure 2 for the degree $n = 5$.

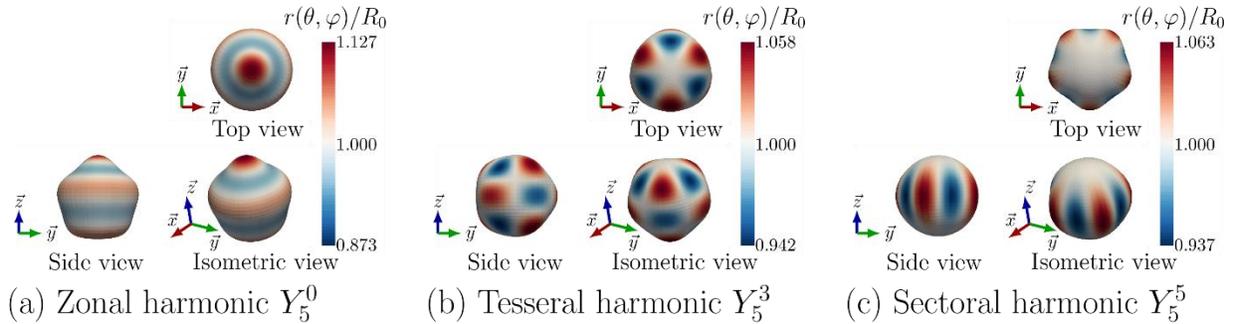

(a) Zonal harmonic $Y_5^0$  (b) Tesseral harmonic $Y_5^3$  (c) Sectoral harmonic $Y_5^5$

Figure 2. Numerical shape deformations of a bubble experiencing (a) the zonal harmonic $Y_5^0$, (b) the tesseral harmonic $Y_5^3$ and (c) the sectoral harmonic $Y_5^5$. Every isometric view of the spherical harmonics is displayed in top (upper) and side (lower) views.

In figure 2(a), the bubble undergoes the zonal harmonic $Y_5^0$. The bubble interface is axisymmetric, the bubble contour looks spherical from the top view and the shape from the side view is a Legendre polynomial ($m=0$). Figure 2(b) illustrates the case of the tesseral harmonic $Y_5^3$. The bubble interface is complex to describe, at first, as angular deviations from the sphere appear along both the elevation and azimuthal directions. A closer look reveals that the bubble interface possesses $n-m=2$ nodal lines along the elevation (see figure 2(b), side view), and $m=3$ meridian nodal lines (see figure 2(b), top view). Along the azimuthal direction, and equivalently along the elevation, two successive extrema are out of phase from both side of the nodal lines. Figure 2(c) displays a bubble experiencing the sectoral harmonic $Y_5^5$. The top-view contour of a sectoral harmonic of degree $n$ corresponds to a $n$-lobe deformation and is easily recognizable. The side view of a sectoral harmonic is close to the spherical shape. In fact, the side-view contour resembles a bell-shape function, with a higher amplitude at the equator $\theta = \pi/2$ and a decreasing oscillation amplitude when reaching the poles. It is worth mentioning that the so-called top and side views are experimentally ruled by the existence of a preferential direction for the triggering of a given shape oscillation at the bubble interface. For instance, in the case of a substrate-attached deformable body, the normal to the substrate is always the preferential direction for the triggering of asymmetric oscillations, as observed in Fauconnier et al. (2022) for wall-attached bubbles and in Chang et al. (2015) for sessile drops.

In the present derivation of the bubble-induced microstreaming, self-interacting asymmetric oscillations are only considered. We recall here that the main mechanism for generating shape oscillations at the bubble interface is the Faraday instability occurring at the subharmonic of the driving frequency. Therefore, interactions of shape oscillations with the radial (spherical) oscillations of the bubble cannot lead to fluid mean flows, as the radial mode oscillates at the driving frequency. In addition, near the instability threshold of a given shape oscillation, we also assume that a single instability is triggered without considering the energy transfer to secondary shape oscillations, as theoretically discussed by Shawn (2006) and experimentally observed by Guédra *et al.* (2016). An analysis of the intermodal interactions and their impact on the microstreaming pattern is performed in Regnault *et al.* (2021). The analysis of self-interacting axisymmetric modes is well-known theoretically (Inserra *et al.* 2020) and has been observed from a side view in a plane containing the bubble symmetry axis $z$ using acoustically-trapped bubbles, being far from any boundary (Cleve *et al.* 2019). The case of self-interacting asymmetric oscillations was considered by Fauconnier *et al.* (2022). The authors investigated experimentally the top-view microstreaming flow induced by a wall-attached bubble experiencing various asymmetric oscillations. The predominant shape oscillations were selected using a spectral analysis of the displacement of the bubble interface, and the modal content of the bubble could safely be associated to the observed fluid pattern. For some experimental cases, the secondary excited shape oscillations were so weak that the microstreaming pattern could be confidently related to the self-interaction of the main triggered surface oscillation. Such patterns resulting from self-interacting asymmetric oscillations are reproduced in figure 3 adopted from Fauconnier *et al.* (2022). Figure 3(a) shows the top-view microstreaming pattern induced by the self-interacting zonal harmonic $Y_4^0$. The pattern is exclusively radial, since the oscillating bubble interface exhibits no azimuthal dependence (figure 2(a), top view). Figure 3(b) represents the top-view microstreaming pattern induced by the self-interacting tesseral harmonic $Y_4^2$. The pattern is characterized by $4m = 8$ lobes, arranged by pairs around the $z$ axis orthogonal to the image, each pair being located between two displacement nodes of the bubble interface (the location of the four meridian lines is clearly visible through the light shades inside the bubble). Figure 3(c) shows the top-view microstreaming pattern induced by the self-interacting sectoral harmonic $Y_3^3$. The microstreaming pattern displays $4m = 12$ lobes that are also arranged in pairs. Because sectoral modes are devoid of nodal parallels and exhibit an azimuthal shape that corresponds to the $\cos(m\phi)$ function along the equator, $2m = 6$ extrema of displacements are observed on the bubble interface. Therefore, two vortices are expected between two successive nodal meridians, leading to the observed 12-lobe pattern.

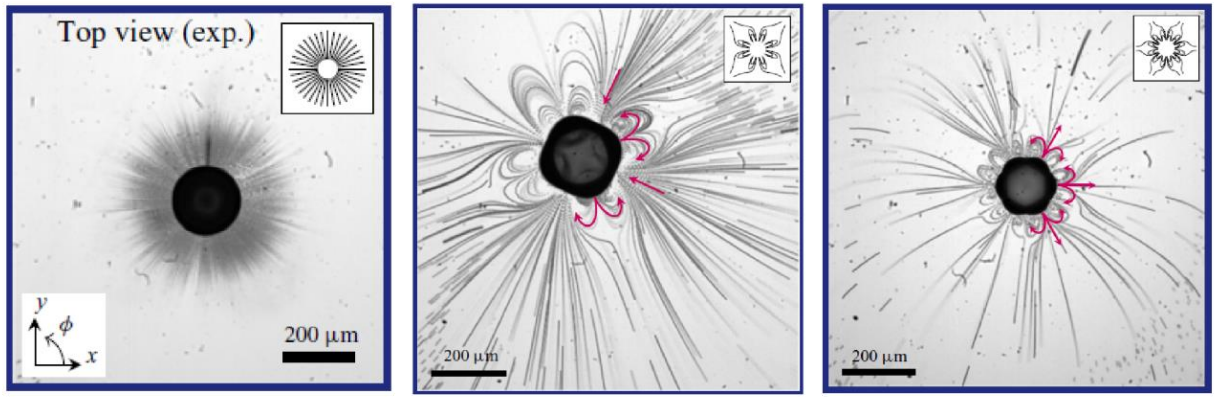

(a) Zonal harmonic $Y_4^0$  (b) Tesseral harmonic $Y_4^2$  (c) Sectoral harmonic $Y_3^3$

Figure 3. Microstreaming patterns associated to the three classes of spherical harmonics: (a) zonal, (b) tesseral and (c) sectoral harmonics, experienced by a wall-attached bubble and observed with microscope in a top-view configuration. Images are reproduced (see the copyright form) from Fauconnier *et al.* (2022).

For all these experimental microstreaming patterns, the top-view configuration avoids the analysis of the full three-dimensional signature of fluid flows induced by asymmetric oscillations, as the seeded particles tracking the fluid motion can escape from the focal plane of observation. The dependence in elevation for the above-mentioned pattern is therefore not ensured and can only be guessed using physical arguments. The three-dimensional nature of the fluid flow was investigated by Marin *et al.* (2015) and Bolanos-Jimenez *et al.* (2017) using an astigmatism particle tracking velocimetry (APTV) technique. However, the analysis was performed on a wall-attached bubble exhibiting an axisymmetric flow around the perpendicular to the wall. In the following, we illustrate the three-dimensional nature of the flow surrounding asymmetric bubbles using the present modeling.

### 3.2. *Signature for axisymmetric, zonal ($m = 0$) harmonics*

Axisymmetric shape oscillations are the widely investigated cases because of the ease of triggering them in various experimental configurations. The lack of the dependence of the zonal harmonics in the azimuthal direction significantly facilitates their mathematical analysis, as already discussed in the Introduction. Therefore, we first analyze the case of a zonal (axisymmetric) harmonic, namely the case ($n = 3, m = 0$). The bubble equilibrium radius is set to 73.8µm, as in the experimental case (figure 3(c)) displaying a shape oscillation of degree $n = 3$. This value is close to the resonant radius of the $n$-th shape oscillation, as derived by Lamb (1916):

$$R_{res}^n = \sqrt[3]{\frac{(n-1)(n+1)(n+2)\sigma}{\rho\omega_d^2/4}}, \qquad (2.73)$$

where $\sigma$ is the surface tension, $\rho$ is the liquid density and $\omega_d = 2\pi f_d$ is the angular frequency of acoustic driving. The value of $\omega_d$ differs from the frequency of the shape oscillation when the surface deformation is induced by a parametric instability. It should hence be noted that, in all the following numerical simulations of microstreaming induced by a shape oscillation, the angular frequency $\omega$ used in the theoretical derivation is assumed equal to half the driving frequency: here $\omega = \omega_d/2$. For the axisymmetric shape oscillations of degree $n = 3$, the resonant radius is $R_{res}^n = 65.9$ µm for an air bubble in water with $\rho = 1000$ kg/m³, $\sigma = 0.0727$ N/m at $f_d = 30.5$ kHz. The value of the frequency corresponds to the experimental driving frequency in Fauconnier *et al.* (2022). We recall that the resonant radius (2.73) is theoretically independent of order $m$ of the spherical harmonics. The amplitudes of the two modes $s^{(n,\pm m)}$ are set equal to 15µm ($|s^{(n,\pm m)}/R_0| \sim 0.2$), so that the case of a stationary surface oscillation is investigated. It is worth mentioning that we have chosen the value of the oscillation amplitude that corresponds roughly to measured amplitudes of asymmetric oscillations in Fauconnier *et al.* (2022). This value leads to not a very small value of the dimensionless parameter $|s^{(n,\pm m)}/R_0|$, which, however, in practice is small enough to ensure the convergence of the first-order solutions (see Section 2.1). What is more, from the point of view of the qualitative behaviour of acoustic streaming, the value of the oscillation amplitude is of no importance in our simulations because mathematically the streaming velocity is directly proportional to the oscillation amplitude squared. Hence, a change in the oscillation amplitude does not change, qualitatively, the obtained results. Numerical simulations are performed for an asymmetric deformation oscillating at half the driving frequency, i.e. for $f = 15.25$ kHz, as the shape oscillations are supposed to be triggered by the Faraday parametric instability. The dimensionless thickness of the viscous boundary layer is therefore $\gamma = \delta_v/R_0 \sim 0.06 \ll 1$. The bubble interface and the resulting microstreaming are presented in figure 4. This display will be used in all investigated numerical cases of standing-wave patterns. Figure 4(a) shows the deformation of the bubble interface on the axisymmetric mode $Y_3^0$, with a similar display as in figure 2 (top, side and three-dimensional view). Keeping the same orientation of the cartesian frame, figures 4(b-d) present the microstreaming pattern derived from the Lagrangian velocity field. The streamlines have been obtained using the library *pyvista* from Python programming. The top and side views of the microstreaming are obtained by calculating the streamlines in a narrow area (in depth) along the whole equator (for the top view, figure 4(b)) or along a whole meridian (side view, figure 4(c)). The procedure is the following. The considered equator/meridian is segmented in 48 particle clouds, termed as source, homogeneously distributed along the perimeter. A source particle cloud is located at the radial distance $1.01 * R_0$ from the center of the bubble, for a given angular position $(\theta, \phi)$. Each source contains 10 particles. The streamlines are generated by using each particle as a starting location. For the sake of readability, the three-dimensional streamlines are computed in figure 4(d) only inside a constrained range of the longitude $\phi$ of the 3D space surrounding the bubble, and by calculating 400 streamlines in the investigated region. Note that, because of these different numerical calculations for the streamlines, the velocity range displayed in the color bars may differ from a view to another. The derived streamlines are in full agreement with the theoretical predictions (Doinikov *et al.* 2019a) and experimental observations (Cleve *et al.* 2019; Fauconnier *et al.* 2022) of the microstreaming induced by an axisymmetric oscillation. The side view exhibits a cross-like shape with small recirculation loops in the vicinity of the bubble interface. It is worth mentioning that, while theoretically all streamlines are closed vortical trajectories, some of the vortices are so extended spatially that they appear as straight lines in figure 4(c). In fact, these streamlines will close at a large distance from the center of the bubble. Increasing the number of calculated streamlines would fill the region between the two orthogonal branches of the cross-like shape in figure 4(c), for instance. Due to the absence of azimuthal dependence, the streamlines from the top view are purely radial. Note that the small recirculation loops along the meridian are not observed because of the cutting of the microstreaming field along the equator. This situation is similar to the one encountered when looking at a streaming field in the focal plane of a microscope (figure 2(a)). The rotational direction of the flow can be perceived from a close look at the streamlines in the side and three-dimensional views: it is known that the fluid particles are propelled away from the anti-nodes (where the streamlines start) and attracted back to the nodes of displacement of the bubble interface (where the streamlines finish). This fact will be very useful in the following to discern the three-dimensional pattern of the flow, in the case of asymmetric modes.

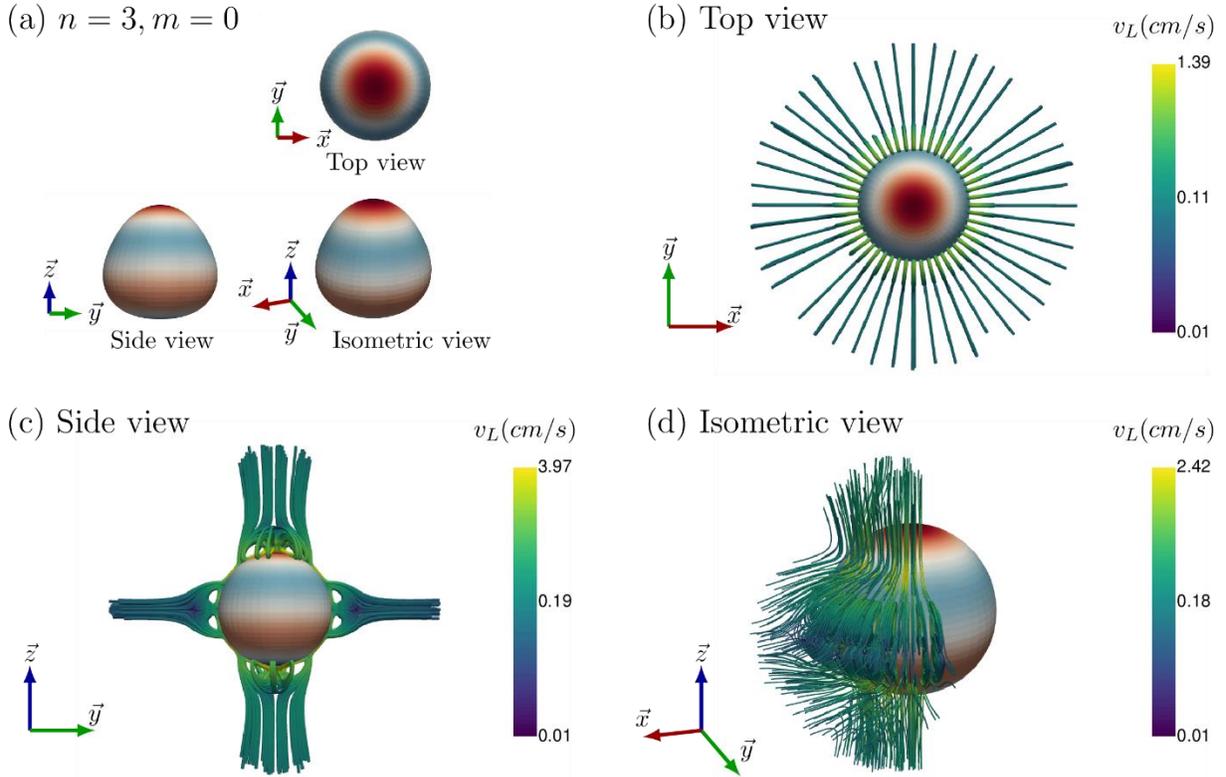

Figure 4. Presentation of the microstreaming pattern observed for a bubble (equilibrium radius 73.8 µm) exhibiting zonal oscillations on the mode ($n = 3, m = 0$). (a) numerical contours of the bubble interface in the top, side and three-dimensional view; (b) top, (c) side and (d) isometric views of the resulting flow.

### 3.3. *Signature for various asymmetric modes with the same degree $n$*

For the same bubble equilibrium radius and mode amplitudes, figures 5-7 show the resulting microstreaming patterns for the class of tesseral ($Y_3^1$, figure 5, and $Y_3^2$, figure 6) and sectoral ($Y_3^3$, figure 7) harmonics of the same degree $n = 3$.

The tesseral mode $Y_3^1$ possesses $m = 1$ meridian nodal lines (figure 5(a), top view) and $n - m = 2$ nodal parallels (figure 5(a), side view). From the top-view configuration, the bubble contour exhibits two nodes and two anti-nodes of displacement. The resulting top-view pattern is characterized by a cross shape (figure 5(b)), similar to the quadrupole pattern generated by a solid-body translation oscillation without shape deformation (Longuet-Higgins 1998; Doinikov *et al.* 2019b). The main difference with this axisymmetric quadrupole pattern relies in the out-of-plane component of the velocity field, as streamlines go away from the equatorial plane and from a closed loop that reaches the nearest displacement node on the bubble interface. The closeness of this top-view asymmetric pattern with the translation-induced quadrupole shape confirms the importance of capturing both the acoustic (high-frequency) dynamics of the bubble interface and the fluid flow, in order to avoid misinterpretation of the physical origin of the streaming pattern. Particularly, from the top view, the tesseral $m = 1$ harmonic, whatever the degree $n$, always exhibits an up-and-down oscillation around the $x$ axis that resembles the translational oscillation of the bubble along the $y$ axis. From the side view, the bubble interface displays six extrema of displacement. It results in a recognizable 12-lobe flow pattern (figure 5(c)), where the lobes are arranged in pairs and are oriented from the extremum of the bubble interface towards the nearest displacement node. Combining these two views and the fact that the fluid particles move from the antinode towards the nearest displacement node on the bubble interface, the isometric view of the pattern can be guessed in figure 5(d).

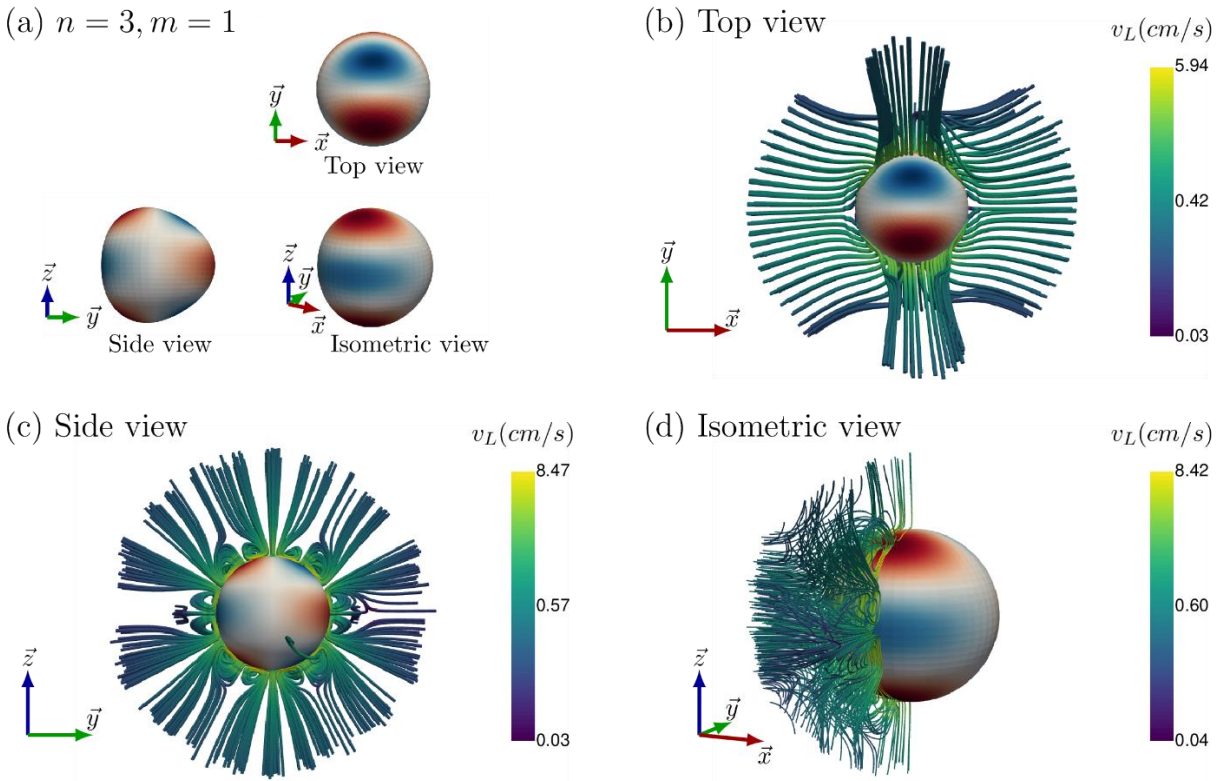

Figure 5. Presentation of the microstreaming pattern observed for a bubble (equilibrium radius 73.8 μm) exhibiting tesseral oscillations on the mode $(n = 3, m = 1)$. (a) numerical contours of the bubble interface in the top, side and three-dimensional view; (b) top, (c) side and (d) isometric views of the resulting flow.

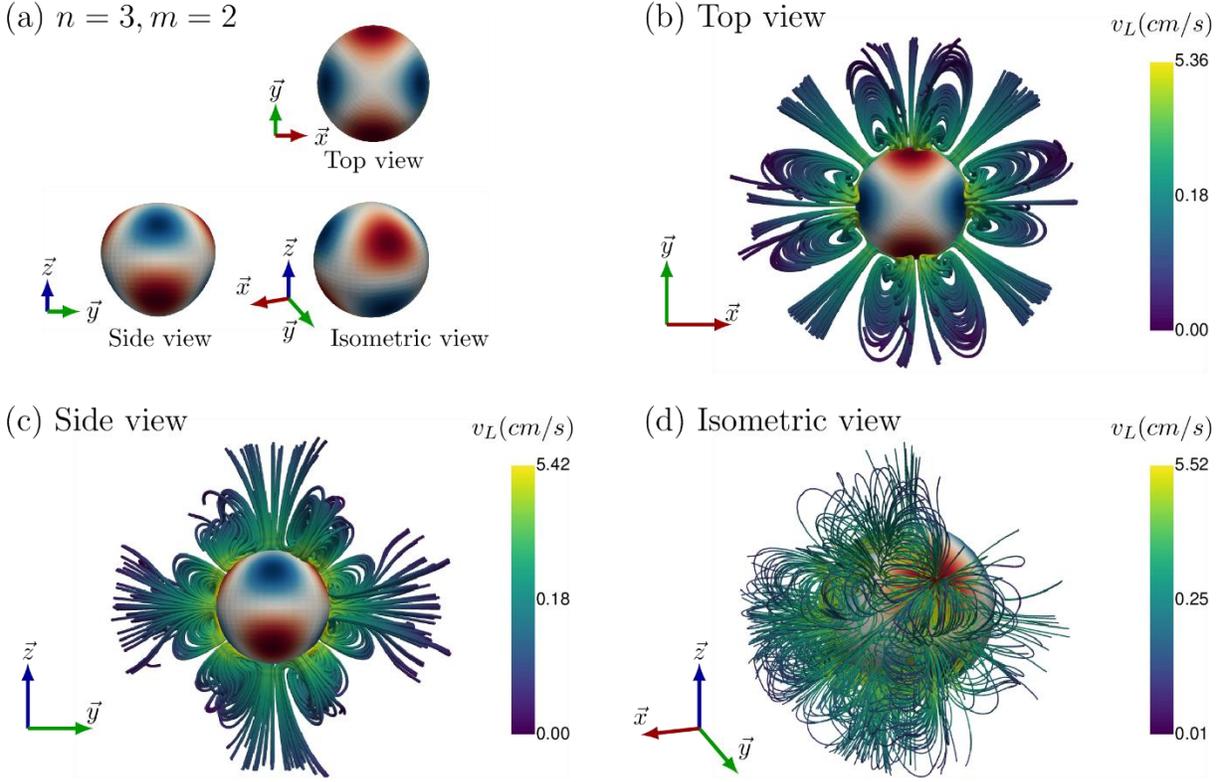

Figure 6. Presentation of the microstreaming pattern observed for a bubble (equilibrium radius 73.8 µm) exhibiting tesseral oscillations on the mode $(n = 3, m = 2)$. (a) numerical contours of the bubble interface in the top, side and three-dimensional view; (b) top, (c) side and (d) isometric views of the resulting flow.

The tesseral mode $Y_3^2$ possesses $m = 2$ meridian nodal lines (figure 6(a), top view) and $n - m = 1$ nodal parallels (figure 6(a), side view). From the top-view configuration, the bubble contour exhibits four nodes and four anti-nodes of displacement. Please note that, because the nodal parallel is located on the equator of the bubble interface, the so-called top-view nodes and anti-nodes are located out of the equatorial plane. The resulting top-view pattern (figure 6(b)) is characterized by a lobe-type pattern with $4m = 8$ recirculation loops arranged in pairs. The overall structure of the top-view streaming is in good agreement with the experimental top-view pattern displayed in figure 3(b) for another tesseral oscillation whose order also equals $m = 2$. The microstreaming pattern from the side view (figure 6(c)) is harder to interpret as the bubble interface displays four extrema of displacement whose location is dependent on the orientation of the bubble around the $z$ axis. Here eight lobes are arranged in pairs and are superimposed to a cross-type pattern. A closer look at the pair of the lobes located at the poles ($\theta = 0$ and $\theta = \pi$) reveals that the recirculation loops are in fact splitted into two loops. This artefact is due to the slicing of the streamlines along the plane containing meridians. In fact, the isometric view (figure 6(d)) indicates how the streamlines start from the displacement anti-node located in the upper right corner in figure 6(d), and reach the nearest displacement nodes that form a triangular shape surrounding this anti-node. It results in a seemingly axisymmetric vortex around the normal at the bubble interface passing through the displacement anti-node.

The results for an asymmetric bubble experiencing sectoral oscillations on the mode $Y_3^3$ are shown in figure 7. Being devoid of nodal parallels, the maximal longitudinal displacement occurs at the equatorial plane, where the bubble interface exhibits as many azimuthal lobes as the modal degree $n$ to which it belongs. The amplitude of the oscillation displacement at a given acoustic phase decreases along the line going from the equator to the poles of the bubble. For $n = 3$, it results in $2n = 6$ extrema for the displacement of the interface along the equator (figure 7(a)), an azimuthal shape that is easily recognizable from the top-view observation. From the side view, the bubble contour is almost spherical, with an outgrowth located at the equator. Due to the (relative) simplicity of the bubble deformation along the elevation and the azimuthal direction, the resulting microstreaming pattern is easy to analyze from the top, side, and even in the isometric view. The top view is characterized by a $4n = 12$-lobe flower-like pattern, where the lobes are arranged in pairs around every extrema of the bubble interface displacement. This pattern is in good qualitative agreement with the experimental microstreaming pattern shown in figure 3(c), both for the number of recirculation loops and for the spatial extension of the vortices. The side view resembles a cross-like shape (figure 7(c)), typical of a quadrupole pattern. This pattern is the analog of the top-view streaming resulting from the tesseral mode $m = 1$ (figure 5(b)) when switching from the top- to the side-view observation. Indeed, sectoral oscillations from the side view exhibit an out-of-phase motion along a meridian from one side of the bubble to the opposite one around the $z$ axis, so that the oscillation seems like the one of a solid-body translation oscillation. Because the sectoral harmonic $Y_3^3$ only possesses meridian nodal lines, the vortices observed from the top-view configuration spread with a decreasing amplitude towards the poles of the bubble (figure 7(d)).

The numerical streaming patterns obtained for the self-interacting tesseral (figure 6) or sectoral (figure 7) oscillations are found to be in good agreement with the patterns (figure 3) obtained by Fauconnier *et al.* (2022). Unlike the experimental observations of the above article, the present theory considers neither the presence of a wall in the bubble vicinity, nor the tethering of the bubble on a surface. It was shown in Fauconnier *et al.* (2020) that the contact angle between the bubble interface and the substrate lies in the range 40-60° with no dependence on the bubble equilibrium radius. It is also indicated that the contact angle remains identical after the activation of the ultrasound driving sequence. For this range of contact angle values, the bubble tethering does not influence strongly the dynamics of the oscillation and the majority of the existing asymmetric oscillations for a given degree *n* are triggered. Concerning the resulting acoustic microstreaming, it is known that the mathematical modeling of the flow surrounding a free bubble, being far from any boundary, can be in fairly good agreement with the experimental flows in the form of a large-scale streaming induced by a substrate-attached bubble (Marmottant & Hilgenfledt (2003)). Here, in the case of asymmetric oscillations, the vortices with the highest velocity magnitude are confined in the close vicinity of the bubble interface (figure 7(b)) and extend much less than the large-scale quadrupole-like streaming. It can then be inferred that the influence of the bubble tethering on the resulting flow can be less pronounced in comparison to the case of an axisymmetric flow. In addition, it should be mentioned that our theory does not consider the effect of buoyancy, which may cause the deviation of experimental streamlines from theoretical Lagrangian streamlines. This fact should be taken into account when comparing experimental results with our theory.

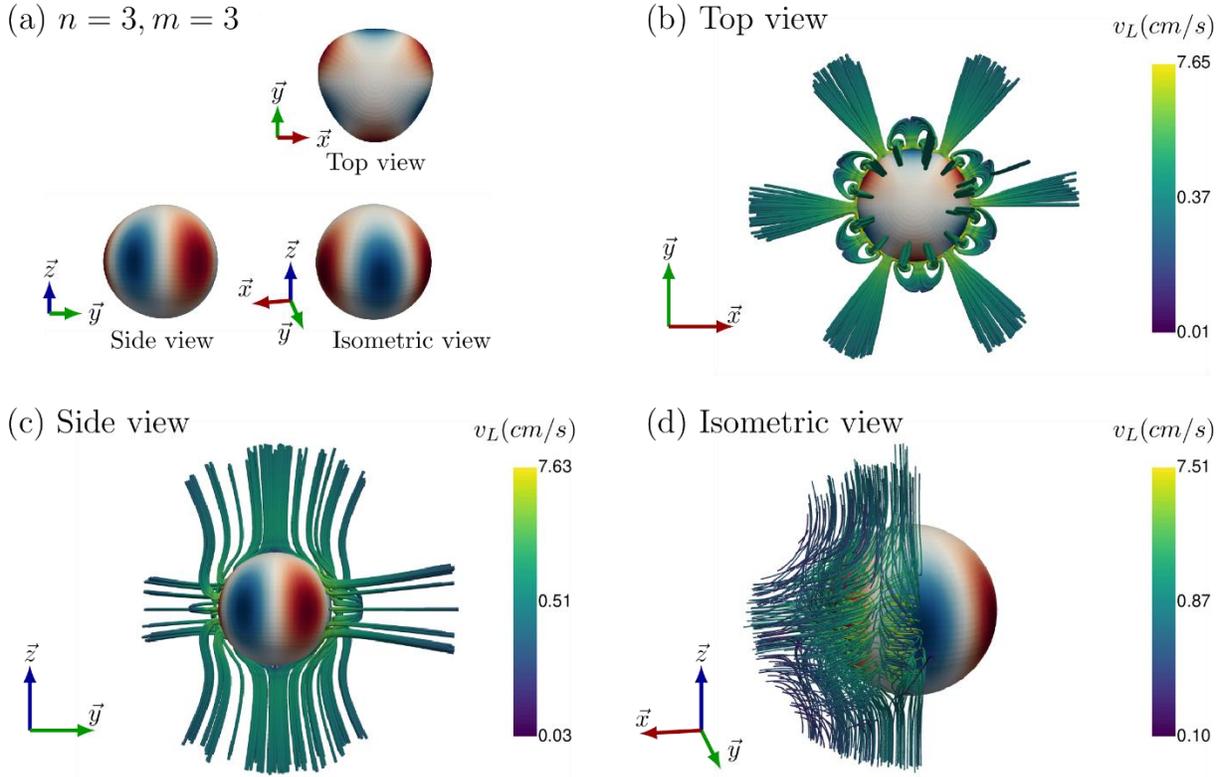

Figure 7. Presentation of the microstreaming pattern observed for a bubble (equilibrium radius 73.8 µm) exhibiting sectoral oscillations on the mode ($n = 3, m = 3$). (a) numerical contours of the bubble interface in the top, side and three-dimensional view; (b) top, (c) side and (d) isometric views of the resulting flow.

3.4. *Extension to other spherical harmonics of any degree n*

The previously-investigated case of the microstreaming induced by the three classes of spherical harmonics for the same degree $n = 3$ can help us in predicting the pattern for any asymmetric modes, whatever the degree $n$ and order $m$. The simplest case concerns bubbles exhibiting zonal oscillations $Y_n^0$, for which theoretical predictions exist (Inserra *et al.* 2020). Due to the absence of azimuthal dependence for the bubble displacement, the top-view pattern of a zonal-induced microstreaming is unique and consists in a purely radial flow. Due to the axi-symmetry of the zonal harmonics, slicing the three-dimensional microstreaming pattern in a plane containing the $z$ axis always provides the same pattern, which consists in a large-scale cross pattern with small recirculation zones in the vicinity of the bubble interface. This general signature is identical whatever the degree $n$ of the investigated zonal harmonics. Only the number of small vortices in the vicinity of the bubble interface increases as the degree $n$ increases.

In general, a bubble exhibiting an asymmetric oscillation on a tesseral harmonic of order $m$, whatever the degree $n$, will exhibit the same top-view patterns as the one shown in figure 5 when $m = 1$ or figure 6 when $m = 2$. Indeed, these patterns are uniquely linked to the number of meridian nodal lines, which equals $m$. For tesseral harmonics with higher order $m$, top-view streaming will be characterized by a lobe-type pattern with $4m$ lobes arranged in pairs. This top-view signature of the pattern resulting from a self-interacting tesseral mode is similar to the shape obtained from another self-interacting spherical harmonics (obtained from a side or a top view). This means that, without knowing the bubble equilibrium radius and hence the degree $n$ ruled by (2.73), the identification of the triggered mode at the bubble interface cannot be inferred when viewing the resulting streaming pattern. The analysis from the side view only is even more complicated, as the orientation of the focal plane around the $z$ axis significantly affects the conformation of the captured flow.

The prediction of the microstreaming pattern induced by the class of sectoral harmonics is far easier to infer in comparison to the tesseral harmonics. All the sectoral harmonics $m = n$ possess the same spatial conformation for the bubble displacement, with a maximal longitudinal displacement occurring at the equatorial plane. Along the equator, as many azimuthal lobes as the modal degree $n$ to which they belong exist. The bubble displacement at the poles always tends to zero, so that the side-view microstreaming pattern always resembles a cross-like shape, whatever the degree $n$ of the sectoral oscillations. This is evidenced in figure 8 where the top- and side-view patterns resulting from self-interacting sectoral oscillations from $n = 2$ to $n = 5$ are shown. Here the bubble equilibrium radii are chosen that they correspond to the resonant radius of a given degree $n$: $R_{res}^2 = 46$ μm ($\gamma = \delta_v/R_0 \sim 0.1$), $R_{res}^3 = 68$ μm ($\gamma = \delta_v/R_0 \sim 0.06$), $R_{res}^4 = 90$ μm ($\gamma = \delta_v/R_0 \sim 0.05$) and $R_{res}^5 = 110$ μm ($\gamma = \delta_v/R_0 \sim 0.04$). The amplitudes of the two modes $s^{(n,\pm n)}$ are set equal to 15μm, resulting in a stationary sectoral oscillation. The top-view microstreaming is characterized by a lobe-type pattern where the lobes are arranged in pairs, and their number equals two times the number of extrema of the bubble displacement along the equator. Therefore, a $4n$-lobe shape is expected from the top view, as shown in figure 8. This signature is in full agreement with experimental observations of top-view microstreaming patterns induced by sectoral oscillation for various degrees $n$ in Fauconnier *et al.* (2022).

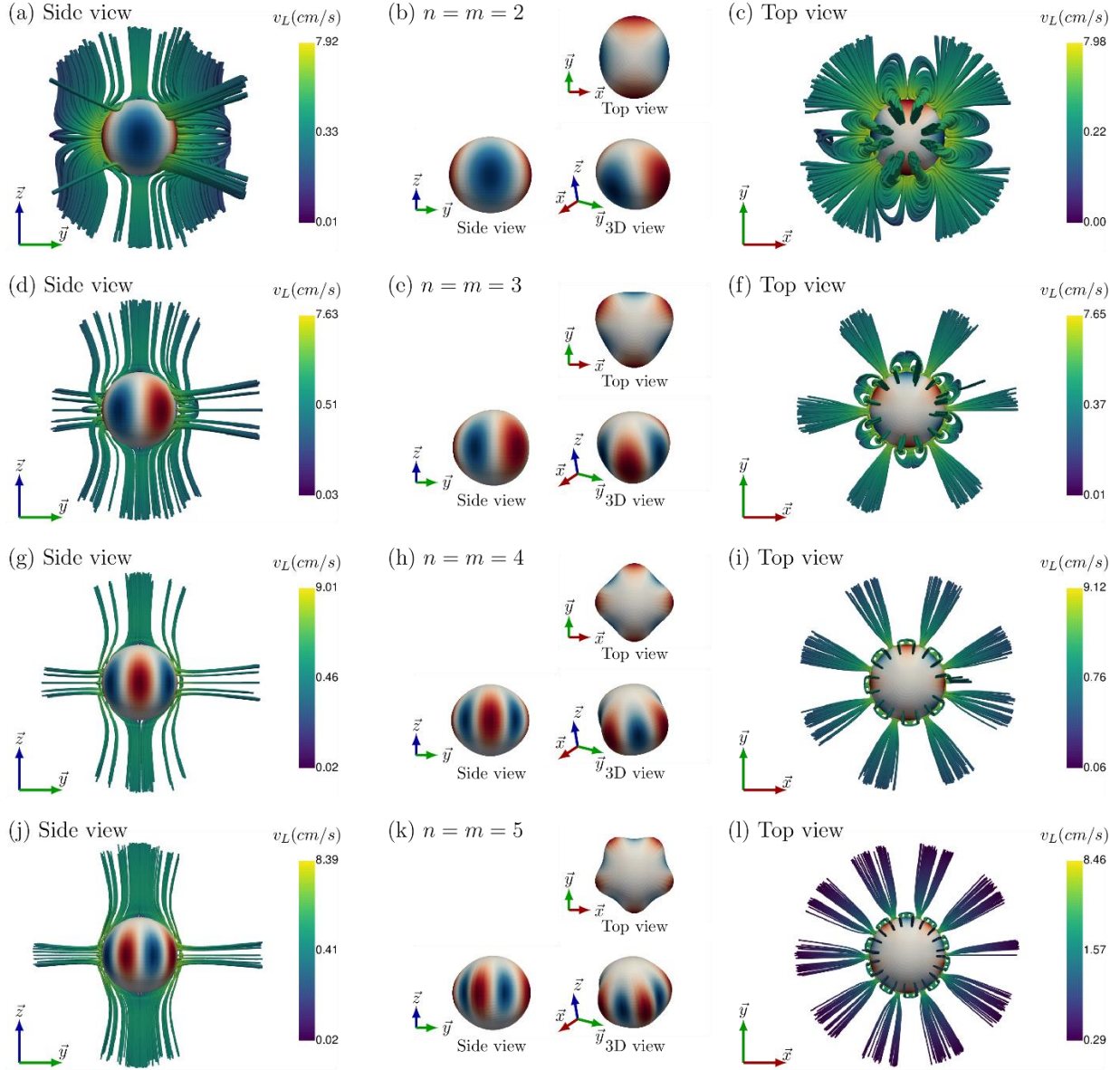

Figure 8. Presentation of the microstreaming pattern observed for various sectoral oscillations with the degree $n$ ranging from 2 to 5. Every line corresponds to a given degree $n$. (left column): the projected side-view microstreaming pattern; (middle column): the deformation of the bubble interface exhibiting the sectoral oscillation; (right column): the top-view microstreaming pattern.

3.5 *Influence of the thickness of the viscous boundary layer on the streaming pattern*

The acoustic microstreaming is driven by the streaming inside the oscillatory boundary layer around the bubble, while nonlinear second-order effects of the hydrodynamic equations are responsible for extending the streaming field much further than the viscous boundary layer. This explains why Elder (1959) reported different types of streaming patterns whose appearance depended mostly on the bubble surface velocity and the fluid viscosity. For a given bubble equilibrium radius and driving frequency, the influence of the thickness of the viscous boundary layer was investigated for varying values of the liquid viscosity in Doinikov *et al.* (2019a) and Inserra *et al.* (2020b) in the case of axisymmetric shape oscillations. For the translational oscillation of a solid body, Li *et al.* (2023) reveal changes in the structure of the streaming pattern for varying driving frequency while keeping the body size and the liquid viscosity constant. Here we investigate the influence of the value of the dimensionless thickness of the viscous boundary layer on the microstreaming pattern. We have previously considered the case (figure 7) of a sectoral oscillation $n = m = 3$ at the frequency $f = 15.25$ kHz for a bubble with the equilibrium radius 73.8 µm in water (dynamic viscosity $\eta = 10^{-6}$ Pa s). It corresponds to the dimensionless thickness of the viscous boundary layer $\gamma = \delta_v/R_0 \sim 0.06$. In order to modify the thickness of the viscous boundary layer $\delta_v$, the liquid viscosity is varied while keeping the bubble equilibrium radius and the oscillation frequency constant, so that the triggered shape oscillation given by (2.73) remains unchanged. For the same parameters as in figure 7, figure 9 presents the microstreaming patterns when the liquid viscosity is varied from $\eta = 10^{-6}$ Pa s ($\gamma = 0.06$, figure 9(a)) to $\eta = 4*10^{-6}$ Pa s ($\gamma = 0.12$, figure 9(b)) and $\eta = 16*10^{-6}$ Pa s ($\gamma = 0.24$, figure 9(c)).

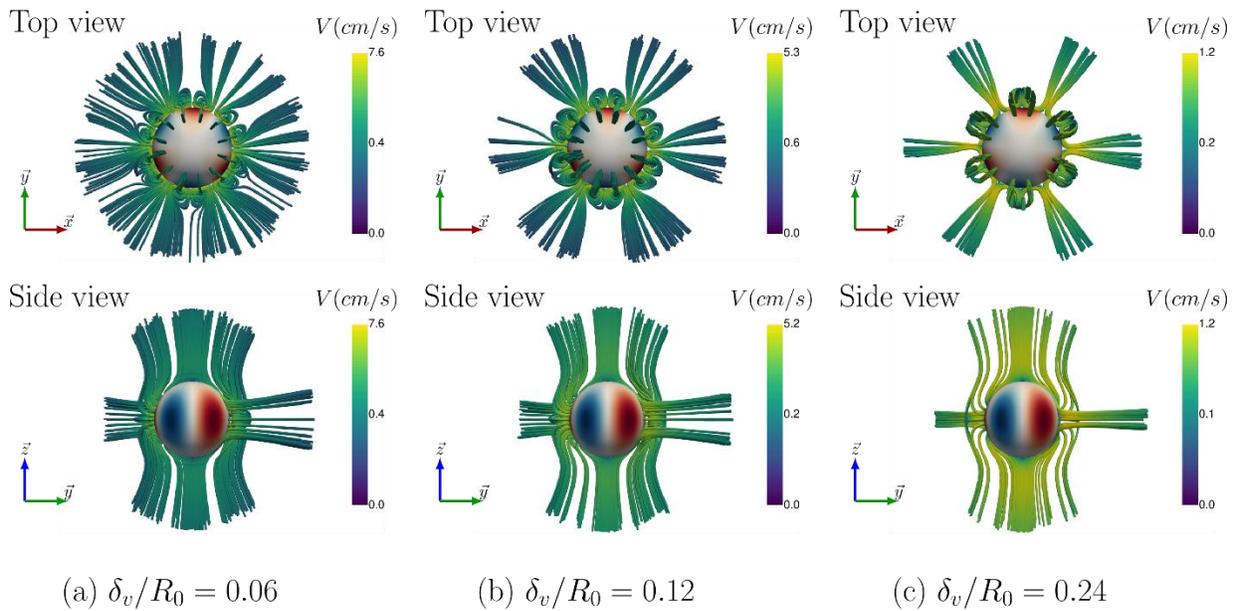

(a) $\delta_v/R_0 = 0.06$       (b) $\delta_v/R_0 = 0.12$       (c) $\delta_v/R_0 = 0.24$

Figure 9. Evolution of the microstreaming pattern for the sectoral oscillation $n = m = 3$ for varying values of the dimensionless thickness of the viscous boundary layer. The liquid dynamic viscosities and resulting dimensionless thickness of the viscous boundary layer are (a) $\eta = 10^{-6}$ Pa s, $\gamma = 0.06$; (b) $\eta = 4 * 10^{-6}$ Pa s, $\gamma = 0.12$ and (c) $\eta = 16 * 10^{-6}$ Pa s, $\gamma = 0.24$. The top- and side views are only displayed.

When the size of the viscous boundary layer is varied by four times, the overall structure of the streaming pattern remains unchanged: the top view is characterized by a 12-lobe flower-like pattern and the side view looks like a cross-like (quadrupolar) shape. When the ratio $\gamma = \delta_v/R_0$ is increased, the magnitude of the maximum streaming velocity decreases, while the size of the small vortices visible in the top view increases slightly. More pronounced modifications of the microstreaming pattern can be expected for large values of the dimensionless parameter $\gamma$, but such values would correspond to unrealistic experimental conditions (too large kinematic viscosity or a bubble size / driving frequency not matching (2.73) anymore).

### 3.6 *Sign reversal of the flow at the bubble zenith*

Substrate-attached microbubbles are commonly used as microactuators for the trapping and manipulation of small particles or biological bodies in their vicinity. The hydrodynamic flow field generated by the bubble oscillation can rotate single microparticles within microchannels (Ahmed *et al.* 2016), precisely manipulate millimetric fish eggs (Lee *et al.* 2012) or manipulate and rupture vesicles (Marmottant *et al.* 2008). In most microfluidic applications, a substrate-attached microbubble experiences a combination of spherical and translational oscillations, at the origin of a large-scale microstreaming pattern constituted of two symmetric vortices (in the plane of observation, from a side view) atop the bubble. When particles are injected near the oscillating bubble, they are expelled from the top bubble interface and move along a close loop back to the substrate. This motion is often called fountain-like streaming, as the fluid particles are expelled from the bubble interface to the surrounding medium at the zenith ($\theta = 0$) of the bubble. A similar fountain behavior is obtained when axisymmetric modes are triggered at the bubble interface, because the zenith of the bubble interface is always a displacement antinode, whatever the degree $n$ of the zonal harmonics (see the first column of figure 10(a)). In the case of wall-bounded semicylindrical bubbles, Wang *et al.* (2013) and Rallabandi *et al.* (2014) have observed an inversion of the flow direction atop the bubble, identifying an anti-fountain behavior of the vortices near the pole. This flow inversion was shown to result from the interaction between the radial oscillation and an axisymmetric shape oscillation (mixed-mode streaming). When asymmetric modes are triggered, the location of the maximum displacement on the bubble interface evolves as the order $m$ increases, tending to a maximum at the equator of the bubble in the case of a sectoral mode. Figure 10(a) illustrates the three-dimensional bubble shapes for various degrees $n$ ranging from 2 to 5. In addition, the value of the colatitude $\theta_{max}$ at which the location of the maximum displacement on the bubble interface occurs, whatever the value of the azimuthal angle $\phi$, is shown in figure 10(b). When the order $m$ is smaller than $(n+1)/2$, $\theta_{max}$ is ranging from 0 to $\pi/4$, and the maximum displacement at the bubble interface is located in the upper part of the bubble, between the zenith and the bisector $\theta_{max} = \pi/4$. For higher values of $m$, $\theta_{max}$ gets closer to $\pi/2$ and the maximum displacement tends to the equatorial plane.

For the investigated asymmetric modes, the direction of the resulting microstreaming has been evaluated atop the bubble, at the radial distance $r = 3R_0$ and for $\theta = 0$. Figure 10(c) shows the amplitude of the radial component of the Lagrangian velocity at this location. A sign reversal of the velocity amplitude occurs when the order $m$ reaches $(n+1)/2$, in agreement with the evolution of the colatitude $\theta_{max}$. In the case of zonal oscillations, the fluid particles are expelled (positive radial component of the velocity) from the zenith of the bubble, resulting in a fountain-like behavior. In the case of sectoral oscillations, the analysis of the deformation of the bubble interface reveals that the maximum displacement occurs at the equator, while the zenith ($\theta = 0$) of the bubble becomes a displacement node. The fluid particles are therefore expelled from the equator and are attracted from the zenith of the bubble to the bubble interface. This results in an anti-fountain behavior, which has not been observed experimentally so far in the case of self-interacting mode streaming. It is worth noting that the observation of the flow direction atop the bubble indirectly provides information on the triggered shape oscillation, as the direction of the flow is correlated to the order $m$ of the asymmetric oscillation. If the bubble equilibrium radius is known, the degree $n$ of the shape oscillations is obtained from the Lamb spectrum (2.73), and the range of possible triggered order $m$ can be deducted from the observation of a fountain or an anti-fountain flow.

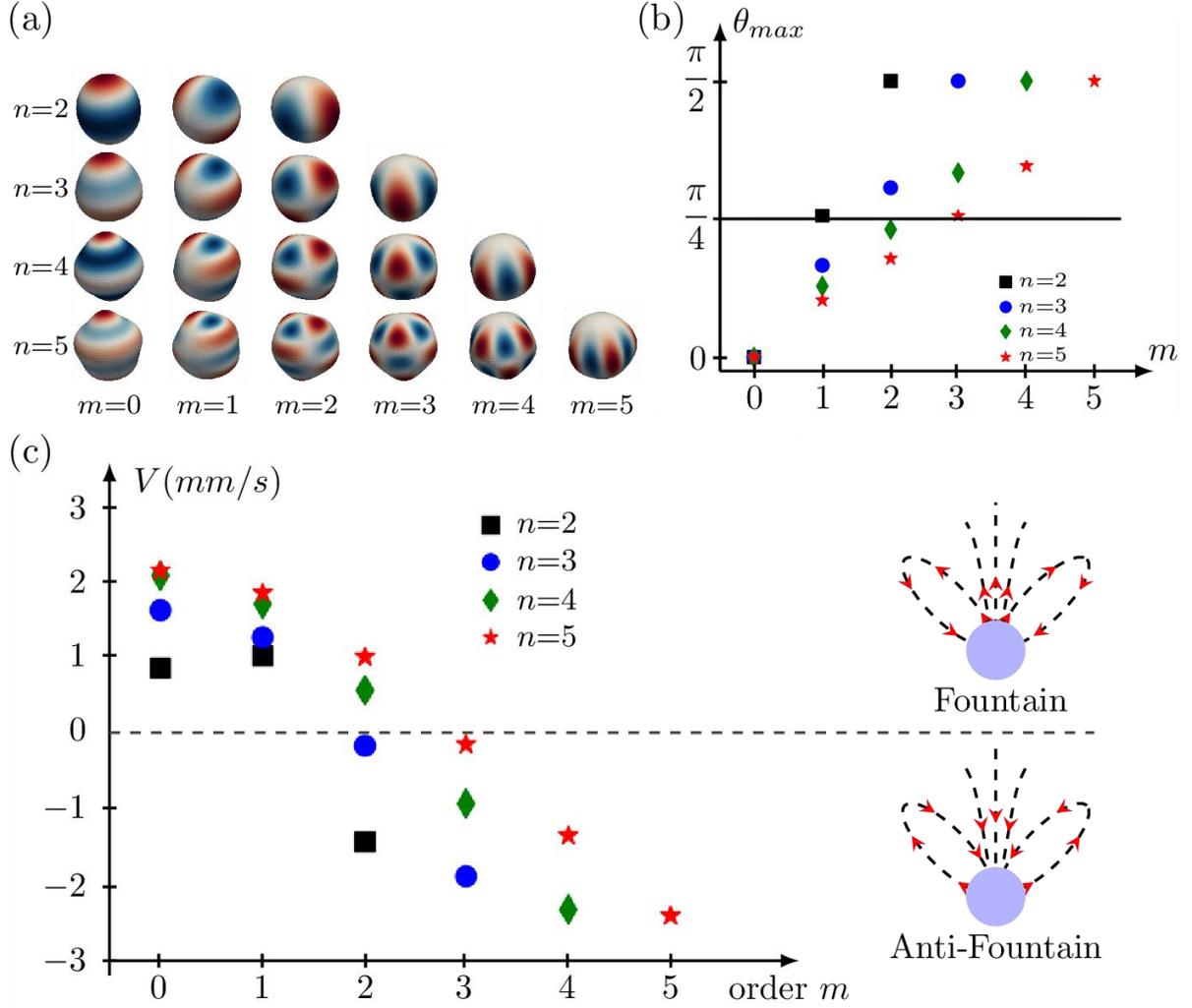

Figure 10. Investigation of the orientation of the flow at the zenith of the bubble ($\theta = 0$). (a) Three-dimensional view of the bubble interface oscillating on asymmetric modes with $n \in [2\ 5]$; (b) Evolution of the value of the colatitude $\theta_{max}$ at which the displacement of the bubble interface is maximum, as a function of the order $m$ for various degrees $n$; (c) Evolution of the amplitude of the radial velocity at the zenith ($\theta = 0$) of the bubble, at the distance $r = 3R_0$ as a function of the order $m$ for various degrees $n$. A positive value corresponds to the fountain-like streaming, while a negative one is typical of the anti-fountain behavior.

### 3.7 *The case of travelling surface waves*

Up to now the stationary shape oscillations were only considered, meaning that the amplitude of the two modes $s^{(n,\pm m)}$ was equal. A slight change in the amplitude of the surface perturbation propagating in one direction along the azimuthal angle $\phi$ in comparison to the one propagating in the opposite direction will result in a quasi-stationary surface wave. Such a bubble will experience a rotating asymmetric surface wave, whose angular velocity will depend on the difference in the two amplitudes of modes $(n, \pm m)$. This has already been reported once in literature by Mekki-Berrada *et al.* (2016), who measured a constant angular velocity of about 0.5 revolution / second for every modal degree $n$, in the case of bubbles flattened between two elastic walls. The underlying cause for the triggering of a rotating shape oscillation remains so far unknown.

The case of travelling surface waves is investigated in figure 11, for different ratios of the amplitudes of the two propagating modes $(n, \pm m)$. The case of the sectoral oscillation $n = m = 3$ is investigated, for which the microstreaming pattern obtained in the case of a stationary wave is presented in figure 7. Only the top view and isometric views of the streaming pattern are displayed in figure 11, and the amplitude of the $(n, m)$ wave is kept equal to 15µm. When the amplitude of the $(n, -m)$ wave is modified to 14.9µm, i.e. when $s^{(n,-m)}/s^{(n,m)} = 0.99$, streamlines are curved along the azimuthal direction (figure 11(a), top view) and the maximum amplitude of the Lagrangian velocity reaches 8.3cm/s. From figure 7, it is worth noting that the maximum amplitude of the Lagrangian velocity reaches 7.65cm/s in the case of a stationary wave, i.e. when $s^{(n,-m)}/s^{(n,m)} = 1$. The increase in difference between the amplitudes of the two counter-propagating surface waves results in a stronger azimuthal component of the velocity field. The amplitude of the streaming flow can increase by a factor of 2 (figure 11(b)) when $s^{(n,-m)}/s^{(n,m)} = 0.93$ or a factor of 4 (figure 11(c)) when $s^{(n,-m)}/s^{(n,m)} = 0.66$, in comparison to the stationary case.

The morphology of the swirling flow can be explained as follows. When the amplitudes $s^{(n,\pm m)}$ of a given shape oscillation are set equal, then a standing shape oscillation is formed on the bubble interface. This means that displacement antinodes (starting points of the streamlines) and nodes (end points of the streamlines) remain at the same location with time. The resulting flow is constituted of vortical lines that are spatially stationary. When the amplitudes $s^{(n,\pm m)}$ of the two components of a shape oscillation differ, a quasi-propagating wave is triggered in a preferential direction on the bubble interface. "Quasi-propagating" refers to the co-existence of a purely standing wave at the bubble interface and of a purely propagating wave. Hence the locations of the displacement nodes and antinodes constantly move along the bubble interface with time. This prevents the spatial stationarity of the flow, so a convective-like flow is generated around the bubble. This flow follows the direction of propagation of the propagative surface wave and is characterized by a strong azimuthal component of the flow velocity, and a large-scale extension at the equator. However, why this flow is stronger in amplitude in comparison to the one induced by the standing wave alone is difficult to explain. Nevertheless, these results highlight the possibility to drastically enhance the streaming efficiency around bubbles if the bubble rotation can be controlled.

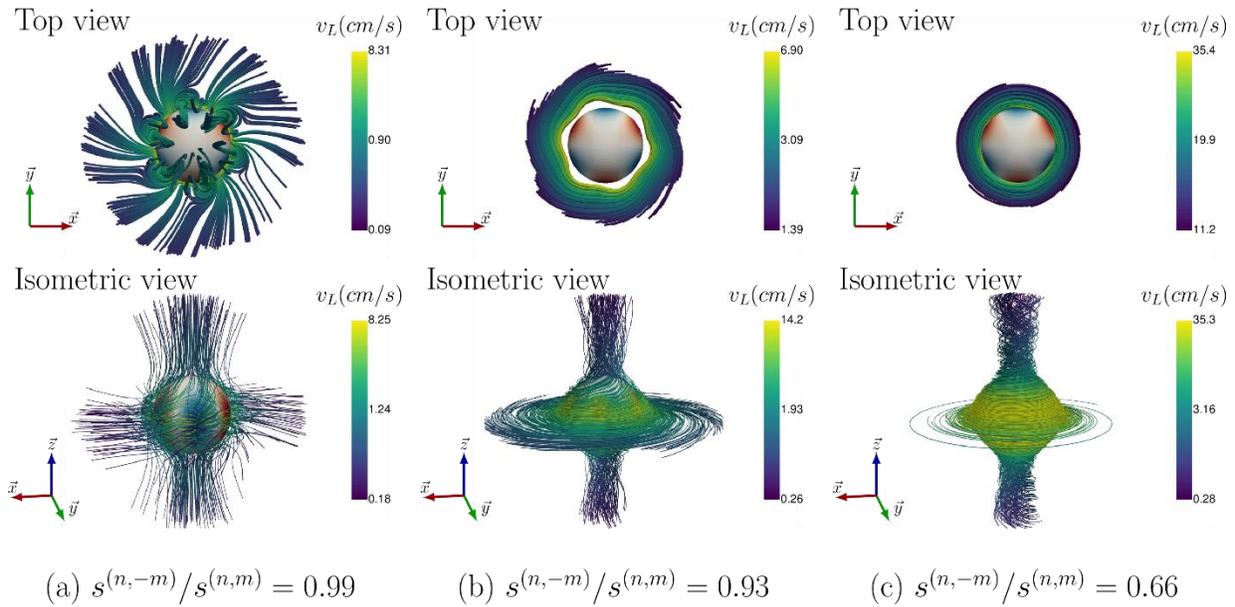

(a) $s^{(n,-m)}/s^{(n,m)} = 0.99$   (b) $s^{(n,-m)}/s^{(n,m)} = 0.93$   (c) $s^{(n,-m)}/s^{(n,m)} = 0.66$

Figure 11. Evolution of the microstreaming pattern in the case of a quasi-stationary wave, for the sectoral oscillation $n = m = 3$. Three ratios of the amplitudes of the two modes $s^{(n,-m)}/s^{(n,m)}$ are investigated for (a) $s^{(n,-m)}/s^{(n,m)} = 0.99$; (b) $s^{(n,-m)}/s^{(n,m)} = 0.93$ and (c) $s^{(n,-m)}/s^{(n,m)} = 0.66$. The top- and isometric views are only displayed.

## 4. Conclusion

We present for the first time a theoretical modeling of acoustic microstreaming induced by a bubble experiencing asymmetric oscillations. The modeling is based on the decomposition of the first- and second-order vorticity field into the poloidal and toroidal fields, allowing the exact analytical derivation of the Lagrangian streaming induced by the bubble. This derivation is valid whatever the bubble size and the liquid viscosity, without any limitation on the thickness of the viscous boundary layer at the bubble interface. By representing the shape of the bubble interface in terms of the set of the orthonormal spherical harmonics, the signatures of the microstreaming patterns are investigated as a function of different classes of triggered spherical harmonics (zonal, tesseral or sectoral oscillations). Keeping in mind that the streamlines start from the displacement antinode at the bubble interface and close at the displacement nodes, the three-dimensional asymmetric patterns are shown to be easily predicted once the asymmetric shape deformation of the bubble is known. For a given class of spherical harmonics of degree $n$, it is shown that the sign reversal in the flow orientation atop the bubble is obtained when the order $m$ evolves between 0 and $n$: zonal harmonics result in the fountain-like streaming while the anti-fountain flow occurs in the case of sectoral oscillations. In between, the flow reverses for a typical value of the order $m$ that allows one to help in the identification of the triggered asymmetric mode by means of the indirect observation of the flow orientation. The case of travelling surface waves along the bubble interface reveals the possibility of a drastic enhancement of the streaming strength. This observation highlights the interest in the controlling of oscillations of rotating bubbles for practical use.

**Funding.** This work was funded by l'Agence Nationale de la Recherche under the project ANR-22-CE92-0062 and supported by the LabEx CeLyA of the University of Lyon (Grants No. ANR-10-LABX-0060 and No. ANR-11-IDEX-0007).

**Declaration of interest.** The authors report no conflict of interest.

**Author ORCIDs.**
Claude Inserra http://orcid.org/0000-0001-5145-7068.

**Appendix A. Calculation of $a_{nm}$, $b_{nm}$ and $c_{nm}$.**

To find $a_{nm}$, $b_{nm}$ and $c_{nm}$, we apply boundary conditions at the bubble surface.

The first boundary condition requires that the normal component of $\mathbf{v}_1^{(n,m)}$ at $r = R_0$ be equal to the respective normal component of the bubble surface velocity. This condition is written by

$$v_{1r}^{(n,m)}\bigg|_{r=R_0} = \frac{dr_s^{(n,m)}}{dt} = -i\omega s^{(n,m)} e^{-i\omega t} Y_n^m(\theta,\phi), \tag{A1}$$

where $v_{1r}^{(n,m)}$ is the normal component of $\mathbf{v}_1^{(n,m)}$ and $r_s^{(n,m)}$ is given by the second term of (2.1).

The second boundary condition assumes slippage on the bubble surface, which means that the tangential stress generated by the liquid motion should be zero on the bubble surface. This condition is written by

$$\sigma_{r\theta} = \eta\left(\frac{1}{r}\frac{\partial v_{1r}^{(n,m)}}{\partial \theta} + \frac{\partial v_{1\theta}^{(n,m)}}{\partial r} - \frac{v_{1\theta}^{(n,m)}}{r}\right) = 0 \text{ at } r = R_0, \tag{A2}$$

$$\sigma_{r\phi} = \eta\left(\frac{\partial v_{1\phi}^{(n,m)}}{\partial r} + \frac{1}{r\sin\theta}\frac{\partial v_{1r}^{(n,m)}}{\partial \phi} - \frac{v_{1\phi}^{(n,m)}}{r}\right) = 0 \text{ at } r = R_0, \tag{A3}$$

where $\sigma_{r\theta}$ and $\sigma_{r\phi}$ are the tangential components of the liquid stress in spherical coordinates (Landau & Lifshitz 1987). It should be mentioned that in the case that there are impurities and surfactants on the bubble surface and in the case of a contrast agent bubble having a shell, the boundary condition of slippage should be replaced by the no-slip boundary condition, i.e., by the condition of adhesion of liquid particles to the bubble surface. In the case of a bubble with a shell, the behavior of the shell material should also be taken into consideration. In view of its complexity, this case requires a separate study. The general algorithm of calculations remains the same but all constants related to the boundary conditions at the bubble surface should be re-calculated.

Substitution of (2.28) into (2.20) yields

$$\mathbf{v}_1^{(n,m)} = v_{1r}^{(n,m)}\mathbf{e}_r + v_{1\theta}^{(n,m)}\mathbf{e}_\theta + v_{1\phi}^{(n,m)}\mathbf{e}_\phi, \tag{A4}$$

$$v_{1r}^{(n,m)} = e^{-i\omega t} f_{nm}(r) Y_n^m(\theta,\phi), \tag{A5}$$

$$v_{1\theta}^{(n,m)} = e^{-i\omega t}\left[\frac{a_{nm} k_v h_n^{(1)}(k_v r)}{\sin\theta}\frac{\partial Y_n^m(\theta,\phi)}{\partial \phi} + g_{nm}(r)\frac{\partial Y_n^m(\theta,\phi)}{\partial \theta}\right], \tag{A6}$$

$$v_{1\phi}^{(n,m)} = e^{-i\omega t}\left[-a_{nm}k_v h_n^{(1)}(k_v r)\frac{\partial Y_n^m(\theta,\phi)}{\partial \theta} + \frac{g_{nm}(r)}{\sin\theta}\frac{\partial Y_n^m(\theta,\phi)}{\partial \phi}\right], \tag{A7}$$

where $f_{nm}(r)$ and $g_{nm}(r)$ are given by

$$f_{nm}(r) = \frac{(n+1)c_{nm}}{r^{n+2}} + \frac{n(n+1)b_{nm}h_n^{(1)}(k_v r)}{k_v r}, \tag{A8}$$

$$g_{nm}(r) = -\frac{c_{nm}}{r^{n+2}} + \frac{b_{nm}}{k_v r}\left[(n+1)h_n^{(1)}(k_v r) - k_v r h_{n+1}^{(1)}(k_v r)\right]. \tag{A9}$$

Substituting (A5) and (A8) into (A1), one obtains

$$c_{nm} + \frac{nR_0^{n+2}h_n^{(1)}(\bar{x})}{\bar{x}}b_{nm} = -\frac{i\omega R_0^{n+2}s^{(n,m)}}{n+1}, \tag{A10}$$

where $\bar{x} = k_v R_0$.

Substitution of (A5) – (A7) into (A2) and (A3) yields

$$a_{nm}k_v\left[\bar{x}h_n^{(1)\prime}(\bar{x}) - h_n^{(1)}(\bar{x})\right]\frac{1}{\sin\theta}\frac{\partial Y_n^m(\theta,\phi)}{\partial \phi} + \left[f_{nm}(R_0) - g_{nm}(R_0) + R_0 g_{nm}'(R_0)\right]\frac{\partial Y_n^m(\theta,\phi)}{\partial \theta} = 0, \tag{A11}$$

$$a_{nm}k_v\left[h_n^{(1)}(\bar{x}) - \bar{x}h_n^{(1)\prime}(\bar{x})\right]\frac{\partial Y_n^m(\theta,\phi)}{\partial \theta} + \left[f_{nm}(R_0) - g_{nm}(R_0) + R_0 g_{nm}'(R_0)\right]\frac{1}{\sin\theta}\frac{\partial Y_n^m(\theta,\phi)}{\partial \phi} = 0, \tag{A12}$$

where

$$g_{nm}'(r) = \frac{(n+2)c_{nm}}{r^{n+3}} + \frac{b_{nm}}{k_v r^2}\left\{(n+1)\left[k_v r h_n^{(1)\prime}(k_v r) - h_n^{(1)}(k_v r)\right] - (k_v r)^2 h_{n+1}^{(1)\prime}(k_v r)\right\}. \tag{A13}$$

Multiplying (A11) by $\sin^{-1}\theta\,\partial Y_n^m(\theta,\phi)/\partial \phi$ and subtracting (A12) multiplied by $\partial Y_n^m(\theta,\phi)/\partial \theta$, one obtains

$$a_{nm}\left[\bar{x}h_n^{(1)\prime}(\bar{x}) - h_n^{(1)}(\bar{x})\right]\left\{\left[\frac{1}{\sin\theta}\frac{\partial Y_n^m(\theta,\phi)}{\partial \phi}\right]^2 + \left[\frac{\partial Y_n^m(\theta,\phi)}{\partial \theta}\right]^2\right\} = 0. \tag{A14}$$

It follows from (A14) that

$$a_{nm} = 0. \tag{A15}$$

Accordingly, (A6) and (A7) take the form:

$$v_{1\theta}^{(n,m)} = e^{-i\omega t} g_{nm}(r) \frac{\partial Y_n^m(\theta,\phi)}{\partial \theta}, \tag{A16}$$

$$v_{1\phi}^{(n,m)} = e^{-i\omega t} \frac{g_{nm}(r)}{\sin\theta} \frac{\partial Y_n^m(\theta,\phi)}{\partial \phi}. \tag{A17}$$

Multiplying (A11) by $\partial Y_n^m(\theta,\phi)/\partial\theta$ and adding to (A12) multiplied by $\sin^{-1}\theta\, \partial Y_n^m(\theta,\phi)/\partial\phi$, one obtains

$$\left[ f_{nm}(R_0) - g_{nm}(R_0) + R_0 g'_{nm}(R_0) \right] \left\{ \left[ \frac{\partial Y_n^m(\theta,\phi)}{\partial \theta} \right]^2 + \left[ \frac{1}{\sin\theta} \frac{\partial Y_n^m(\theta,\phi)}{\partial \phi} \right]^2 \right\} = 0. \tag{A18}$$

It follows from (A18) that

$$f_{nm}(R_0) - g_{nm}(R_0) + R_0 g'_{nm}(R_0) = 0. \tag{A19}$$

Substituting (A8) and (A9) into (A19) and using the identity

$$h_{n+1}^{(1)}(x) = \frac{n}{x} h_n^{(1)}(x) - h_n^{(1)\prime}(x), \tag{A20}$$

one obtains

$$\frac{2(n+2)}{R_0^{n+2}} c_{nm} + \frac{b_{nm}}{\bar{x}} \left[ (n^2 + n - 2) h_n^{(1)}(\bar{x}) + \bar{x}^2 h_n^{(1)\prime\prime}(\bar{x}) \right] = 0. \tag{A21}$$

Combining (A10) and (A21), one finds

$$c_{nm} = s^{(n,m)} \frac{i\omega R_0^{n+2} [(2 - n - n^2) h_n^{(1)}(\bar{x}) - \bar{x}^2 h_n^{(1)\prime\prime}(\bar{x})]}{(n+1)[\bar{x}^2 h_n^{(1)\prime\prime}(\bar{x}) - (n^2 + 3n + 2) h_n^{(1)}(\bar{x})]}, \tag{A22}$$

$$b_{nm} = s^{(n,m)} \frac{2i(n+2)\bar{x}\omega}{(n+1)[\bar{x}^2 h_n^{(1)\prime\prime}(\bar{x}) - (n^2 + 3n + 2) h_n^{(1)}(\bar{x})]}, \quad n \geq 1. \tag{A23}$$

**Appendix B. Solving the equations of acoustic streaming for the mode (*n*,*m*)**

To solve (2.52) and (2.53), we first need to calculate their right-hand sides.

The calculation of $\mathbf{v}_1^{(n,m)} \cdot \nabla \mathbf{v}_1^{(n,m)}$ yields

$$\mathbf{v}_1^{(n,m)} \cdot \nabla \mathbf{v}_1^{(n,m)} = \left( v_{1r}^{(n,m)} \frac{\partial}{\partial r} + \frac{v_{1\theta}^{(n,m)}}{r} \frac{\partial}{\partial \theta} + \frac{v_{1\phi}^{(n,m)}}{r\sin\theta} \frac{\partial}{\partial \phi} \right) (v_{1r}^{(n,m)} \mathbf{e}_r + v_{1\theta}^{(n,m)} \mathbf{e}_\theta + v_{1\phi}^{(n,m)} \mathbf{e}_\phi)$$

$$= \mathbf{e}_r \left( v_{1r}^{(n,m)} \frac{\partial v_{1r}^{(n,m)}}{\partial r} + \frac{v_{1\theta}^{(n,m)}}{r} \frac{\partial v_{1r}^{(n,m)}}{\partial \theta} + \frac{v_{1\phi}^{(n,m)}}{r\sin\theta} \frac{\partial v_{1r}^{(n,m)}}{\partial \phi} - \frac{(v_{1\theta}^{(n,m)})^2 + (v_{1\phi}^{(n,m)})^2}{r} \right)$$

$$+\boldsymbol{e}_\theta\left(v_{1r}^{(n,m)}\frac{\partial v_{1\theta}^{(n,m)}}{\partial r}+\frac{v_{1\theta}^{(n,m)}}{r}\frac{\partial v_{1\theta}^{(n,m)}}{\partial \theta}+\frac{v_{1\phi}^{(n,m)}}{r\sin\theta}\frac{\partial v_{1\theta}^{(n,m)}}{\partial \phi}+\frac{v_{1r}^{(n,m)}v_{1\theta}^{(n,m)}}{r}-\frac{\cos\theta}{r\sin\theta}(v_{1\phi}^{(n,m)})^2\right)$$

$$+\boldsymbol{e}_\phi\left(v_{1r}^{(n,m)}\frac{\partial v_{1\phi}^{(n,m)}}{\partial r}+\frac{v_{1\theta}^{(n,m)}}{r}\frac{\partial v_{1\phi}^{(n,m)}}{\partial \theta}+\frac{v_{1\phi}^{(n,m)}}{r\sin\theta}\frac{\partial v_{1\phi}^{(n,m)}}{\partial \phi}+\frac{v_{1r}^{(n,m)}v_{1\phi}^{(n,m)}}{r}+\frac{\cos\theta}{r\sin\theta}v_{1\theta}^{(n,m)}v_{1\phi}^{(n,m)}\right). \quad (B1)$$

Substituting (A5), (A16) and (A17) into (B1) and averaging it over time, one obtains

$$\left\langle \boldsymbol{v}_1^{(n,m)}\cdot\nabla\boldsymbol{v}_1^{(n,m)}\right\rangle = \frac{1}{2}\mathrm{Re}\left\{\boldsymbol{e}_r\left[f'_{nm}f^*_{nm}Y_n^m Y_n^{m*}+\frac{(f_{nm}-g_{nm})g^*_{nm}}{r}\left(\frac{\partial Y_n^m}{\partial \theta}\frac{\partial Y_n^{m*}}{\partial \theta}+\frac{1}{\sin^2\theta}\frac{\partial Y_n^m}{\partial \phi}\frac{\partial Y_n^{m*}}{\partial \phi}\right)\right]\right.$$

$$+\boldsymbol{e}_\theta\left[f^*_{nm}\left(g'_{nm}+\frac{g_{nm}}{r}\right)Y_n^{m*}\frac{\partial Y_n^m}{\partial \theta}+\frac{|g_{nm}|^2}{r}\left(\frac{\partial Y_n^{m*}}{\partial \theta}\frac{\partial^2 Y_n^m}{\partial \theta^2}+\frac{1}{\sin^2\theta}\frac{\partial Y_n^{m*}}{\partial \phi}\frac{\partial^2 Y_n^m}{\partial \theta\partial \phi}-\frac{\cos\theta}{\sin^3\theta}\frac{\partial Y_n^{m*}}{\partial \phi}\frac{\partial Y_n^m}{\partial \phi}\right)\right]$$

$$\left.+\frac{\boldsymbol{e}_\phi}{\sin\theta}\left[f^*_{nm}\left(g'_{nm}+\frac{g_{nm}}{r}\right)Y_n^{m*}\frac{\partial Y_n^m}{\partial \phi}+\frac{|g_{nm}|^2}{r}\left(\frac{\partial Y_n^{m*}}{\partial \theta}\frac{\partial^2 Y_n^m}{\partial \theta\partial \phi}+\frac{1}{\sin^2\theta}\frac{\partial Y_n^{m*}}{\partial \phi}\frac{\partial^2 Y_n^m}{\partial \phi^2}\right)\right]\right\}, \quad (B2)$$

For simplicity, we drop the arguments of the functions in (B2).

The calculation of $\nabla\times\left\langle \boldsymbol{v}_1^{(n,m)}\cdot\nabla\boldsymbol{v}_1^{(n,m)}\right\rangle$, after some rearrangements, results in

$$\nabla\times\left\langle \boldsymbol{v}_1^{(n,m)}\cdot\nabla\boldsymbol{v}_1^{(n,m)}\right\rangle = \frac{1}{2}\mathrm{Re}\left\{\boldsymbol{e}_r\frac{f^*_{nm}(g_{nm}+rg'_{nm})}{r^2\sin\theta}\left(\frac{\partial Y_n^m}{\partial \phi}\frac{\partial Y_n^{m*}}{\partial \theta}-\frac{\partial Y_n^{m*}}{\partial \phi}\frac{\partial Y_n^m}{\partial \theta}\right)\right.$$

$$-\frac{(f^*_{nm}g_{nm}+rf^*_{nm}g'_{nm})'}{r}\left(\frac{\boldsymbol{e}_\theta}{\sin\theta}Y_n^{m*}\frac{\partial Y_n^m}{\partial \phi}-\boldsymbol{e}_\phi Y_n^{m*}\frac{\partial Y_n^m}{\partial \theta}\right)$$

$$+\frac{f_{nm}f'^*_{nm}+f^*_{nm}f'_{nm}}{2r}\left[\frac{\boldsymbol{e}_\theta}{\sin\theta}\frac{\partial}{\partial \phi}(Y_n^m Y_n^{m*})-\boldsymbol{e}_\phi\frac{\partial}{\partial \theta}(Y_n^m Y_n^{m*})\right]$$

$$+\frac{1}{2r}\left[\frac{f_{nm}g^*_{nm}+f^*_{nm}g_{nm}-2g_{nm}g^*_{nm}}{r}-(g_{nm}g^*_{nm})'\right]$$

$$\left.\times\left[\frac{\boldsymbol{e}_\theta}{\sin\theta}\frac{\partial}{\partial \phi}\left(\frac{\partial Y_n^m}{\partial \theta}\frac{\partial Y_n^{m*}}{\partial \theta}+\frac{1}{\sin^2\theta}\frac{\partial Y_n^m}{\partial \phi}\frac{\partial Y_n^{m*}}{\partial \phi}\right)-\boldsymbol{e}_\phi\frac{\partial}{\partial \theta}\left(\frac{\partial Y_n^m}{\partial \theta}\frac{\partial Y_n^{m*}}{\partial \theta}+\frac{1}{\sin^2\theta}\frac{\partial Y_n^m}{\partial \phi}\frac{\partial Y_n^{m*}}{\partial \phi}\right)\right]\right\}. \quad (B3)$$

Comparison of (2.52) and (B3) leads to

$$\sum_{k=1}^{\infty}\sum_{l=-k}^{k}k(k+1)\left[P_{kl}^{(n,m)//}(r)-\frac{k(k+1)P_{kl}^{(n,m)}(r)}{r^2}\right]Y_k^l(\theta,\phi)$$

$$=\frac{1}{2\nu}\mathrm{Re}\left\{\frac{f^*_{nm}(g_{nm}+rg'_{nm})}{\sin\theta}\left(\frac{\partial Y_n^m}{\partial \phi}\frac{\partial Y_n^{m*}}{\partial \theta}-\frac{\partial Y_n^{m*}}{\partial \phi}\frac{\partial Y_n^m}{\partial \theta}\right)\right\}. \quad (B4)$$

With the help of (D8) and (D21), the right-hand side of (B4) is transformed so that (B4) takes the form:

$$\sum_{k=1}^{\infty}\sum_{l=-k}^{k} k(k+1)\left[P_{kl}^{(n,m)//}(r) - \frac{k(k+1)P_{kl}^{(n,m)}(r)}{r^2}\right]Y_k^l(\theta,\phi) = \frac{1}{2\nu}\text{Re}\sum_{k=1}^{\infty}\sum_{l=-k}^{k} imf_{nm}^*(g_{nm} + rg'_{nm})$$

$$\times\left[nC_{(n+1)m}(B_{kl}^{(n+1)mnm} + B_{kl}^{nm(n+1)m}) - (n+1)C_{nm}(B_{kl}^{(n-1)mnm} + B_{kl}^{nm(n-1)m})\right]Y_k^l(\theta,\phi), \quad (B5)$$

where $C_{nm}$ and $B_{kl}^{n_1 m_1 n_2 m_2}$ are constants that are calculated by (D11) and (D22), respectively. In fact, (B5) only involves terms with $l=0$ as for $m_1 = m_2$ $B_{kl}^{n_1 m_1 n_2 m_2} = 0$ for $l \neq 0$. This fact is in agreement with the right-hand side of (B4), which is independent of $\phi$.

Keeping only nonzero terms, one obtains from (B5),

$$P_{k0}^{(n,m)//}(r) - \frac{k(k+1)}{r^2}P_{k0}^{(n,m)}(r) = F_k^{(n,m)}(r), \quad (B6)$$

where

$$F_k^{(n,m)}(r) = \frac{m}{2k(k+1)\nu}\text{Re}\left\{if_{nm}^*(r)\left[g_{nm}(r) + rg'_{nm}(r)\right]\right\}$$

$$\times\left[nC_{(n+1)m}\left(B_{k0}^{(n+1)mnm} + B_{k0}^{nm(n+1)m}\right) - (n+1)C_{nm}\left(B_{k0}^{(n-1)mnm} + B_{k0}^{nm(n-1)m}\right)\right]. \quad (B7)$$

The calculation of the $r$-component of $\nabla\times\nabla\times\langle v_1^{(n,m)}\cdot\nabla v_1^{(n,m)}\rangle$ gives

$$\mathbf{e}_r\cdot\left[\nabla\times\nabla\times\langle v_1^{(n,m)}\cdot\nabla v_1^{(n,m)}\rangle\right] = \text{Re}\left\{\frac{f_{nm}f_{nm}^{/*} + f_{nm}^* f_{nm}'}{4r^2}L^2\left[Y_n^m Y_n^{m*}\right]\right.$$

$$+\frac{f_{nm}g_{nm}^* + f_{nm}^* g_{nm} - 2g_{nm}g_{nm}^* - r(g_{nm}g_{nm}^*)'}{4r^3}L^2\left[\frac{\partial Y_n^m}{\partial\theta}\frac{\partial Y_n^{m*}}{\partial\theta} + \frac{m^2 Y_n^m Y_n^{m*}}{\sin^2\theta}\right]$$

$$\left.+\frac{(f_{nm}^* g_{nm} + rf_{nm}^* g'_{nm})'}{2r^2}\left[\frac{\partial Y_n^m}{\partial\theta}\frac{\partial Y_n^{m*}}{\partial\theta} + \frac{m^2 Y_n^m Y_n^{m*}}{\sin^2\theta} - n(n+1)Y_n^m Y_n^{m*}\right]\right\}, \quad (B8)$$

where $L^2$ stands for the square of the orbital angular momentum operator (Varshalovich *et al.* 1988),

$$L^2 = -\frac{1}{\sin\theta}\frac{\partial}{\partial\theta}\sin\theta\frac{\partial}{\partial\theta} - \frac{1}{\sin^2\theta}\frac{\partial^2}{\partial\phi^2}. \quad (B9)$$

By using (D5), it can be shown that the following identity holds:

$$-\frac{1}{2}L^2\left[Y_n^m Y_n^{m*}\right] = \frac{\partial Y_n^m}{\partial\theta}\frac{\partial Y_n^{m*}}{\partial\theta} + \frac{m^2 Y_n^m Y_n^{m*}}{\sin^2\theta} - n(n+1)Y_n^m Y_n^{m*}. \quad (B10)$$

Substitution of (B10) into (B8) yields

$$\boldsymbol{e}_r \cdot \left[ \nabla \times \nabla \times \left\langle \boldsymbol{v}_1^{(n,m)} \cdot \nabla \boldsymbol{v}_1^{(n,m)} \right\rangle \right] = \frac{1}{8r^3} \operatorname{Re} \left\{ L^4 \left[ Y_n^m Y_n^{m*} \right] \left[ 2 g_{nm} g_{nm}^* - f_{nm} g_{nm}^* - f_{nm}^* g_{nm} + r(g_{nm} g_{nm}^*)' \right] \right.$$

$$+ 2L^2 \left[ Y_n^m Y_n^{m*} \right] \left\{ r \left[ f_{nm} f_{nm}'^* + f_{nm}^* f_{nm}' - (f_{nm}^* g_{nm} + r f_{nm}^* g_{nm}')' \right] \right.$$

$$\left. + n(n+1) \left[ f_{nm} g_{nm}^* + f_{nm}^* g_{nm} - 2 g_{nm} g_{nm}^* - r(g_{nm} g_{nm}^*)' \right] \right\} . \tag{B11}$$

By using (D17) and the identity $L^2 Y_k^l = k(k+1) Y_k^l$, one obtains

$$\boldsymbol{e}_r \cdot \left[ \nabla \times \nabla \times \left\langle \boldsymbol{v}_1^{(n,m)} \cdot \nabla \boldsymbol{v}_1^{(n,m)} \right\rangle \right] = \frac{1}{8r^3} \operatorname{Re} \sum_{k=1}^{\infty} \sum_{l=-k}^{k} k(k+1) A_{kl}^{nmnm} Y_k^l (\theta, \phi)$$

$$\times \left\{ 2r \left[ f_{nm} f_{nm}'^* + f_{nm}^* f_{nm}' - (f_{nm}^* g_{nm} + r f_{nm}^* g_{nm}')' \right] \right.$$

$$\left. + [k(k+1) - 2n(n+1)] \left[ 2 g_{nm} g_{nm}^* - f_{nm} g_{nm}^* - f_{nm}^* g_{nm} + r(g_{nm} g_{nm}^*)' \right] \right\}, \tag{B12}$$

where the constant coefficients $A_{kl}^{nmnm}$ are calculated by (D20). In fact, (B12) only involves terms with $l = 0$ as $A_{kl}^{nmnm} = 0$ for $l \neq 0$.

Comparing (2.53) with (B12) and keeping only nonzero terms, one obtains

$$T_{k0}^{(n,m)\prime\prime}(r) - \frac{k(k+1)}{r^2} T_{k0}^{(n,m)}(r) = G_k^{(n,m)}(r), \tag{B13}$$

where

$$G_k^{(n,m)}(r) = \frac{A_{k0}^{nmnm}}{4\nu r} \operatorname{Re} \left\{ r f_{nm}^*(r) \left[ 2 f_{nm}'(r) - 2 g_{nm}'(r) - r g_{nm}''(r) \right] - r f_{nm}'^*(r) \left[ g_{nm}(r) + r g_{nm}'(r) \right] \right.$$

$$\left. + [k(k+1) - 2n(n+1)] g_{nm}^*(r) \left[ g_{nm}(r) + r g_{nm}'(r) - f_{nm}(r) \right] \right\}. \tag{B14}$$

The derivatives, which appear in (B14), are calculated by

$$f_{nm}'(r) = -\frac{(n+1)(n+2) c_{nm}}{r^{n+3}} + \frac{n(n+1) b_{nm}}{k_\nu r^2} \left[ (n-1) h_n^{(1)}(k_\nu r) - k_\nu r h_{n+1}^{(1)}(k_\nu r) \right], \tag{B15}$$

$$g_{nm}'(r) = \frac{(n+2) c_{nm}}{r^{n+3}} + \frac{b_{nm}}{k_\nu r^2} \left\{ [n^2 - 1 - (k_\nu r)^2] h_n^{(1)}(k_\nu r) + k_\nu r h_{n+1}^{(1)}(k_\nu r) \right\}, \tag{B16}$$

$$g_{nm}''(r) = -\frac{(n+2)(n+3) c_{nm}}{r^{n+4}} + \frac{b_{nm}}{k_\nu r^3} \left\{ 2(1-n^2) h_n^{(1)}(k_\nu r) + k_\nu r [n^2 - 1 - (k_\nu r)^2] h_n^{(1)\prime}(k_\nu r) \right.$$

$$\left. - k_\nu r h_{n+1}^{(1)}(k_\nu r) + (k_\nu r)^2 h_{n+1}^{(1)\prime}(k_\nu r) \right\}. \tag{B17}$$

Equations (B6) and (B13) are solved by the method of variation of parameters (Boyce & DiPrima 2001). According to this method, a solution to (B6) is given by

$$P_{k0}^{(n,m)}(r) = r^{k+1}C_{1k}^{(n,m)}(r) + r^{-k}C_{2k}^{(n,m)}(r), \tag{B18}$$

where $C_{1k}^{(n,m)}(r)$ and $C_{2k}^{(n,m)}(r)$ obey the following equations:

$$r^{k+1}C_{1k}^{(n,m)\prime}(r) + r^{-k}C_{2k}^{(n,m)\prime}(r) = 0, \tag{B19}$$

$$(k+1)r^k C_{1k}^{(n,m)\prime}(r) - kr^{-k-1}C_{2k}^{(n,m)\prime}(r) = F_k^{(n,m)}(r). \tag{B20}$$

Solving (B19) and (B20) gives

$$C_{1k}^{(n,m)}(r) = \bar{C}_{1k}^{(n,m)} + \frac{1}{2k+1}\int_{R_0}^{r} s^{-k} F_k^{(n,m)}(s)ds, \tag{B21}$$

$$C_{2k}^{(n,m)}(r) = \bar{C}_{2k}^{(n,m)} - \frac{1}{2k+1}\int_{R_0}^{r} s^{k+1} F_k^{(n,m)}(s)ds, \tag{B22}$$

where $\bar{C}_{1k}^{(n,m)}$ and $\bar{C}_{2k}^{(n,m)}$ are constants. From the condition that the acoustic streaming vanishes at infinity, one finds

$$\bar{C}_{1k}^{(n,m)} = -\frac{1}{2k+1}\int_{R_0}^{\infty} s^{-k} F_k^{(n,m)}(s)ds, \tag{B23}$$

while $\bar{C}_{2k}^{(n,m)}$ is calculated by boundary conditions at the bubble surface; see below.

Equation (B13) is similar to (B6), so its solution can be written by analogy,

$$T_{k0}^{(n,m)}(r) = r^{k+1}C_{3k}^{(n,m)}(r) + r^{-k}C_{4k}^{(n,m)}(r), \tag{B24}$$

$$C_{3k}^{(n,m)}(r) = \bar{C}_{3k}^{(n,m)} + \frac{1}{2k+1}\int_{R_0}^{r} s^{-k} G_k^{(n,m)}(s)ds, \tag{B25}$$

$$C_{4k}^{(n,m)}(r) = \bar{C}_{4k}^{(n,m)} - \frac{1}{2k+1}\int_{R_0}^{r} s^{k+1} G_k^{(n,m)}(s)ds, \tag{B26}$$

$$\bar{C}_{3k}^{(n,m)} = -\frac{1}{2k+1}\int_{R_0}^{\infty} s^{-k} G_k^{(n,m)}(s)ds, \tag{B27}$$

where $\bar{C}_{4k}^{(n,m)}$ is a constant to be calculated by boundary conditions at the bubble surface; see below.

Going on with the calculation of $v_E^{(n,m)}$, one obtains from (2.50),

$$v_E^{(n,m)} = \nabla\times\left[e_r \sum_{k=1}^{\infty}\sum_{l=-k}^{k} P_{kl}^{(n,m)}(r)Y_k^l(\theta,\phi)\right] + e_r \sum_{k=1}^{\infty}\sum_{l=-k}^{k} T_{kl}^{(n,m)}(r)Y_k^l(\theta,\phi) + \nabla\Phi^{(n,m)}(r,\theta,\phi). \tag{B28}$$

It follows from (2.44) that

$$\Delta\Phi^{(n,m)}(r,\theta,\phi) = -\sum_{k=1}^{\infty}\sum_{l=-k}^{k}\nabla\cdot\left[T_{kl}^{(n,m)}(r)Y_k^l(\theta,\phi)e_r\right] = -\sum_{k=1}^{\infty}\sum_{l=-k}^{k}\left[T_{kl}^{(n,m)\prime}(r) + 2r^{-1}T_{kl}^{(n,m)}(r)\right]Y_k^l(\theta,\phi).$$

(B29)

In fact, (B29) only involves terms with $l=0$ as $T_{kl}^{(n,m)}(r)=0$ for $l\neq 0$.

We assume that

$$\Phi^{(n,m)}(r,\theta,\phi) = \sum_{k=1}^{\infty}\sum_{l=-k}^{k}\Phi_{kl}^{(n,m)}(r)Y_k^l(\theta,\phi).$$ (B30)

Substituting (B30) into (B29), using (B24) and keeping only nonzero terms, one has

$$\Phi_{k0}^{(n,m)\prime\prime}(r) + \frac{2}{r}\Phi_{k0}^{(n,m)\prime}(r) - \frac{k(k+1)}{r^2}\Phi_{k0}^{(n,m)}(r) = H_k^{(n,m)}(r),$$ (B31)

where

$$H_k^{(n,m)}(r) = -T_{k0}^{(n,m)\prime}(r) - 2r^{-1}T_{k0}^{(n,m)}(r) = -(k+3)r^k C_{3k}^{(n,m)}(r) + (k-2)r^{-k-1}C_{4k}^{(n,m)}(r).$$ (B32)

Equation (B31) is solved by the method of variation of parameters, which results in

$$\Phi_{k0}^{(n,m)}(r) = r^k C_{5k}^{(n,m)}(r) + r^{-k-1}C_{6k}^{(n,m)}(r),$$ (B33)

where $C_{5k}^{(n,m)}(r)$ and $C_{6k}^{(n,m)}(r)$ obey the following equations:

$$r^k C_{5k}^{(n,m)\prime}(r) + r^{-k-1}C_{6k}^{(n,m)\prime}(r) = 0,$$ (B34)

$$kr^{k-1}C_{5k}^{(n,m)\prime}(r) - (k+1)r^{-k-2}C_{6k}^{(n,m)\prime}(r) = H_k^{(n,m)}(r).$$ (B35)

Solving (B34) and (B35) gives

$$C_{5k}^{(n,m)}(r) = \bar{C}_{5k}^{(n,m)} + \frac{1}{2k+1}\int_{R_0}^{r} s^{1-k}H_k^{(n,m)}(s)ds,$$ (B36)

$$C_{6k}^{(n,m)}(r) = \bar{C}_{6k}^{(n,m)} - \frac{1}{2k+1}\int_{R_0}^{r} s^{k+2}H_k^{(n,m)}(s)ds.$$ (B37)

The constant $\bar{C}_{6k}^{(n,m)}$ is calculated by boundary conditions at the bubble surface; see below.

From the condition that the acoustic streaming vanishes at infinity, one finds

$$\bar{C}_{5k}^{(n,m)} = -\frac{1}{2k+1}\int_{R_0}^{\infty} s^{1-k}H_k^{(n,m)}(s)ds.$$ (B38)

Substituting (B32) into (B38) and using (B26), one obtains

$$\bar{C}_{5k}^{(n,m)} = -\frac{(k-2)\bar{C}_{4k}^{(n,m)}}{(4k^2-1)R_0^{2k-1}} + I_k^{(n,m)}, \tag{B39}$$

where

$$I_k^{(n,m)} = \frac{k+3}{2k+1}\int_{R_0}^{\infty} rC_{3k}^{(n,m)}(r)dr + \frac{k-2}{(2k+1)^2}\int_{R_0}^{\infty}\left[\int_{R_0}^{r} s^{k+1}G_k^{(n,m)}(s)ds\right]\frac{dr}{r^{2k}}. \tag{B40}$$

Substitution of (B30) into (B28) yields

$$\boldsymbol{v}_E^{(n,m)} = \sum_{k=1}^{\infty}\sum_{l=-k}^{k}\left\{\boldsymbol{e}_r\left[T_{kl}^{(n,m)}(r) + \Phi_{kl}^{(n,m)\prime}(r)\right]Y_k^l(\theta,\phi)\right.$$

$$\left. + \frac{\Phi_{kl}^{(n,m)}(r)}{r}\left[\boldsymbol{e}_\theta\frac{\partial Y_k^l(\theta,\phi)}{\partial\theta} + \frac{\boldsymbol{e}_\phi}{\sin\theta}\frac{\partial Y_k^l(\theta,\phi)}{\partial\phi}\right] + \frac{P_{kl}^{(n,m)}(r)}{r}\left[\frac{\boldsymbol{e}_\theta}{\sin\theta}\frac{\partial Y_k^l(\theta,\phi)}{\partial\phi} - \boldsymbol{e}_\phi\frac{\partial Y_k^l(\theta,\phi)}{\partial\theta}\right]\right\}. \tag{B41}$$

Keeping only nonzero terms (those with $l = 0$) and using (D1) and (D9), one obtains

$$v_{Er}^{(n,m)} = \frac{1}{2\sqrt{\pi}}\sum_{k=1}^{\infty}\sqrt{2k+1}\left[T_{k0}^{(n,m)}(r) + \Phi_{k0}^{(n,m)\prime}(r)\right]P_k(\cos\theta), \tag{B42}$$

$$v_{E\theta}^{(n,m)} = \frac{1}{2\sqrt{\pi}r}\sum_{k=1}^{\infty}\sqrt{2k+1}\Phi_{k0}^{(n,m)}(r)P_k^1(\cos\theta), \tag{B43}$$

$$v_{E\phi}^{(n,m)} = -\frac{1}{2\sqrt{\pi}r}\sum_{k=1}^{\infty}\sqrt{2k+1}P_{k0}^{(n,m)}(r)P_k^1(\cos\theta). \tag{B44}$$

Equation (B42) involves $\Phi_{k0}^{(n,m)\prime}(r)$, which is calculated by (B33) and (B34) to be

$$\Phi_{k0}^{(n,m)\prime}(r) = kr^{k-1}C_{5k}^{(n,m)}(r) - (k+1)r^{-k-2}C_{6k}^{(n,m)}(r). \tag{B45}$$

In order to go on with the calculation, we need to apply boundary conditions for acoustic streaming at the bubble surface. To do this, we need to know the Stokes drift velocity (Longuet-Higgins 1998), which is calculated by (Doinikov *et al.* 2019a)

$$\boldsymbol{v}_S^{(n,m)} = \frac{1}{2\omega}\text{Re}\left\{i\boldsymbol{v}_1^{(n,m)}\cdot\nabla\boldsymbol{v}_1^{(n,m)*}\right\}. \tag{B46}$$

Equation (B46) gives

$$v_{Sr}^{(n,m)} = -\frac{1}{2\omega}\text{Im}\left\{v_{1r}^{(n,m)}\frac{\partial v_{1r}^{(n,m)*}}{\partial r} + \frac{v_{1\theta}^{(n,m)}}{r}\frac{\partial v_{1r}^{(n,m)*}}{\partial\theta} + \frac{v_{1\phi}^{(n,m)}}{r\sin\theta}\frac{\partial v_{1r}^{(n,m)*}}{\partial\phi}\right\}, \tag{B47}$$

$$v_{S\theta}^{(n,m)} = -\frac{1}{2\omega}\text{Im}\left\{v_{1r}^{(n,m)}\frac{\partial v_{1\theta}^{(n,m)*}}{\partial r} + \frac{v_{1\theta}^{(n,m)}}{r}\frac{\partial v_{1\theta}^{(n,m)*}}{\partial\theta} + \frac{v_{1\phi}^{(n,m)}}{r\sin\theta}\frac{\partial v_{1\theta}^{(n,m)*}}{\partial\phi} + \frac{v_{1r}^{(n,m)*}v_{1\theta}^{(n,m)}}{r}\right\}, \tag{B48}$$

$$v_{S\phi}^{(n,m)} = -\frac{1}{2\omega} \text{Im} \left\{ v_{1r}^{(n,m)} \frac{\partial v_{1\phi}^{(n,m)*}}{\partial r} + \frac{v_{1\theta}^{(n,m)}}{r} \frac{\partial v_{1\phi}^{(n,m)*}}{\partial \theta} + \frac{v_{1\phi}^{(n,m)}}{r \sin \theta} \frac{\partial v_{1\phi}^{(n,m)*}}{\partial \phi} + \frac{v_{1\phi}^{(n,m)} v_{1r}^{(n,m)*}}{r} + \frac{\cos \theta v_{1\phi}^{(n,m)} v_{1\theta}^{(n,m)*}}{r \sin \theta} \right\}.$$

(B49)

Substituting (A5), (A16) and (A17) into (B47) – (B49) and using (D6), one obtains

$$v_{Sr}^{(n,m)} = -\frac{1}{2\omega} \text{Re} \left\{ i f_{nm}^{*} f_{nm}^{/} Y_n^m Y_n^{m*} + \frac{i f_{nm} g_{nm}^{*}}{r} \left( \frac{\partial Y_n^m}{\partial \theta} \frac{\partial Y_n^{m*}}{\partial \theta} + \frac{m^2 Y_n^m Y_n^{m*}}{\sin^2 \theta} \right) \right\},$$

(B50)

$$v_{S\theta}^{(n,m)} = -\frac{1}{4\omega} \text{Re} \left\{ i \left[ f_{nm}^{*} g_{nm}^{/} - \frac{\left( f_{nm}^{*} - n(n+1) g_{nm}^{*} \right) g_{nm}}{r} \right] \frac{\partial \left( Y_n^m Y_n^{m*} \right)}{\partial \theta} \right\},$$

(B51)

$$v_{S\phi}^{(n,m)} = \frac{m}{2\omega} \text{Re} \left\{ f_{nm}^{*} \left( g_{nm}^{/} - \frac{g_{nm}}{r} \right) \frac{Y_n^m Y_n^{m*}}{\sin \theta} \right.$$

$$\left. + \frac{g_{nm} g_{nm}^{*}}{r \sin \theta} \left( \frac{\partial Y_n^m}{\partial \theta} \frac{\partial Y_n^{m*}}{\partial \theta} + \frac{m^2 Y_n^m Y_n^{m*}}{\sin^2 \theta} - \frac{2 \cos \theta}{\sin \theta} Y_n^m \frac{\partial Y_n^{m*}}{\partial \theta} \right) \right\}.$$

(B52)

With the help of (B10) and (D17), (B50) is represented by

$$v_{Sr}^{(n,m)} = -\frac{1}{2\omega} \text{Re} \left\{ i \left[ f_{nm}^{*} f_{nm}^{/} + \frac{n(n+1) f_{nm} g_{nm}^{*}}{r} \right] Y_n^m Y_n^{m*} - \frac{i f_{nm} g_{nm}^{*}}{2r} L^2 \left[ Y_n^m Y_n^{m*} \right] \right\}$$

$$= \frac{1}{2\omega} \text{Re} \sum_{k=1}^{\infty} \sum_{l=-k}^{k} i A_{kl}^{nmnm} \left[ \frac{k(k+1) - 2n(n+1)}{2r} f_{nm} g_{nm}^{*} - f_{nm}^{*} f_{nm}^{/} \right] Y_k^l(\theta, \phi).$$

(B53)

Keeping only nonzero terms and using (D1), one obtains

$$v_{Sr}^{(n,m)} = \frac{1}{4\sqrt{\pi}\omega} \sum_{k=1}^{\infty} \sqrt{2k+1} A_{k0}^{nmnm} \text{Re} \left\{ i f_{nm}^{*}(r) \left[ \frac{2n(n+1) - k(k+1)}{2r} g_{nm}(r) - f_{nm}^{/}(r) \right] \right\} P_k(\cos \theta).$$

(B54)

Substitution of (D17) into (B51) yields

$$v_{S\theta}^{(n,m)} = \frac{1}{4\omega} \text{Re} \sum_{k=1}^{\infty} \sum_{l=-k}^{k} i A_{kl}^{nmnm} \left\{ \frac{g_{nm} \left[ f_{nm}^{*} - n(n+1) g_{nm}^{*} \right]}{r} - f_{nm}^{*} g_{nm}^{/} \right\} \frac{\partial Y_k^l(\theta, \phi)}{\partial \theta}.$$

(B55)

Keeping only nonzero terms and using (D1) and (D9), one obtains

$$v_{S\theta}^{(n,m)} = \frac{1}{8\sqrt{\pi}\omega} \text{Re} \left\{ i f_{nm}^{*}(r) \left[ \frac{g_{nm}(r)}{r} - g_{nm}^{/}(r) \right] \right\} \sum_{k=1}^{\infty} \sqrt{2k+1} A_{k0}^{nmnm} P_k^1(\cos \theta).$$

(B56)

With the help of (B10), (B52) is represented by

$$v_{S\phi}^{(n,m)} = \frac{m}{2\omega}\mathrm{Re}\left\{\left[f_{nm}^*\left(g_{nm}' - \frac{g_{nm}}{r}\right) + \frac{n(n+1)|g_{nm}|^2}{r}\right]\frac{Y_n^m Y_n^{m*}}{\sin\theta} + \frac{|g_{nm}|^2}{r}\frac{\partial}{\partial\theta}\left(\frac{Y_n^{m*}}{\sin\theta}\frac{\partial Y_n^m}{\partial\theta}\right)\right\}. \tag{B57}$$

Note that $v_{S\phi}^{(n,m)} = 0$ for $m=0$. For $n=m=1$, with the help of (D1), (B57) is represented by

$$v_{S\phi}^{(1,1)} = E^{(1,1)}(r)P_1^1(\cos\theta), \tag{B58}$$

where

$$E^{(1,1)}(r) = \frac{3}{16\pi\omega r}\mathrm{Re}\left\{f_{11}^*(r)\left[g_{11}(r) - rg_{11}'(r)\right] - |g_{11}(r)|^2\right\}. \tag{B59}$$

For $n>1$, by using (D23), (D24), (D1), (D3) and (D15), keeping only nonzero terms, (B57) is represented by

$$v_{S\phi}^{(n,m)} = \sum_{k=1}^{\infty} E_k^{(n>1,m)}(r)P_k^1(\cos\theta), \tag{B60}$$

where

$$E_k^{(n>1,m)}(r) = \frac{m\sqrt{2k+1}}{4\sqrt{\pi}\omega r}\left\{\left[nC_{(n+1)m}B_{k0}^{(n+1)mnm} - (n+1)C_{nm}B_{k0}^{(n-1)mnm}\right]|g_{nm}(r)|^2\right.$$
$$-\sqrt{\frac{(2n+1)(n-m)!}{(n+m)!}}\mathrm{Re}\{f_{nm}^*(r)[g_{nm}(r) - rg_{nm}'(r)] - n(n+1)|g_{nm}(r)|^2\}$$
$$\left.\times \frac{1}{\sqrt{k(k+1)}}\sum_{s=1}^{[\frac{n-m+2}{2}]}\sqrt{\frac{(2n-4s+3)(n+m-2s)!}{(n-m-2s+2)!}}A_{k(-1)}^{(n-2s+1)(m-1)nm}\right\}. \tag{B61}$$

Equations (B58) and (B60) can be combined as follows:

$$v_{S\phi}^{(n,m)} = \sum_{k=1}^{\infty} E_k^{(n,m)}(r)P_k^1(\cos\theta), \tag{B62}$$

where

$$E_k^{(n,m)}(r) = \begin{cases} E^{(1,1)}(r)\delta_{1k}, & n=1 \\ E_k^{(n>1,m)}(r), & n>1 \end{cases} \tag{B63}$$

We can now apply the boundary conditions for the acoustic streaming at the bubble surface. They are given by

$$v_{Lr}^{(n,m)} = 0 \text{ at } r = R_0, \tag{B64}$$

$$\sigma_{Lr\theta}^{(n,m)} = \eta\left(\frac{1}{r}\frac{\partial v_{Lr}^{(n,m)}}{\partial \theta} + \frac{\partial v_{L\theta}^{(n,m)}}{\partial r} - \frac{v_{L\theta}^{(n,m)}}{r}\right) = 0 \text{ at } r = R_0, \tag{B65}$$

$$\sigma_{Lr\phi}^{(n,m)} = \eta\left(\frac{\partial v_{L\phi}^{(n,m)}}{\partial r} + \frac{1}{r\sin\theta}\frac{\partial v_{Lr}^{(n,m)}}{\partial \phi} - \frac{v_{L\phi}^{(n,m)}}{r}\right) = 0 \text{ at } r = R_0, \tag{B66}$$

where $v_L^{(n,m)} = v_E^{(n,m)} + v_S^{(n,m)}$ is the Lagrangian streaming velocity and $\sigma_{Lr\theta}^{(n,m)}$ and $\sigma_{Lr\phi}^{(n,m)}$ are the tangential components of the stress produced by $v_L^{(n,m)}$. We use (B64) – (B66) in order to calculate the constants $\bar{C}_{2k}^{(n,m)}$, $\bar{C}_{4k}^{(n,m)}$ and $\bar{C}_{6k}^{(n,m)}$.

Substituting (B42) and (B54) into (B64), one obtains

$$T_{k0}^{(n,m)}(R_0) + \Phi_{k0}^{(n,m)\prime}(R_0) = Q_k^{(n,m)}, \tag{B67}$$

where

$$Q_k^{(n,m)} = \frac{A_{k0}^{nmnm}}{2\omega} \text{Re}\left\{if_{nm}^*(R_0)\left[f_{nm}'(R_0) + \frac{k(k+1) - 2n(n+1)}{2R_0} g_{nm}(R_0)\right]\right\}. \tag{B68}$$

To calculate $\sigma_{Lr\theta}^{(n,m)}$, we use (B42), (B43), (B54) and (B56),

$$\sigma_{Lr\theta}^{(n,m)} = \frac{\eta}{2\sqrt{\pi}} \sum_{k=1}^{\infty} \sqrt{2k+1} P_k^1(\cos\theta) \Bigg\{ \frac{T_{k0}^{(n,m)}(r)}{r} + \frac{2\Phi_{k0}^{(n,m)\prime}(r)}{r} - \frac{2\Phi_{k0}^{(n,m)}(r)}{r^2}$$

$$+ \frac{A_{k0}^{nmnm}}{4\omega r} \text{Re}\left\{if_{nm}^{\prime*}(r)\left[g_{nm}(r) - rg_{nm}'(r)\right]\right.$$

$$+ if_{nm}^*(r)\left[\frac{2n(n+1) - k(k+1) - 2}{r} g_{nm}(r) + 2g_{nm}'(r) - rg_{nm}''(r) - 2f_{nm}'(r)\right]\Bigg\}\Bigg\}. \tag{B69}$$

Substitution of (B69) into (B65) yields

$$T_{k0}^{(n,m)}(R_0) + 2\Phi_{k0}^{(n,m)\prime}(R_0) - \frac{2}{R_0}\Phi_{k0}^{(n,m)}(R_0) = S_k^{(n,m)}, \tag{B70}$$

where

$$S_k^{(n,m)} = \frac{A_{k0}^{nmnm}}{4\omega} \text{Re}\left\{if_{nm}^{\prime*}(R_0)\left[R_0 g_{nm}'(R_0) - g_{nm}(R_0)\right]\right.$$

$$+ if_{nm}^*(R_0)\left[\frac{2 + k(k+1) - 2n(n+1)}{R_0} g_{nm}(R_0) - 2g_{nm}'(R_0) + R_0 g_{nm}''(R_0) + 2f_{nm}'(R_0)\right]\Bigg\}. \tag{B71}$$

Combining (B67) and (B70) and using (B24), (B33), (B39) and (B45), one finds

$$\bar{C}_{4k}^{(n,m)} = \frac{(2k-1)R_0^k}{(k+1)(2k+1)}\left[2(k+2)Q_k^{(n,m)} - (k+1)S_k^{(n,m)} - (k+3)R_0^{k+1}\bar{C}_{3k}^{(n,m)} - 2(2k+1)R_0^{k-1}I_k^{(n,m)}\right],$$

$$\tag{B72}$$

$$\overline{C}_{6k}^{(n,m)} = \frac{R_0^2}{k+1}\left[\frac{(k+1)(3k-1)}{4k^2-1}\overline{C}_{4k}^{(n,m)} + R_0^{2k+1}\overline{C}_{3k}^{(n,m)} + kR_0^{2k-1}I_k^{(n,m)} - R_0^k Q_k^{(n,m)}\right]. \quad (B73)$$

To calculate $\sigma_{Lr\phi}^{(n,m)}$, we use (B44) and (B62),

$$\sigma_{Lr\phi}^{(n,m)} = \frac{\eta}{r^2}\sum_{k=1}^{\infty}P_k^1(\cos\theta)\left\{\frac{\sqrt{2k+1}}{2\sqrt{\pi}}\left[2P_{k0}^{(n,m)}(r) - rP_{k0}^{(n,m)/}(r)\right] - rE_k^{(n,m)}(r) + r^2 E_k^{(n,m)/}(r)\right\}, \quad (B74)$$

where $P_{k0}^{(n,m)/}(r)$ and $E_k^{(n,m)/}(r)$ are calculated by

$$P_{k0}^{(n,m)/}(r) = (k+1)r^k C_{1k}^{(n,m)}(r) - kr^{-k-1}C_{2k}^{(n,m)}(r), \quad (B75)$$

$$E_k^{(n,m)/}(r) = \begin{cases} E^{(1,1)/}(r)\delta_{1k}, & n=1 \\ E_k^{(n>1,m)/}(r), & n>1 \end{cases}, \quad (B76)$$

$$E^{(1,1)/}(r) = \frac{3}{16\pi\omega r^2}\mathrm{Re}\left\{f_{11}^*(r)\left[rg_{11}'(r) - r^2 g_{11}''(r) - g_{11}(r)\right]\right.$$
$$\left. + rf_{11}'^*(r)\left[g_{11}(r) - rg_{11}'(r)\right] + g_{11}^*(r)[g_{11}(r) - 2rg_{11}'(r)]\right\}, \quad (B77)$$

$$E_k^{(n>1,m)/}(r) = -\frac{E_k^{(n>1,m)}(r)}{r} + \frac{m\sqrt{2k+1}}{4\sqrt{\pi}\omega r}\left\{2\mathrm{Re}\left\{g_{nm}'(r)g_{nm}^*(r)\right\}\left[nC_{(n+1)m}B_{k0}^{(n+1)mnm} - (n+1)C_{nm}B_{k0}^{(n-1)mnm}\right]\right.$$
$$\left. - \sqrt{\frac{(2n+1)(n-m)!}{(n+m)!}}\mathrm{Re}\{f_{nm}'^*(r)[g_{nm}(r) - rg_{nm}'(r)] - rf_{nm}^*(r)g_{nm}''(r) - 2n(n+1)g_{nm}'(r)g_{nm}^*(r)\}\right.$$
$$\left. \times \frac{1}{\sqrt{k(k+1)}}\sum_{s=1}^{[\frac{(n-m+2)}{2}]}\sqrt{\frac{(2n-4s+3)(n+m-2s)!}{(n-m-2s+2)!}}A_{k(-1)}^{(n-2s+1)(m-1)nm}\right\}. \quad (B78)$$

Substituting (B74) into (B66) and using (B18) and (B75), one finds

$$\overline{C}_{2k}^{(n,m)} = \frac{R_0^{k+1}}{k+2}\left\{(k-1)R_0^k\overline{C}_{1k}^{(n,m)} + \frac{2\sqrt{\pi}}{\sqrt{2k+1}}\left[E_k^{(n,m)}(R_0) - R_0 E_k^{(n,m)/}(R_0)\right]\right\}. \quad (B79)$$

## Appendix C. Solving the equations of acoustic streaming for the cross terms

To solve (2.64) and (2.65), we first need to calculate their right-hand sides.

The calculation of $\mathbf{v}_1^{(n,m)}\cdot\nabla\mathbf{v}_1^{(n,-m)}$ yields

$$\mathbf{v}_1^{(n,m)}\cdot\nabla\mathbf{v}_1^{(n,-m)} = \left(v_{1r}^{(n,m)}\frac{\partial}{\partial r} + \frac{v_{1\theta}^{(n,m)}}{r}\frac{\partial}{\partial\theta} + \frac{v_{1\phi}^{(n,m)}}{r\sin\theta}\frac{\partial}{\partial\phi}\right)(v_{1r}^{(n,-m)}\mathbf{e}_r + v_{1\theta}^{(n,-m)}\mathbf{e}_\theta + v_{1\phi}^{(n,-m)}\mathbf{e}_\phi)$$

$$= \boldsymbol{e}_r \left( v_{1r}^{(n,m)} \frac{\partial v_{1r}^{(n,-m)}}{\partial r} + \frac{v_{1\theta}^{(n,m)}}{r} \frac{\partial v_{1r}^{(n,-m)}}{\partial \theta} + \frac{v_{1\phi}^{(n,m)}}{r\sin\theta} \frac{\partial v_{1r}^{(n,-m)}}{\partial \phi} - \frac{v_{1\theta}^{(n,m)} v_{1\theta}^{(n,-m)} + v_{1\phi}^{(n,m)} v_{1\phi}^{(n,-m)}}{r} \right)$$

$$+ \boldsymbol{e}_\theta \left( v_{1r}^{(n,m)} \frac{\partial v_{1\theta}^{(n,-m)}}{\partial r} + \frac{v_{1\theta}^{(n,m)}}{r} \frac{\partial v_{1\theta}^{(n,-m)}}{\partial \theta} + \frac{v_{1\phi}^{(n,m)}}{r\sin\theta} \frac{\partial v_{1\theta}^{(n,-m)}}{\partial \phi} + \frac{v_{1\theta}^{(n,m)} v_{1r}^{(n,-m)}}{r} - \frac{\cos\theta\, v_{1\phi}^{(n,m)} v_{1\phi}^{(n,-m)}}{r\sin\theta} \right)$$

$$+ \boldsymbol{e}_\phi \left( v_{1r}^{(n,m)} \frac{\partial v_{1\phi}^{(n,-m)}}{\partial r} + \frac{v_{1\theta}^{(n,m)}}{r} \frac{\partial v_{1\phi}^{(n,-m)}}{\partial \theta} + \frac{v_{1\phi}^{(n,m)}}{r\sin\theta} \frac{\partial v_{1\phi}^{(n,-m)}}{\partial \phi} + \frac{v_{1\phi}^{(n,m)} v_{1r}^{(n,-m)}}{r} + \frac{\cos\theta\, v_{1\phi}^{(n,m)} v_{1\theta}^{(n,-m)}}{r\sin\theta} \right). \quad (C1)$$

Substituting (2.30) – (2.32) and (2.41) – (2.43) into (C1) and averaging it over time, one obtains

$$\left\langle \boldsymbol{v}_1^{(n,m)} \cdot \nabla \boldsymbol{v}_1^{(n,-m)} \right\rangle$$

$$= \frac{(-1)^m}{2} \operatorname{Re} \left\{ \boldsymbol{e}_r \varepsilon \left\{ V_n V_n^{\prime *} \left( Y_n^m \right)^2 + \frac{W_n (V_n^* - W_n^*)}{r} \left[ \left( \frac{\partial Y_n^m}{\partial \theta} \right)^2 + \frac{1}{\sin^2\theta} \left( \frac{\partial Y_n^m}{\partial \phi} \right)^2 \right] \right\} \right.$$

$$+ \boldsymbol{e}_\theta \varepsilon \left\{ \left( V_n W_n^{\prime *} + \frac{W_n V_n^*}{r} \right) Y_n^m \frac{\partial Y_n^m}{\partial \theta} + \frac{|W_n|^2}{r} \left[ \frac{\partial Y_n^m}{\partial \theta} \frac{\partial^2 Y_n^m}{\partial \theta^2} + \frac{1}{\sin^2\theta} \frac{\partial Y_n^m}{\partial \phi} \frac{\partial^2 Y_n^m}{\partial \theta \partial \phi} - \frac{\cos\theta}{\sin^3\theta} \left( \frac{\partial Y_n^m}{\partial \phi} \right)^2 \right] \right\}$$

$$\left. + \frac{\boldsymbol{e}_\phi \varepsilon}{\sin\theta} \left\{ \left( V_n W_n^{\prime *} + \frac{W_n V_n^*}{r} \right) Y_n^m \frac{\partial Y_n^m}{\partial \phi} + \frac{|W_n|^2}{r} \left( \frac{\partial Y_n^m}{\partial \theta} \frac{\partial^2 Y_n^m}{\partial \theta \partial \phi} + \frac{1}{\sin^2\theta} \frac{\partial Y_n^m}{\partial \phi} \frac{\partial^2 Y_n^m}{\partial \phi^2} \right) \right\} \right\}, \quad (C2)$$

where $\varepsilon = s^{(n,m)} s^{(n,-m)*}$. For simplicity, we drop the arguments of the functions in (C2).

The calculation of the curl of (C2) results in

$$\nabla \times \left\langle \boldsymbol{v}_1^{(n,m)} \cdot \nabla \boldsymbol{v}_1^{(n,-m)} \right\rangle = \frac{(-1)^m}{2} \operatorname{Re} \left\{ \frac{\boldsymbol{e}_\theta \varepsilon}{r\sin\theta} \left\{ \left[ 2 V_n V_n^{\prime *} - \left( r V_n W_n^{\prime *} + W_n V_n^* \right)' \right] Y_n^m \frac{\partial Y_n^m}{\partial \phi} \right. \right.$$

$$\left. + \left[ \frac{2 W_n (V_n^* - W_n^*)}{r} - (W_n W_n^*)' \right] \left( \frac{\partial Y_n^m}{\partial \theta} \frac{\partial^2 Y_n^m}{\partial \theta \partial \phi} + \frac{1}{\sin^2\theta} \frac{\partial Y_n^m}{\partial \phi} \frac{\partial^2 Y_n^m}{\partial \phi^2} \right) \right\}$$

$$+ \frac{\boldsymbol{e}_\phi \varepsilon}{r} \left\{ \left[ \left( r V_n W_n^{\prime *} + W_n V_n^* \right)' - 2 V_n V_n^{\prime *} \right] Y_n^m \frac{\partial Y_n^m}{\partial \theta} \right.$$

$$\left. \left. + \left[ (W_n W_n^*)' - \frac{2 W_n (V_n^* - W_n^*)}{r} \right] \left[ \frac{\partial Y_n^m}{\partial \theta} \frac{\partial^2 Y_n^m}{\partial \theta^2} + \frac{1}{\sin^2\theta} \frac{\partial Y_n^m}{\partial \phi} \frac{\partial^2 Y_n^m}{\partial \theta \partial \phi} - \frac{\cos\theta}{\sin^3\theta} \left( \frac{\partial Y_n^m}{\partial \phi} \right)^2 \right] \right\} \right\}. \quad (C3)$$

Since (C3) does not contain a radial component, (2.64) leads to

$$P_{kl}^{(\times)\prime\prime}(r) - \frac{k(k+1) P_{kl}^{(\times)}(r)}{r^2} = 0. \quad (C4)$$

A solution to (C4) is given by

$$P_{kl}^{(\times)}(r) = \bar{C}_{1k}^{(\times)} r^{k+1} + \bar{C}_{2k}^{(\times)} r^{-k},  \tag{C5}$$

where $\bar{C}_{1k}^{(\times)}$ and $\bar{C}_{2k}^{(\times)}$ are constants. From the condition that the acoustic streaming vanishes at infinity, it follows that

$$\bar{C}_{1k}^{(\times)} = 0,  \tag{C6}$$

while $\bar{C}_{2k}^{(\times)}$ is calculated by boundary conditions at the bubble surface; see below.

The calculation of the $r$-component of $\nabla \times \nabla \times \langle v_1^{(n,m)} \cdot \nabla v_1^{(n,-m)} \rangle$ results in

$$\begin{aligned}
\boldsymbol{e}_r \cdot \left[ \nabla \times \nabla \times \langle v_1^{(n,m)} \cdot \nabla v_1^{(n,-m)} \rangle \right] &= \frac{(-1)^m}{4r^2} \mathrm{Re}\left\{ \varepsilon \left[ 2V_n V_n'^* - \left(rV_n W_n'^* + W_n V_n^*\right)' \right] L^2\left[(Y_n^m)^2\right] \right. \\
&\quad \left. + \varepsilon \left[ \frac{2W_n(V_n^* - W_n^*)}{r} - (W_n W_n^*)' \right] L^2\left[ \left(\frac{\partial Y_n^m}{\partial \theta}\right)^2 + \frac{1}{\sin^2\theta}\left(\frac{\partial Y_n^m}{\partial \phi}\right)^2 \right] \right\},
\end{aligned} \tag{C7}$$

where the operator $L^2$ is given by (B9).

The $r$-component of $\nabla \times \nabla \times \langle v_1^{(n,-m)} \cdot \nabla v_1^{(n,m)} \rangle$ is calculated from (C7) by swapping $m$ with $-m$. Doing so and using (D3), one obtains

$$\begin{aligned}
\boldsymbol{e}_r \cdot &\left[ \nabla \times \nabla \times \langle v_1^{(n,m)} \cdot \nabla v_1^{(n,-m)} + v_1^{(n,-m)} \cdot \nabla v_1^{(n,m)} \rangle \right] \\
&= \frac{(-1)^m}{4r^2} \mathrm{Re}\left\{ \varepsilon \left[ 2V_n V_n'^* + 2V_n^* V_n' - \left(rV_n W_n'^* + W_n V_n^*\right)' - \left(rV_n^* W_n' + W_n^* V_n\right)' \right] L^2\left[(Y_n^m)^2\right] \right. \\
&\quad \left. + 2\varepsilon \left[ \frac{V_n W_n^* + V_n^* W_n - 2W_n W_n^*}{r} - (W_n W_n^*)' \right] L^2\left[ \left(\frac{\partial Y_n^m}{\partial \theta}\right)^2 - \frac{m^2(Y_n^m)^2}{\sin^2\theta} \right] \right\}.
\end{aligned} \tag{C8}$$

By using (D5), it can be shown that the following identity holds:

$$\left(\frac{\partial Y_n^m}{\partial \theta}\right)^2 - \frac{m^2(Y_n^m)^2}{\sin^2\theta} = n(n+1)(Y_n^m)^2 - \frac{1}{2}L^2\left[(Y_n^m)^2\right]. \tag{C9}$$

Substitution of (C9) into (C8) yields

$$\begin{aligned}
\boldsymbol{e}_r \cdot &\left[ \nabla \times \nabla \times \langle v_1^{(n,m)} \cdot \nabla v_1^{(n,-m)} + v_1^{(n,-m)} \cdot \nabla v_1^{(n,m)} \rangle \right] = \frac{(-1)^m}{4r^2} \mathrm{Re}\left\{ \varepsilon L^2\left[(Y_n^m)^2\right]\left\{ 2V_n V_n'^* + 2V_n^* V_n' \right. \right. \\
&\left. \left. - \left(rV_n W_n'^* + V_n^* W_n + rV_n^* W_n' + V_n W_n^*\right)' + 2n(n+1)\left[ \frac{V_n W_n^* + V_n^* W_n - 2W_n W_n^*}{r} - (W_n W_n^*)' \right] \right\} \right.
\end{aligned}$$

$$+\varepsilon L^4 \left[\left(Y_n^m\right)^2\right]\left[(W_n W_n^*)' - \frac{V_n W_n^* + V_n^* W_n - 2W_n W_n^*}{r}\right]\right\}. \tag{C10}$$

Substituting (C10) into (2.65), expressing $\left(Y_n^m\right)^2$ by (D25) and using the identity $L^2 Y_k^l = k(k+1)Y_k^l$, one obtains

$$T_{kl}^{(\times)''}(r) - \frac{k(k+1)}{r^2}T_{kl}^{(\times)}(r) = D_{kl}^{nmnm} G_k^{(\times)}(r), \tag{C11}$$

where the constant coefficients $D_{kl}^{nmnm}$ are calculated by (D27) and $G_k^{(\times)}(r)$ is given by

$$G_k^{(\times)}(r) = \frac{(-1)^m \varepsilon}{2\nu}\mathrm{Re}\left\{V_n^*(r)\left[2V_n'(r) - 2W_n'(r) - rW_n''(r)\right] - V_n'^*(r)\left[W_n(r) + rW_n'(r)\right]\right.$$
$$\left. + \frac{2n(n+1) - k(k+1)}{r}W_n^*(r)\left[V_n(r) - W_n(r) - rW_n'(r)\right]\right\} \tag{C12}$$

A solution to (C11) is given by

$$T_{kl}^{(\times)}(r) = D_{kl}^{nmnm}\left[r^{k+1}C_{3k}^{(\times)}(r) + r^{-k}C_{4k}^{(\times)}(r)\right], \tag{C13}$$

where

$$C_{3k}^{(\times)}(r) = \bar{C}_{3k}^{(\times)} + \frac{1}{2k+1}\int_{R_0}^{r} s^{-k}G_k^{(\times)}(s)ds, \tag{C14}$$

$$C_{4k}^{(\times)}(r) = \bar{C}_{4k}^{(\times)} - \frac{1}{2k+1}\int_{R_0}^{r} s^{k+1}G_k^{(\times)}(s)ds, \tag{C15}$$

the constant $\bar{C}_{3k}^{(\times)}$ is calculated from the condition that the acoustic streaming vanishes at infinity,

$$\bar{C}_{3k}^{(\times)} = -\frac{1}{2k+1}\int_{R_0}^{\infty} s^{-k}G_k^{(\times)}(s)ds, \tag{C16}$$

and the constant $\bar{C}_{4k}^{(\times)}$ is calculated by boundary conditions at the bubble surface; see below.

From (2.63), it follows that

$$\mathbf{v}_E^{(\times)} = \mathrm{Re}\left\{\nabla\times\left[\mathbf{e}_r\sum_{k=1}^{\infty}\sum_{l=-k}^{k}P_{kl}^{(\times)}(r)Y_k^l(\theta,\phi)\right] + \mathbf{e}_r\sum_{k=1}^{\infty}\sum_{l=-k}^{k}T_{kl}^{(\times)}(r)Y_k^l(\theta,\phi) + \nabla\Phi^{(\times)}(r,\theta,\phi)\right\}. \tag{C17}$$

Substitution of (C17) into (2.44) yields

$$\Delta\Phi^{(\times)}(r,\theta,\phi) = -\sum_{k=1}^{\infty}\sum_{l=-k}^{k}\nabla\cdot\left[T_{kl}^{(\times)}(r)Y_k^l(\theta,\phi)\mathbf{e}_r\right] = -\sum_{k=1}^{\infty}\sum_{l=-k}^{k}\left[T_{kl}^{(\times)'}(r) + 2r^{-1}T_{kl}^{(\times)}(r)\right]Y_k^l(\theta,\phi). \tag{C18}$$

We assume that

$$\Phi^{(\times)}(r,\theta,\phi) = \sum_{k=1}^{\infty} \sum_{l=-k}^{k} \Phi_{kl}^{(\times)}(r) Y_k^l(\theta,\phi). \tag{C19}$$

Substituting (C19) into (C18) and using (C13), one obtains

$$\Phi_{kl}^{(\times)\prime\prime}(r) + \frac{2}{r}\Phi_{kl}^{(\times)\prime}(r) - \frac{k(k+1)}{r^2}\Phi_{kl}^{(\times)}(r) = D_{kl}^{nmnm} H_k^{(\times)}(r), \tag{C20}$$

where

$$H_k^{(\times)}(r) = -(k+3)r^k C_{3k}^{(\times)}(r) + (k-2)r^{-k-1} C_{4k}^{(\times)}(r). \tag{C21}$$

Equation (C20) is solved by the method of variation of parameters, which results in

$$\Phi_{kl}^{(\times)}(r) = D_{kl}^{nmnm}\left[ r^k C_{5k}^{(\times)}(r) + r^{-k-1} C_{6k}^{(\times)}(r) \right], \tag{C22}$$

where $C_{5k}^{(\times)}(r)$ and $C_{6k}^{(\times)}(r)$ obey the following equations:

$$r^k C_{5k}^{(\times)\prime}(r) + r^{-k-1} C_{6k}^{(\times)\prime}(r) = 0, \tag{C23}$$

$$kr^{k-1} C_{5k}^{(\times)\prime}(r) - (k+1)r^{-k-2} C_{6k}^{(\times)\prime}(r) = H_k^{(\times)}(r). \tag{C24}$$

Solutions to (C23) and (C24) are given by

$$C_{5k}^{(\times)}(r) = \bar{C}_{5k}^{(\times)} + \frac{1}{2k+1}\int_{R_0}^{r} s^{1-k} H_k^{(\times)}(s)ds, \tag{C25}$$

$$C_{6k}^{(\times)}(r) = \bar{C}_{6k}^{(\times)} - \frac{1}{2k+1}\int_{R_0}^{r} s^{k+2} H_k^{(\times)}(s)ds, \tag{C26}$$

where $\bar{C}_{5k}^{(\times)}$ and $\bar{C}_{6k}^{(\times)}$ are constants. $\bar{C}_{6k}^{(\times)}$ is calculated by boundary conditions at the bubble surface; see below. $\bar{C}_{5k}^{(\times)}$ is calculated from the condition that the acoustic streaming vanishes at infinity, which gives

$$\bar{C}_{5k}^{(\times)} = -\frac{1}{2k+1}\int_{R_0}^{\infty} s^{1-k} H_k^{(\times)}(s)ds. \tag{C27}$$

Substituting (C21) into (C27) and using (C15), one obtains

$$\bar{C}_{5k}^{(\times)} = -\frac{(k-2)\bar{C}_{4k}^{(\times)}}{(4k^2-1)R_0^{2k-1}} + I_k^{(\times)}, \tag{C28}$$

where

$$I_k^{(\times)} = \frac{k+3}{2k+1}\int_{R_0}^{\infty} r C_{3k}^{(\times)}(r)dr + \frac{k-2}{(2k+1)^2}\int_{R_0}^{\infty} r^{-2k}\left[\int_{R_0}^{r} s^{k+1} G_k^{(\times)}(s)ds\right]dr. \tag{C29}$$

Substitution of (C19) into (C17) yields

$$v_{Er}^{(\times)} = \operatorname{Re} \sum_{k=1}^{\infty} \sum_{l=-k}^{k} \left[ T_{kl}^{(\times)}(r) + \Phi_{kl}^{(\times)\prime}(r) \right] Y_k^l(\theta,\phi), \tag{C30}$$

$$v_{E\theta}^{(\times)} = \operatorname{Re} \sum_{k=1}^{\infty} \sum_{l=-k}^{k} \left[ \frac{\Phi_{kl}^{(\times)}(r)}{r} \frac{\partial Y_k^l(\theta,\phi)}{\partial \theta} + \frac{P_{kl}^{(\times)}(r)}{r \sin\theta} \frac{\partial Y_k^l(\theta,\phi)}{\partial \phi} \right], \tag{C31}$$

$$v_{E\phi}^{(\times)} = \operatorname{Re} \sum_{k=1}^{\infty} \sum_{l=-k}^{k} \left[ \frac{\Phi_{kl}^{(\times)}(r)}{r \sin\theta} \frac{\partial Y_k^l(\theta,\phi)}{\partial \phi} - \frac{P_{kl}^{(\times)}(r)}{r} \frac{\partial Y_k^l(\theta,\phi)}{\partial \theta} \right]. \tag{C32}$$

The function $\Phi_{kl}^{(\times)\prime}(r)$, which appears in (C30), is calculated by (C22) and (C23) to be

$$\Phi_{kl}^{(\times)\prime}(r) = D_{kl}^{nmnm} \left[ kr^{k-1} C_{5k}^{(\times)}(r) - (k+1) r^{-k-2} C_{6k}^{(\times)}(r) \right]. \tag{C33}$$

In order to go on with the calculation, we need to apply the boundary conditions for the acoustic streaming at the bubble surface. To do this, we need to know the Stokes drift velocity, which is calculated by

$$\boldsymbol{v}_S^{(\times)} = \frac{1}{2\omega} \operatorname{Re} \left\{ i\boldsymbol{v}_1^{(n,m)} \cdot \nabla \boldsymbol{v}_1^{(n,-m)*} + i\boldsymbol{v}_1^{(n,-m)} \cdot \nabla \boldsymbol{v}_1^{(n,m)*} \right\}. \tag{C34}$$

The expression $\operatorname{Re}\left\{ i\boldsymbol{v}_1^{(n,m)} \cdot \nabla \boldsymbol{v}_1^{(n,-m)*} \right\}$ is calculated by

$$\operatorname{Re}\left\{ i\boldsymbol{v}_1^{(n,m)} \cdot \nabla \boldsymbol{v}_1^{(n,-m)*} \right\}$$
$$= \operatorname{Re}\left\{ \boldsymbol{e}_r i \left( v_{1r}^{(n,m)} \frac{\partial v_{1r}^{(n,-m)*}}{\partial r} + \frac{v_{1\theta}^{(n,m)}}{r} \frac{\partial v_{1r}^{(n,-m)*}}{\partial \theta} + \frac{v_{1\phi}^{(n,m)}}{r\sin\theta} \frac{\partial v_{1r}^{(n,-m)*}}{\partial \phi} - \frac{v_{1\theta}^{(n,m)} v_{1\theta}^{(n,-m)*}}{r} - \frac{v_{1\phi}^{(n,m)} v_{1\phi}^{(n,-m)*}}{r} \right) \right.$$
$$+ \boldsymbol{e}_\theta i \left( v_{1r}^{(n,m)} \frac{\partial v_{1\theta}^{(n,-m)*}}{\partial r} + \frac{v_{1\theta}^{(n,m)}}{r} \frac{\partial v_{1\theta}^{(n,-m)*}}{\partial \theta} + \frac{v_{1\phi}^{(n,m)}}{r\sin\theta} \frac{\partial v_{1\theta}^{(n,-m)*}}{\partial \phi} + \frac{v_{1\theta}^{(n,m)} v_{1r}^{(n,-m)*}}{r} - \frac{\cos\theta\, v_{1\phi}^{(n,m)} v_{1\phi}^{(n,-m)*}}{r\sin\theta} \right)$$
$$\left. + \boldsymbol{e}_\phi i \left( v_{1r}^{(n,m)} \frac{\partial v_{1\phi}^{(n,-m)*}}{\partial r} + \frac{v_{1\theta}^{(n,m)}}{r} \frac{\partial v_{1\phi}^{(n,-m)*}}{\partial \theta} + \frac{v_{1\phi}^{(n,m)}}{r\sin\theta} \frac{\partial v_{1\phi}^{(n,-m)*}}{\partial \phi} + \frac{v_{1\phi}^{(n,m)} v_{1r}^{(n,-m)*}}{r} + \frac{\cos\theta\, v_{1\phi}^{(n,m)} v_{1\theta}^{(n,-m)*}}{r\sin\theta} \right) \right\}. \tag{C35}$$

Substituting (2.30) – (2.32) and (2.41) – (2.43) into (C35) and using (C9), one obtains

$$\operatorname{Re}\left\{ i\boldsymbol{v}_1^{(n,m)} \cdot \nabla \boldsymbol{v}_1^{(n,-m)*} \right\}$$
$$= (-1)^m \operatorname{Re}\left\{ \boldsymbol{e}_r i\varepsilon \left\{ \left[ V_n V_n^{\prime *} + \frac{n(n+1) W_n (V_n^* - W_n^*)}{r} \right] \left( Y_n^m \right)^2 - \frac{W_n (V_n^* - W_n^*)}{2r} L^2 \left[ \left( Y_n^m \right)^2 \right] \right\} \right.$$
$$\left. + \boldsymbol{e}_\theta \frac{i\varepsilon}{2} \frac{\partial}{\partial\theta} \left\{ \left[ V_n W_n^{\prime *} + \frac{W_n \left( V_n^* + n(n+1) W_n^* \right)}{r} \right] \left( Y_n^m \right)^2 - \frac{W_n W_n^*}{2r} L^2 \left[ \left( Y_n^m \right)^2 \right] \right\} \right.$$

$$+\mathbf{e}_\phi \frac{i\varepsilon}{2\sin\theta} \frac{\partial}{\partial \phi} \left\{ \left[ V_n W_n'^* + \frac{W_n \left( V_n^* + n(n+1) W_n^* \right)}{r} \right] \left( Y_n^m \right)^2 - \frac{W_n W_n^*}{2r} L^2 \left[ \left( Y_n^m \right)^2 \right] \right\}. \quad (C36)$$

The expression $\operatorname{Re}\{i\mathbf{v}_1^{(n,-m)} \cdot \nabla \mathbf{v}_1^{(n,m)*}\}$ is calculated from (C36) by swapping $m$ with $-m$. Doing so and using (D3), one obtains

$$v_{Sr}^{(\times)} = \frac{(-1)^m}{2\omega} \operatorname{Re}\left\{ i\varepsilon \left[ \left[ V_n V_n'^* - V_n^* V_n' - \frac{n(n+1)\left( V_n W_n^* - V_n^* W_n \right)}{r} \right] \left( Y_n^m \right)^2 \right. \right.$$

$$\left. \left. + \frac{V_n W_n^* - V_n^* W_n}{2r} L^2 \left[ \left( Y_n^m \right)^2 \right] \right] \right\}, \quad (C37)$$

$$v_{S\theta}^{(\times)} = \frac{(-1)^m}{4\omega} \operatorname{Re}\left\{ i\varepsilon \left[ V_n W_n'^* - V_n^* W_n' - \frac{V_n W_n^* - V_n^* W_n}{r} \right] \frac{\partial \left( Y_n^m \right)^2}{\partial \theta} \right\}, \quad (C38)$$

$$v_{S\phi}^{(\times)} = \frac{(-1)^m}{4\omega} \operatorname{Re}\left\{ i\varepsilon \left[ V_n W_n'^* - V_n^* W_n' - \frac{V_n W_n^* - V_n^* W_n}{r} \right] \frac{1}{\sin\theta} \frac{\partial \left( Y_n^m \right)^2}{\partial \phi} \right\}. \quad (C39)$$

With the help of (D25) and the identity $L^2 Y_k^l = k(k+1) Y_k^l$, (C37) – (C39) are transformed to

$$v_{Sr}^{(\times)} = \operatorname{Re} \sum_{k=1}^{\infty} S_k^{(\times)}(r) \sum_{l=-k}^{k} D_{kl}^{nmnm} Y_k^l(\theta,\phi), \quad (C40)$$

$$v_{S\theta}^{(\times)} = \operatorname{Re}\left\{ U_{nm}(r) \sum_{k=1}^{\infty} \sum_{l=-k}^{k} D_{kl}^{nmnm} \frac{\partial Y_k^l(\theta,\phi)}{\partial \theta} \right\}, \quad (C41)$$

$$v_{S\phi}^{(\times)} = \operatorname{Re}\left\{ i U_{nm}(r) \sum_{k=1}^{\infty} \sum_{l=-k}^{k} l D_{kl}^{nmnm} \frac{Y_k^l(\theta,\phi)}{\sin\theta} \right\}, \quad (C42)$$

where

$$S_k^{(\times)}(r) = \frac{(-1)^m \varepsilon}{\omega} \operatorname{Im}\left\{ V_n^*(r) \left[ V_n'(r) - \frac{n(n+1)}{r} W_n(r) \right] + \frac{k(k+1)}{2r} V_n^*(r) W_n(r) \right\}, \quad (C43)$$

$$U_{nm}(r) = \frac{(-1)^m \varepsilon}{2\omega} \operatorname{Im}\left\{ V_n^*(r) \left[ W_n'(r) - \frac{W_n(r)}{r} \right] \right\}. \quad (C44)$$

Equations (C41) and (C42) are convenient to use in the boundary conditions at the bubble surface. However, for the numerical calculation of $v_{S\theta}^{(\times)}$ and $v_{S\phi}^{(\times)}$, it is convenient to transform (C41) and (C42).

With the help of (D9), (C41) is transformed to

$$v_{S\theta}^{(\times)} = \frac{1}{2}\text{Re}\{U_{nm}(r)$$

$$\times \sum_{k=1}^{\infty}\sum_{l=-k}^{k} D_{kl}^{nmnm}\left[\sqrt{k(k+1)-l(l+1)}Y_k^{l+1}(\theta,\phi)e^{-i\phi} - \sqrt{k(k+1)-l(l-1)}Y_k^{l-1}(\theta,\phi)e^{i\phi}\right]\}. \quad \text{(C45)}$$

With the help of (D16), (C42) is transformed to

$$v_{S\phi}^{(\times)} = -\text{Re}\left\{iU_{nm}(r)e^{i\phi}\sum_{k=1}^{\infty}\sum_{l=-k}^{k} l\sqrt{\frac{(2k+1)(k-l)!}{(k+l)!}}D_{kl}^{nmnm}\right.$$

$$\left.\times \sum_{s=1}^{[(k-l+2)/2]}\sqrt{\frac{(2k-4s+3)(k+l-2s)!}{(k-l-2s+2)!}}Y_{k-2s+1}^{l-1}(\theta,\phi)\right\}. \quad \text{(C46)}$$

We can now apply the boundary conditions for the acoustic streaming at the bubble surface. They are given by

$$v_{Lr}^{(\times)} = 0 \text{ at } r = R_0, \quad \text{(C47)}$$

$$\sigma_{Lr\theta}^{(\times)} = \eta\left(\frac{1}{r}\frac{\partial v_{Lr}^{(\times)}}{\partial \theta} + \frac{\partial v_{L\theta}^{(\times)}}{\partial r} - \frac{v_{L\theta}^{(\times)}}{r}\right) = 0 \text{ at } r = R_0, \quad \text{(C48)}$$

$$\sigma_{Lr\phi}^{(\times)} = \eta\left(\frac{\partial v_{L\phi}^{(\times)}}{\partial r} + \frac{1}{r\sin\theta}\frac{\partial v_{Lr}^{(\times)}}{\partial \phi} - \frac{v_{L\phi}^{(\times)}}{r}\right) = 0 \text{ at } r = R_0, \quad \text{(C49)}$$

where $\boldsymbol{v}_L^{(\times)} = \boldsymbol{v}_E^{(\times)} + \boldsymbol{v}_S^{(\times)}$ is the Lagrangian streaming velocity and $\sigma_{Lr\theta}^{(\times)}$ and $\sigma_{Lr\phi}^{(\times)}$ are the tangential components of the stress produced by $\boldsymbol{v}_L^{(\times)}$. We use (C47) – (C49) in order to calculate the constants $\overline{C}_{2k}^{(\times)}$, $\overline{C}_{4k}^{(\times)}$ and $\overline{C}_{6k}^{(\times)}$.

Substituting (C30) and (C40) into (C47) and using (C13), (C28) and (C33), one obtains

$$\frac{(k+1)(3k-1)}{4k^2-1}\overline{C}_{4k}^{(\times)} - (k+1)R_0^{-2}\overline{C}_{6k}^{(\times)} = -R_0^{2k+1}\overline{C}_{3k}^{(\times)} - kR_0^{2k-1}I_k^{(\times)} - R_0^k S_k^{(\times)}(R_0). \quad \text{(C50)}$$

The calculation of $\sigma_{Lr\theta}^{(\times)}$ and $\sigma_{Lr\phi}^{(\times)}$ results in

$$\sigma_{Lr\theta}^{(\times)} = \eta\,\text{Re}\sum_{k=1}^{\infty}\sum_{l=-k}^{k}\left\{D_{kl}^{nmnm}\left[r^k C_{3k}^{(\times)}(r) + r^{-k-1}C_{4k}^{(\times)}(r) + 2(k-1)r^{k-2}C_{5k}^{(\times)}(r) - 2(k+2)r^{-k-3}C_{6k}^{(\times)}(r)\right.\right.$$

$$\left.\left. + \frac{S_k^{(\times)}(r) - U_{nm}(r) + rU_{nm}'(r)}{r}\right]\frac{\partial Y_k^l(\theta,\phi)}{\partial \theta} - il(k+2)\overline{C}_{2k}^{(\times)}r^{-k-2}\frac{Y_k^l(\theta,\phi)}{\sin\theta}\right\}, \quad \text{(C51)}$$

$$\sigma_{Lr\phi}^{(\times)} = \eta \operatorname{Re} \sum_{k=1}^{\infty} \sum_{l=-k}^{k} \left\{ D_{kl}^{nmnm} \left[ r^k C_{3k}^{(\times)}(r) + r^{-k-1} C_{4k}^{(\times)}(r) + 2(k-1) r^{k-2} C_{5k}^{(\times)}(r) - 2(k+2) r^{-k-3} C_{6k}^{(\times)}(r) \right. \right.$$

$$\left. \left. + \frac{S_k^{(\times)}(r) - U_{nm}(r) + r U_{nm}'(r)}{r} \right] \frac{ilY_k^l}{\sin\theta} + (k+2) \bar{C}_{2k}^{(\times)} r^{-k-2} \frac{\partial Y_k^l}{\partial \theta} \right\}. \tag{C52}$$

Equations (C51) and (C52) show that (C48) and (C49) are satisfied if $\bar{C}_{2k}^{(\times)} = 0$. In this case, (C48) and (C49) gives

$$\frac{2k^2 + 6k - 5}{4k^2 - 1} \bar{C}_{4k}^{(\times)} - 2(k+2) R_0^{-2} \bar{C}_{6k}^{(\times)}$$

$$= -R_0^{2k+1} \bar{C}_{3k}^{(\times)} - 2(k-1) R_0^{2k-1} I_k^{(\times)} - R_0^k \left[ S_k^{(\times)}(R_0) - U_{nm}(R_0) + R_0 U_{nm}'(R_0) \right]. \tag{C53}$$

Combining (C50) and (C53), one obtains

$$\bar{C}_{4k}^{(\times)} = -\frac{(k+3)(2k-1)}{(k+1)(2k+1)} R_0^{2k+1} \bar{C}_{3k}^{(\times)} - \frac{2(2k-1)}{k+1} R_0^{2k-1} I_k^{(\times)}$$

$$- \frac{2k-1}{2k+1} R_0^k \left[ \frac{k+3}{k+1} S_k^{(\times)}(R_0) + U_{nm}(R_0) - R_0 U_{nm}'(R_0) \right], \tag{C54}$$

$$\bar{C}_{6k}^{(\times)} = \frac{3k-1}{4k^2-1} R_0^2 \bar{C}_{4k}^{(\times)} + \frac{R_0^{k+2}}{k+1} \left[ R_0^{k+1} \bar{C}_{3k}^{(\times)} + k R_0^{k-1} I_k^{(\times)} + S_k^{(\times)}(R_0) \right], \tag{C55}$$

where

$$U_{nm}'(r) = \frac{(-1)^m \varepsilon}{2\omega} \operatorname{Im} \left\{ V_n'^*(r) \left[ W_n'(r) - \frac{W_n(r)}{r} \right] + V_n^*(r) \left[ W_n''(r) - \frac{W_n'(r)}{r} + \frac{W_n(r)}{r^2} \right] \right\}. \tag{C56}$$

Since $\bar{C}_{1k}^{(\times)} = \bar{C}_{2k}^{(\times)} = 0$ and hence $P_{kl}^{(\times)}(r) \equiv 0$, $v_{E\theta}^{(\times)}$ and $v_{E\phi}^{(\times)}$, given by (C31) and (C32), are represented by using (D9) and (D16) as follows:

$$v_{E\theta}^{(\times)} = \operatorname{Re} \sum_{k=1}^{\infty} \sum_{l=-k}^{k} \frac{\Phi_{kl}^{(\times)}(r)}{r} \frac{\partial Y_k^l(\theta,\phi)}{\partial \theta}$$

$$= \operatorname{Re} \sum_{k=1}^{\infty} \sum_{l=-k}^{k} \frac{\Phi_{kl}^{(\times)}(r)}{2r} \left[ \sqrt{k(k+1) - l(l+1)} Y_k^{l+1}(\theta,\phi) e^{-i\phi} - \sqrt{k(k+1) - l(l-1)} Y_k^{l-1}(\theta,\phi) e^{i\phi} \right], \tag{C57}$$

$$v_{E\phi}^{(\times)} = \operatorname{Re} \sum_{k=1}^{\infty} \sum_{l=-k}^{k} \frac{il\Phi_{kl}^{(\times)}(r)}{r} \frac{Y_k^l(\theta,\phi)}{\sin\theta}$$

$$= -\operatorname{Re} \sum_{k=1}^{\infty} \sum_{l=-k}^{k} \frac{il\Phi_{kl}^{(\times)}(r) e^{i\phi}}{r} \sqrt{\frac{(2k+1)(k-l)!}{(k+l)!}} \sum_{s=1}^{[(k-l+2)/2]} \sqrt{\frac{(2k-4s+3)(k+l-2s)!}{(k-l-2s+2)!}} Y_{k-2s+1}^{l-1}(\theta,\phi). \tag{C58}$$

## Appendix D. Mathematical identities used in calculations

In our derivation, the function $Y_n^m(\theta,\phi)$ is defined by the following equations (Varshalovich *et al.* 1988):

$$Y_n^m(\theta,\phi) = \sqrt{\frac{(2n+1)(n-m)!}{4\pi(n+m)!}} e^{im\phi} P_n^m(\cos\theta), \quad 0 \leq m \leq n, \tag{D1}$$

$$P_n^m(\mu) = (-1)^m (1-\mu^2)^{\frac{m}{2}} \frac{d^m P_n(\mu)}{d\mu^m}, \quad m \geq 0, \tag{D2}$$

$$Y_n^{-m}(\theta,\phi) = (-1)^m Y_n^m(\theta,-\phi) = (-1)^m Y_n^{m*}(\theta,\phi), \tag{D3}$$

$$\int_0^{2\pi} d\phi \int_0^{\pi} d\theta \sin\theta Y_{n_1}^{m_1}(\theta,\phi) Y_{n_2}^{m_2*}(\theta,\phi) = \delta_{n_1 n_2} \delta_{m_1 m_2}, \tag{D4}$$

where $P_n(\mu)$ is the Legendre polynomial of degree $n$, $P_n^m(\mu)$ is the associated Legendre polynomial of order $m$ and degree $n$, $\mu = \cos\theta$, $\delta_{nm}$ is the Kronecker delta and the asterisk denotes the complex conjugate.

In the process of calculations, the following identities are used (Varshalovich *et al.* 1988):

$$\frac{\partial^2 Y_n^m(\theta,\phi)}{\partial \theta^2} + \cot\theta \frac{\partial Y_n^m(\theta,\phi)}{\partial \theta} = \frac{m^2 Y_n^m(\theta,\phi)}{\sin^2\theta} - n(n+1) Y_n^m(\theta,\phi), \tag{D5}$$

$$\frac{dP_n(\mu)}{d\mu} = \sum_{k=1}^{[(n+1)/2]} (2n-4k+3) P_{n-2k+1}(\mu), \quad n \geq 1, \tag{D6}$$

$$\int_0^{2\pi} d\phi \int_0^{\pi} d\theta \sin\theta Y_{n_1}^{m_1}(\theta,\phi) Y_{n_2}^{m_2}(\theta,\phi) Y_{n_3}^{m_3*}(\theta,\phi) = \sqrt{\frac{(2n_1+1)(2n_2+1)}{4\pi(2n_3+1)}} C_{n_1 0 n_2 0}^{n_3 0} C_{n_1 m_1 n_2 m_2}^{n_3 m_3}, \tag{D7}$$

$$\sin\theta \frac{\partial Y_n^m(\theta,\phi)}{\partial \theta} = n C_{(n+1)m} Y_{n+1}^m(\theta,\phi) - (n+1) C_{nm} Y_{n-1}^m(\theta,\phi), \quad n \geq 1, \tag{D8}$$

$$\frac{\partial Y_n^m(\theta,\phi)}{\partial \theta} = \frac{1}{2}\sqrt{n(n+1) - m(m+1)} Y_n^{m+1}(\theta,\phi) e^{-i\phi} - \frac{1}{2}\sqrt{n(n+1) - m(m-1)} Y_n^{m-1}(\theta,\phi) e^{i\phi}, \tag{D9}$$

$$\cos\theta Y_n^m(\theta,\phi) = C_{(n+1)m} Y_{n+1}^m(\theta,\phi) + C_{nm} Y_{n-1}^m(\theta,\phi), \tag{D10}$$

where $C_{nm}$ is defined by

$$C_{nm} = \sqrt{\frac{n^2 - m^2}{(2n-1)(2n+1)}}, \tag{D11}$$

[ ] means the integer part of an expression in brackets, and $C_{n_1 m_1 n_2 m_2}^{n_3 m_3} = \langle n_1 m_1 n_2 m_2 | n_3 m_3 \rangle$ are the Clebsch-Gordan coefficients. The Clebsch-Gordan coefficients are zero unless the following conditions are satisfied: $m_3 = m_1 + m_2$, $n_1 + n_2 - n_3 \geq 0$, $n_1 - n_2 + n_3 \geq 0$, $-n_1 + n_2 + n_3 \geq 0$, $|m_1| \leq n_1$, $|m_2| \leq n_2$, $|m_3| \leq n_3$.

In our derivation, we use the fact that an arbitrary function $f(\theta, \phi)$ can be expanded in spherical harmonics by

$$f(\theta, \phi) = \sum_{n=0}^{\infty} \sum_{m=-n}^{n} a_{nm} Y_n^m(\theta, \phi), \tag{D12}$$

where the expansion coefficients are calculated by

$$a_{nm} = \int_0^{2\pi} d\phi \int_0^{\pi} d\theta \sin\theta Y_n^{m*}(\theta, \phi) f(\theta, \phi). \tag{D13}$$

From (D13), by using (D3), one obtains

$$a_{n(-m)} = (-1)^m \left[ \int_0^{2\pi} d\phi \int_0^{\pi} d\theta \sin\theta Y_n^{m*}(\theta, \phi) f^*(\theta, \phi) \right]^*. \tag{D14}$$

If $f(\theta, \phi)$ is a real function, then it follows from (D14) that

$$a_{n(-m)} = (-1)^m a_{nm}^*. \tag{D15}$$

With the help of (D1), (D2) and (D6), one obtains

$$\frac{Y_n^m(\theta, \phi)}{\sin\theta} = \frac{1}{\sqrt{1-\mu^2}} \sqrt{\frac{(2n+1)(n-m)!}{4\pi(n+m)!}} e^{im\phi} (-1)^m (1-\mu^2)^{\frac{m}{2}} \frac{d^m P_n(\mu)}{d\mu^m}$$

$$= -e^{i\phi} \sqrt{\frac{(2n+1)(n-m)!}{4\pi(n+m)!}} e^{i(m-1)\phi} (-1)^{m-1} (1-\mu^2)^{\frac{m-1}{2}} \frac{d^{m-1}}{d\mu^{m-1}} \left( \frac{dP_n(\mu)}{d\mu} \right)$$

$$= -e^{i\phi} \sqrt{\frac{(2n+1)(n-m)!}{4\pi(n+m)!}} e^{i(m-1)\phi} (-1)^{m-1} (1-\mu^2)^{\frac{m-1}{2}} \sum_{k=1}^{[(n+1)/2]} (2n-4k+3) \frac{d^{m-1} P_{n-2k+1}(\mu)}{d\mu^{m-1}}$$

$$= -e^{i\phi} \sqrt{\frac{(2n+1)(n-m)!}{(n+m)!}} \sum_{k=1}^{[(n-m+2)/2]} \sqrt{\frac{(2n-4k+3)(n+m-2k)!}{(n-m-2k+2)!}} Y_{n-2k+1}^{m-1}(\theta, \phi), \quad n, m \geq 1. \tag{D16}$$

Let us expand $Y_{n_1}^{m_1}(\theta, \phi) Y_{n_2}^{m_2 *}(\theta, \phi)$ in spherical harmonics. According to (D12) and (D13), we obtain

$$Y_{n_1}^{m_1}(\theta, \phi) Y_{n_2}^{m_2 *}(\theta, \phi) = \sum_{n=0}^{\infty} \sum_{m=-n}^{n} A_{nm}^{n_1 m_1 n_2 m_2} Y_n^m(\theta, \phi), \tag{D17}$$

where

$$A_{nm}^{n_1 m_1 n_2 m_2} = \int_0^{2\pi} d\phi \int_0^\pi d\theta \sin\theta Y_n^{m*}(\theta,\phi) Y_{n_1}^{m_1}(\theta,\phi) Y_{n_2}^{m_2*}(\theta,\phi). \tag{D18}$$

It follows from (D7) that

$$\int_0^{2\pi} d\phi \int_0^\pi d\theta \sin\theta Y_n^{m*}(\theta,\phi) Y_{n_1}^{m_1}(\theta,\phi) Y_{n_2}^{m_2*}(\theta,\phi) = \sqrt{\frac{(2n+1)(2n_2+1)}{4\pi(2n_1+1)}} C_{n0n_20}^{n_10} C_{nmn_2m_2}^{n_1m_1}. \tag{D19}$$

This equation is nonzero only if $m + m_2 = m_1$ and $|n_1 - n_2| \leq n \leq n_1 + n_2$. Therefore, $A_{nm}^{n_1 m_1 n_2 m_2}$ is calculated by

$$A_{nm}^{n_1 m_1 n_2 m_2} = \begin{cases} \sqrt{\dfrac{(2n+1)(2n_2+1)}{4\pi(2n_1+1)}} C_{n0n_20}^{n_10} C_{nmn_2m_2}^{n_1m_1}, & m = m_1 - m_2 \text{ and } |n_1 - n_2| \leq n \leq n_1 + n_2 \\ 0, & m \neq m_1 - m_2 \text{ or } n < |n_1 - n_2| \text{ or } n > n_1 + n_2 \end{cases}. \tag{D20}$$

Note that $A_{nm}^{n_1 m_1 n_2 m_2}$ is real. It also follows from the properties of the Clebsch-Gordan coefficients that $A_{nm}^{n_1 m_1 n_2 m_2} = 0$ unless $|m_1| \leq n_1$, $|m_2| \leq n_2$ and $|m| \leq n$. This fact should be kept in mind when implementing a numerical code.

The use of (D16) and (D17) gives the following expansion:

$$\frac{Y_{n_1}^{m_1}(\theta,\phi) Y_{n_2}^{m_2*}(\theta,\phi)}{\sin^2\theta} = \sum_{n=0}^\infty \sum_{m=-n}^n B_{nm}^{n_1 m_1 n_2 m_2} Y_n^m(\theta,\phi), \quad n_{1,2}, m_{1,2} \geq 1, \tag{D21}$$

where

$$B_{nm}^{n_1 m_1 n_2 m_2} = \sqrt{\frac{(2n_1+1)(2n_2+1)(n_1-m_1)!(n_2-m_2)!}{(n_1+m_1)!(n_2+m_2)!}}$$

$$\times \sum_{k_1=1}^{[(n_1-m_1+2)/2]} \sum_{k_2=1}^{[(n_2-m_2+2)/2]} \sqrt{\frac{(2n_1-4k_1+3)(2n_2-4k_2+3)(n_1+m_1-2k_1)!(n_2+m_2-2k_2)!}{(n_1-m_1-2k_1+2)!(n_2-m_2-2k_2+2)!}}$$

$$\times A_{nm}^{(n_1-2k_1+1)(m_1-1)(n_2-2k_2+1)(m_2-1)}. \tag{D22}$$

Note that $B_{nm}^{n_1 m_1 n_2 m_2}$ is real. It should be emphasized that (D21) and (D22) are not valid if $m_1 = 0$ or $m_2 = 0$. In our derivation, such terms vanish. However, in order to avoid problems in numerical simulations, one should specify $B_{nm}^{n_1 m_1 n_2 m_2} = 0$ if $m_1 = 0$ or $m_2 = 0$. It also follows from (D21) that $B_{nm}^{n_1 m_1 n_2 m_2} = 0$ if $|m_1| > n_1$ or $|m_2| > n_2$ or $|m| > n$.

In order to transform the angle-dependent functions that appear in (B64), we apply the following equations.

With the help of (D16) and (D17), we obtain the following identity:

$$\frac{Y_{n_1}^{m_1}(\theta,\phi)Y_{n_2}^{m_2*}(\theta,\phi)}{\sin\theta} = -\sqrt{\frac{(2n_1+1)(n_1-m_1)!}{(n_1+m_1)!}}\sum_{n=0}^{\infty}\sum_{m=-n}^{n}Y_n^m(\theta,\phi)e^{i\phi}$$

$$\times \sum_{k=1}^{[(n_1-m_1+2)/2]}\sqrt{\frac{(2n_1-4k+3)(n_1+m_1-2k)!}{(n_1-m_1-2k+2)!}}A_{nm}^{(n_1-2k+1)(m_1-1)n_2 m_2}, \quad n_1, m_1 \geq 1. \quad (D23)$$

The use of (D8) and (D21) gives

$$\frac{1}{\sin\theta}\frac{\partial Y_{n_1}^{m_1}(\theta,\phi)}{\partial\theta}Y_{n_2}^{m_2*}(\theta,\phi) = \sum_{n=0}^{\infty}\sum_{m=-n}^{n}\left[n_1 C_{(n_1+1)m_1}B_{nm}^{(n_1+1)m_1 n_2 m_2} - (n_1+1)C_{n_1 m_1}B_{nm}^{(n_1-1)m_1 n_2 m_2}\right]Y_n^m(\theta,\phi)$$

$$n_1 \geq 1, \; n_2, m_{1,2} \geq 1. \quad (D24)$$

Let us expand $Y_{n_1}^{m_1}(\theta,\phi)Y_{n_2}^{m_2}(\theta,\phi)$ in spherical harmonics. According to (D12) and (D13), we obtain

$$Y_{n_1}^{m_1}(\theta,\phi)Y_{n_2}^{m_2}(\theta,\phi) = \sum_{k=0}^{\infty}\sum_{l=-k}^{k}D_{kl}^{n_1 m_1 n_2 m_2}Y_k^l(\theta,\phi), \quad (D25)$$

where

$$D_{kl}^{n_1 m_1 n_2 m_2} = \int_0^{2\pi}d\phi\int_0^{\pi}d\theta\sin\theta Y_k^{l*}(\theta,\phi)Y_{n_1}^{m_1}(\theta,\phi)Y_{n_2}^{m_2}(\theta,\phi). \quad (D26)$$

It follows from (D7) that

$$D_{kl}^{n_1 m_1 n_2 m_2} = \sqrt{\frac{(2n_1+1)(2n_2+1)}{4\pi(2k+1)}}C_{n_1 0 n_2 0}^{k0}C_{n_1 m_1 n_2 m_2}^{kl}. \quad (D27)$$

The coefficients $D_{kl}^{n_1 m_1 n_2 m_2}$ are nonzero only if the following conditions are satisfied: $l = m_1 + m_2$, $|n_1 - n_2| \leq k \leq n_1 + n_2$, $|m_1| \leq n_1$, $|m_2| \leq n_2$, $|l| \leq k$.

Abramowitz, M. & Stegun, I.A. 1972 *Handbook of Mathematical Functions*. Dover.

Aghakhani, A., Yasa, O., Wrede, P. & Sitti, M. 2020 Acoustically powered surface-slipping mobile microrobots, PNAS **117** (7), 3469-3477.

Ahmed, D., Lu, M., Nourhani, A., Lammert, P.E., Stratton, Z., Muddana, H.S., Crespi V.H. & Huang, T.J. 2015 Selectively manipulable acoustic-powered microswimmers, *Scientific Reports* **5**, 9744.

Ahmed, D., Ozcelik, A., Bojanala, N., Nama, N., Upadhyay, A, Chen, Y., Hanna-Rose, W & Huang, T.J. 2016 Rotational manipulation of single cells and organisms using acoustic waves, *Nature Comm.* **7**, 11085.

Backus, G. 1958 A class of self-sustaining dissipative spherical dynamos, *Ann. Phys.* **4**, 372-447.

Backus, G. 1986 Poloidal and toroidal fields in geomagnetic field modelling. *Reviews of Geophysics* **24,** 75–109.

Bertin, N., Spelman, T.A., Stephan, O., Gredy, L., Bouriau, M., Lauga, E. & Marmottant, P. 2015 Propulsion of bubble-based acoustic microswimmers, *Physical Review Applied* **4**, 064012.

Birkin, P.R., Offin D.G. & Leighton T.G. 2016 An activated fluid stream - New techniques for cold water cleaning, *Ultrason. Sonochem.* **29**, 612-618.

Bolanos-Jimenez, R., Rossi, M., Fernandez Rivas, D., Kähler, C.J. & Marin, A. 2017 Streaming flow by oscillating bubbles: quantitative diagnostics via particle tracking velocimetry, *Journal of Fluid Mechanics* **820**, 529-548.

Boyce, W.E. & DiPrima, R.C. 2001 *Elementary Differential Equations and Boundary Value Problems*. Wiley.

Bullard, E.C. & Gellman, H. 1954 Homogeneous dynamics and terrestrial magnetism, *Philos. Trans. R Soc. London, Ser. A* **247**, 213-278.

Chadwick, P. & Trowbridge, E.A. 1967 Elastic wave fields generated by scalar wave functions, *Proc. Camb. Phil. Soc.* **63**, 1177-1187.

Chandrasekhar, S. 1961 *Hydrodynamics and hydromagnetic stability*, Clarendon Press, Oxford.

Chang, C.T., Bostwick, J.B., Daniel S. & Steen, P.H. 2015 Dynamics of sessile drops. Part 2. Experiment, *Journal of Fluid Mechanics* **768**, 442-467.

Cleve, S., Guedra, M., Mauger, C., Inserra, C. & Blanc-Benon, P. 2019 Microstreaming induced by acoustically trapped, non-spherically oscillating microbubbles, *Journal of Fluid Mechanics* **875**, 597-621.

Davidson, B.J. & Riley, N. 1971 Cavitation microstreaming, *Journal of Sound and Vibration* **15**(2), 217-233.

Dijkink, R.J., van der Dennen, J.P., Ohl, C.D. & Prosperetti, A. 2006 The 'acoustic scallop': a bubble powered actuator. *J. Micromech. Microeng.* **16**, 1653-1659.

Doinikov, A.A., Cleve, S., Regnault, G., Mauger, C. & Inserra, C. 2019*a* Acoustic microstreaming produced by nonspherical oscillations of a gas bubble. I. Case of modes 0 and *m*. *Phys. Rev.* E **100**, 033104.

Doinikov, A.A., Cleve, S., Regnault, G., Mauger, C. & Inserra, C. 2019*b* Acoustic microstreaming produced by nonspherical oscillations of a gas bubble. II. Case of modes 1 and *m*. *Phys. Rev.* E **100**, 033105.

Elder, S. 1959 Cavitation microstreaming. *J. Acoust. Soc. Am.* 31, 54-64.

Elsasser, W.M. 1946 Induction effects in terrestrial magnetism, *Phys. Rev.* **69**, 106-116.

Fan, C.H., Liu, H.L., Ting, C.Y., Lee, Y.H., Huang, C.Y., Ma, Y.J., Wei, K.C., Yen, T.C. & Yeh, C.K. 2014 Submicron-bubble-enhanced focused ultrasound for blood-brain barrier disruption and improved CNS drug delivery, *PLoS ONE* **9** (5), e96327.

Fauconnier, M., Bera, J.C. & Inserra, C. 2020 Nonspherical modes nondegeneracy of a tethered bubble, *Physical Review E* **102**, 033108.

Fauconnier, M., Mauger, C., Bera, J.C. & Inserra, C. 2022 Nonspherical dynamics and microstreaming of a wall-attached microbubble, *Journal of Fluid Mechanics* **935**, A22.

Inserra, C., Regnault, G., Cleve, S., Mauger, C. & Doinikov, A.A. 2020a Acoustic microstreaming produced by nonspherical oscillations of a gas bubble. III. Case of self-interacting modes n-n. *Phys. Rev. E* **101**, 013111.

Inserra, C., Regnault, G., Cleve, S., Mauger, C. & Doinikov, A.A. 2020b Acoustic microstreaming produced by nonspherical oscillations of a gas bubble. IV. Case of modes n and m. *Phys. Rev. E* **102**, 043103.

Kim, W., Kim T-H., Choi, J. & Kim H-Y. 2009 Mechanism of particle removal by megasonic waves, *Appl. Phys. Lett.* **94**, 081908.

Lajoinie, G., De Cock, I., Coussios, C.C., Lentacker, I., Le Gac, S., Stride, E. & Versluis, M. 2016 In vitro methods to study bubble-cell interactions: Fundamentals and therapeutic applications, *Biomicrofluidics* **10**, 011501.

Lamb, H. 1881 On the oscillations of a viscous spheroid, *Proc. London Math. Soc.* **13**, 51-66.

Lamb, H. 1916 *Hydrodynamics*. Cambridge University Press.

Landau, L.D. & Lifshitz E.M. 1987 *Fluid Mechanics.* Pergamon Press.

Lee, K.H., Lee, J.H., Won, J.M., Rhee, K. & Chung, S.K. 2012 Micromanipulation using cavitational microstreaming generated by acoustically oscillating twin bubbles. *Sensors and Actuators A: Physical* **188**, 442–449.

Li, P., Collis, J.F., Brumley, D.R., Schneiders, L. & Sader, J.E. 2023 Structure of the streaming flow generated by a sphere in a fluid undergoing rectilinear oscillation. *Journal of Fluid Mechanics* **974**:A37.

Longuet-Higgins, M.S. 1998 Viscous streaming from an oscillating spherical bubble. *Proc. R. Soc. Lond. Ser.* A **454**, 725–742.

Maksimov, A.O. 2007 Viscous streaming from surface waves on the wall of acoustically-driven gas bubbles, *European Journal of Mechanics B/Fluids* **26**, 28-42.

Maksimov, A. 2020 Splitting of the surface modes for bubble oscillations near a boundary, *Physics of Fluids* **32**, 102104.

Marin, A., Rossi, M., Rallabandi, B., Wang, C., Hilgenfeldt, S. & Kähler, C.J. 2015 Three-dimensional phenomena in microbubble acoustic streaming, *Physical Review Applied* **3**, 041001.

Marmottant, P. & Hilgenfeldt, S. 2003 Controlled vesicle deformation and lysis by single oscillating bubbles. *Nature* **423**, 153-156.

Marmottant, P., Versluis, M., de Jong, N., Hilgenfeldt, S. & Lohse, D. 2006 High-speed imaging of an ultrasound-driven bubble in contact with a wall: "Narcissus" effect and resolved acoustic streaming, *Experiments in Fluids* **41**, 147-153.

Marmottant, P., Biben, T. & Hilgenfeldt, S. 2008 Deformation and rupture of lipid vesicles in the strong shear flow generated by ultrasound-driven microbubbles, *Proc. R. Soc. A* **464**, 1781-1800.

Mason, T.J. 1999 Sonochemistry. Vol. 2, ed. Oxford University Press, New York.

Mekki-Berrada, F., Thibault, P. & Marmottant, P. 2016 Acoustic pulsation of a microbubble confined between elastic walls, *Phys. Fluids* **28**, 032004.

Mie, G. 1908 Beiträge zur Optik trüber Medien, speziell kolloidaler Metallösungen, *Ann. Phys.* 25, 377-445.

Mohanty, S., Zhang, J., McNeill, J.M., Kuenen, T., Linde, F.P., Rouwkema, J. & Misra, S. 2021 Acoustically-actuated bubble-powered rotational micro-propellers, *Sensors and Actuators: B. Chemical* **347**, 130589.

Nyborg, W.L. 1958 Acoustic streaming near a boundary, *The Journal of the Acoustical Society of America* **30**, 329.

Ohl, C.D. & Wolfrum B. 2003 Detachment and sonoporation of adherent Hela-cells by shock wave-induced cavitation, *Biochimica et Biophysica Acta (BBA) - General Subjects* **1624**(1-3), 131-138.

Padmavati, B.S. & Amaranath, T. 2002 A note on decomposition of solenoidal fields, *Appl. Math. Lett.* **15**, 803-805.

Pereno, V., Aron, M., Vince, O., Mannaris, C., Seth, A., de Saint Victor, M., Lajoinie, G., Versluis, M., Coussios, C., Carugo, D. & Stride, E. 2018 Layered acoustofluidic resonators for the simultaneous optical and acoustic characterisation of cavitation dynamics, microstreaming, and biological effects, *Biomicrofluidics* **12**, 034109.

Prosperetti, A. 1977 Viscous effects on perturbed spherical flows. *Quart. Appl. Math*. **34**, 339–352.

Rallabandi, B., Wang, C. & Hilgenfeldt, S. 2014 Two-dimensional streaming flows driven by sessile semicylindrical microbubbles, *Journal of Fluid Mechanics* **739**, 57-71.

Regnault, G., Mauger, C., Blanc-Benon, P., Doinikov, A.A. & Inserra, C. 2021 Signatures of microstreaming patterns induced by non-spherically oscillating bubbles, *J. Acous. Soc. Am.* **150**(2), 1188-1197.

Reuter, F., Lauterborn S., Mettin, R. & Lauterborn W. 2017 Membrane cleaning with ultrasonically driven bubbles, *Ultrason. Sonochem.* **37**, 542-560.

Saint-Michel, B. & Garbin, V. 2020 Acoustic bubble dynamics in a yield-stress fluid, Soft Matter **16**, 10405.

Spelman, T.A. & Lauga, E. 2017 Arbitrary axisymmetric steady streaming: Flow, force and propulsion, *Journal of Engineering Mathematics* **105**, 31-65.

Tho, P., Manasseh, R. & Ooi, A. 2007 Cavitation microstreaming patterns in single and multiple bubble systems, *Journal of Fluid Mechanics* **576**, 191-233.

Varshalovich, D.A., Moskalev, A.N. & Khersonskii, V.K. 1988 *Quantum Theory of Angular Momentum*. World Scientific.

Volk, A., Rossi, M., Rallabandi, B., Kähler, C.J., Hilgenfeldt, S. & Marin, A. 2020 Size-dependent particle migration and trapping in three-dimensional microbubble streaming flows, *Phys. Rev. Fluids* **5**, 114201.

Wang, C., Rallabandi B. & Hilgenfeldt S. 2013 Frequency dependence and frequency control of microbubble streaming flows, *Physics of Fluids* **25**, 022002.

Wu, J. & Nyborg, W.L. 2008 Ultrasound, cavitation bubbles and their interaction with cells, *Advanced Drug Delivery Reviews* **60**(10), 1103-1116.

Zhou, Y., Dai, L. & Jiao, N. 2022 Review of bubble applications in microrobotics: Propulsion, manipulation, and assembly, *Micromachines* **13**, 1068.